\documentclass[conference]{IEEEtran}
\IEEEoverridecommandlockouts
\usepackage{times,url}
\usepackage{epsfig}
\usepackage{graphicx}
\usepackage{amsmath}
\usepackage{amssymb}
\usepackage{multicol,multirow}
\usepackage[font=footnotesize]{caption}
\usepackage{cite}
\usepackage{booktabs}
\usepackage{bigstrut}
\usepackage{floatrow}
\usepackage{rotating}
\usepackage{adjustbox}
\usepackage{afterpage}
\usepackage{xcolor}
\def\BibTeX{{\rm B\kern-.05em{\sc i\kern-.025em b}\kern-.08em
    T\kern-.1667em\lower.7ex\hbox{E}\kern-.125emX}}
\begin{document}

\title{MKIS-Net: A Light-Weight Multi-Kernel Network for Medical Image Segmentation}



\author{\IEEEauthorblockN{1\textsuperscript{st} Tariq M. Khan}
\IEEEauthorblockA{\textit{School of Computer Science and Engineering} \\
\textit{UNSW, Sydney, Australia.}\\
tariq045@gmail.com}
\and
\IEEEauthorblockN{2\textsuperscript{nd} Muhammad Arsalan}
\IEEEauthorblockA{\textit{Division of Electronics and Electrical Engineering} \\
\textit{Dongguk University, Seoul, Korea}\\
engineerarsal5@gmail.com}
\and
\IEEEauthorblockN{3\textsuperscript{rd} Antonio Robles-Kelly}
\IEEEauthorblockA{\textit{Defence Science and Technology Group,} \\
\textit{ Edinburgh, Australia.}\\
antonio.robles-kelly@deakin.edu.au}
\and
\IEEEauthorblockN{4\textsuperscript{th} Erik Meijering}
\IEEEauthorblockA{\textit{School of Computer Science and Engineering} \\
\textit{UNSW, Sydney, Australia.}\\
erik.meijering@unsw.edu.au
}}

\maketitle

\begin{abstract}
Image segmentation is an important task in medical imaging. It constitutes the backbone of a wide variety of clinical diagnostic methods, treatments, and computer-aided surgeries. In this paper, we propose a multi-kernel image segmentation net (MKIS-Net), which uses multiple kernels to create an efficient receptive field and enhance segmentation performance. As a result of its multi-kernel design, MKIS-Net is a light-weight architecture with a small number of trainable parameters. Moreover, these multi-kernel receptive fields also contribute to better segmentation results. We demonstrate the efficacy of MKIS-Net on several tasks including segmentation of retinal vessels, skin lesion segmentation, and chest X-ray segmentation. The performance of the proposed network is quite competitive, and often superior, in comparison to state-of-the-art methods. Moreover, in some cases MKIS-Net has more than an order of magnitude fewer trainable parameters than existing medical image segmentation alternatives and is at least four times smaller than other light-weight architectures.
\end{abstract}


\section{Introduction}
The aim of medical image segmentation is to facilitate the analysis of pathological or anatomical structural changes in patients. This often plays a critical role in computer-aided diagnosis and treatment \cite{lei2020medical}. Popular image segmentation tasks include retinal vessel \cite{khan2020semantically,khan2022width,khan2020exploiting,khan2020residual}, optic disc \cite{tabassum2020cded,khan2020region,imtiaz2021screening}, brain tumor \cite{rehman2019fully}, lung \cite{AITSKOURT2018109}, and skin lesion segmentation \cite{8941944}.

In the last decade, medical segmentation methods have been typically based on deep artificial neural networks. In \cite{7298965} fully convolutional networks (FCNs) are constructed from locally connected layers including convolution, pooling, and upsampling. In their FCN architecture, the authors employed two main components: the downsampling and upsampling stages. The path to the downsampling stage captures contextual information while the upsampling path retrieves spatial information. U-Net \cite{Ronneberger2015} originated from FCNs. The key difference is that each upsampling stage is connected via a concatenation operator to its corresponding downsampling stage. This way, each upsampling stage inherits the details on the corresponding downsampling stage, which leads to better segmentation performance. In SegNet, a novel technique is used for the upsampled encoder output, which involves storing the max-pooling indices at the pooling layer \cite{Badrinarayanan2017}.

While effective, these methods can be quite computationally demanding. Indeed, designing an FCN architecture for more than one medical image segmentation application in a resource constrained environment remains a challenging task. This is especially true since model size and computational requirements in terms of memory and processing power are important factors for real-world deployment of medical image segmentation methods, and medical imaging platforms often have limited computational resources available for complex or computationally costly operations.

To reduce the number of trainable parameters, model compression \cite{Rastegari2016} or binary network weights \cite{courbariaux2016binarized} have been used. On the other hand, light-weight networks with shallow architectures have been applied to real-world scenarios that require real-time prediction and decision tasks \cite{Li_2018_ECCV,howard2017mobilenets}. Some have proposed to use network compression to remove redundancies for a pretrained model through pruning techniques \cite{Li_2018_ECCV}. In \cite{Rastegari2016,courbariaux2016binarized} only a few bits are used to represent learned model weights instead of high-precision floating point numbers. These approaches do modify the network structure and often come at the cost of poor segmentation performance. Light-weight CNNs, on the other hand, are computationally cheap shallow networks with improved efficiency \cite{zhang2017shufflenet, howard2017mobilenets}. In such networks, to reduce the model size, convolutional factorization is often used. For example, depth-wise separable convolutions are employed by both ShuffleNet \cite{zhang2017shufflenet} and MobileNets \cite{howard2017mobilenets}, where 1$\times$1 point-wise and depth-wise convolutions are used instead of standard convolutions. In ERFNet \cite{Romera_2018}, a 2D convolution is decomposed into two 1D-factorized convolutions. Despite their success, convolutional factorisation can undermine the ability of light-weight networks to learn discriminative local structures, leading to performance degradation \cite{8265281}. We have recently proposed the RC-Net \cite{khan2021rc} and the more optimized T-Net \cite{khan2022t} light-weight networks, which use VGG16 as their base net. T-Net extracts low frequency features through the use of small pooling kernels. However, it performs poorly on datasets with high feature variation.

Here, we present a novel network architecture, called multi-kernel image segmentation network (MKIS-Net), which does not adopt a factorisation approach but is rather an FCN that uses a multi-kernel architecture. This produces an effective receptive field at the encoder stage which, in turn, yields better segmentation performance. Like other light-weight architectures, MKIS-Net is a shallow network. In contrast, however, it is devoid of pooling layers. This yields an architecture with a reduced loss of spatial information often associated with a reduction in feature map size.

\begin{figure*}[t]
  \centering
  \includegraphics[scale=0.4]{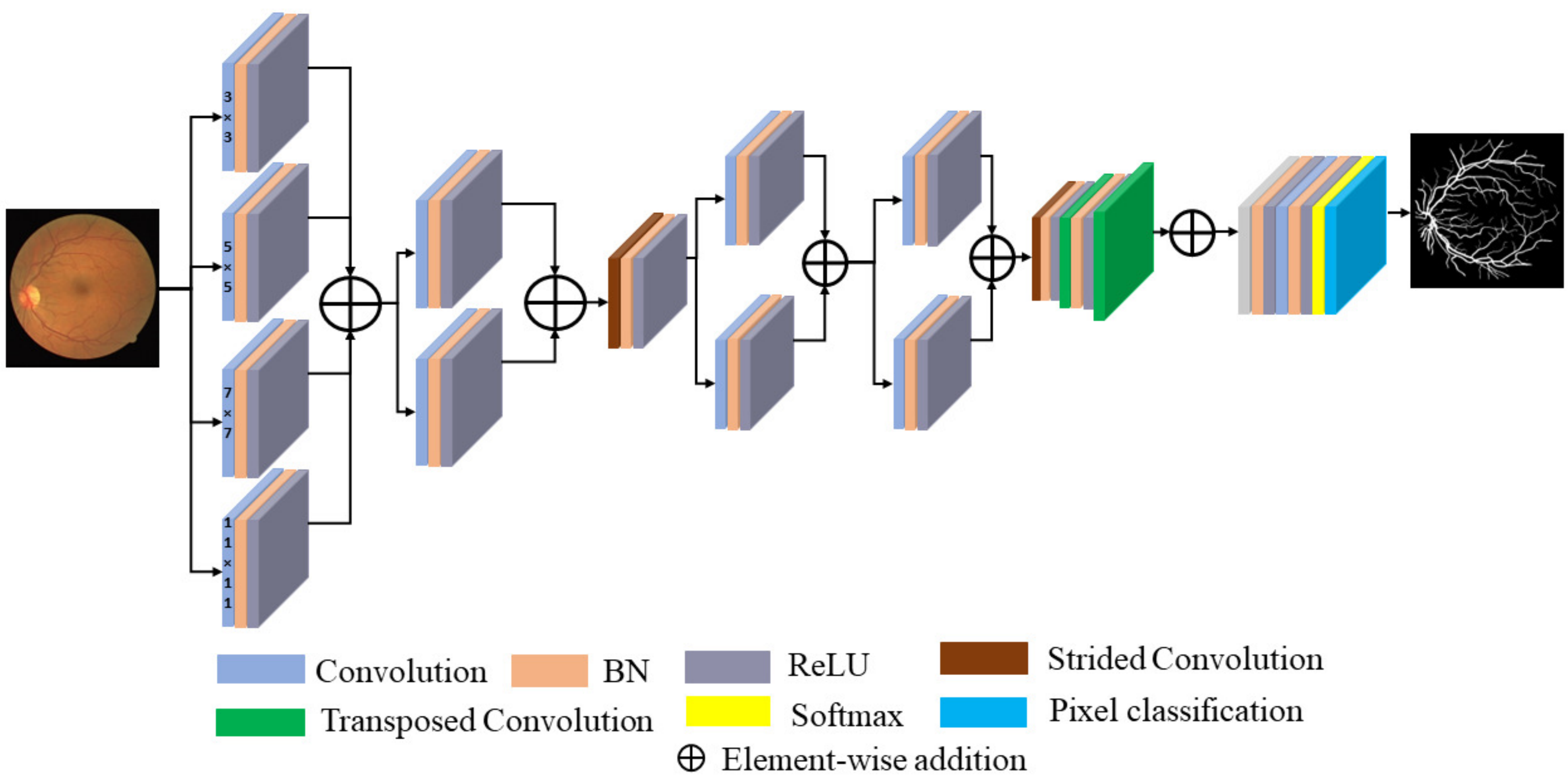}
  \caption{Architecture of the proposed MKIS-Net. It uses four convolutional input layers in parallel, with kernels of size 3$\times$3, 5$\times$5, 7$\times$7, and 11$\times$11 pixels, respectively, as indicated. BN = batch normalization. ReLU = rectified linear unit.}
  \label{ProposedArch}
\end{figure*}

\section{MKIS-Net Architecture}

The MKIS-Net architecture is motivated by the fact that, in the domain of computer-aided diagnosis, accuracy and computational cost go hand in hand. This is because diagnostic platforms often have limited computational resources while requiring high accuracy to be reliable tools to professionals in the field. Thus, our architecture addresses two important needs. Firstly, the need to keep the number of learnable parameters low. Reducing the number of layers and filters reduces the overall computational complexity. Secondly, the need to accurately process high-resolution imagery. This is particularly important in medical imaging, since fine-grained structures, such as vessels or alveoli, play an important role in the diagnosis of multiple conditions.

To motivate our architecture, we note that the feature maps extracted from a CNN kernel can be viewed, particularly in earlier stages of the network, as representative of lower-level features in the images. When these features are sparse and a large number of filters is used, the feature maps may contain overlapping features \cite{Ma_2018_ECCV}, which do not contribute to the performance. Also, frequent usage of pooling layers can cause spatial information loss \cite{8949760}. In the medical-related segmentation literature, the complexity of convolutional networks has therefore typically been reduced by proposing shallow nets, with a reduced number of layers \cite{Howard_2019_ICCV,Ma_2018_ECCV}. Thus, to address the mentioned issues, we designed a shallow multi-kernel architecture that reduces network complexity and avoids the use of pooling layers. In contrast with other light-weight approaches, our MKIS-Net adopts a multi-kernel architecture at each layer. This is motivated by the notion that smaller kernels provide receptive fields that cover smaller regions in the image, whereas larger ones provide a means for larger receptive fields in higher-resolution imagery. The main problem with larger kernels, however, is increased computational cost. Moreover, as a network grows deeper, these costs increase even further with the addition of layers to the network.

The network architecture of MKIS-Net (Fig.\ \ref{ProposedArch}) uses four convolutional input layers in parallel, with kernels of increasing size, 3$\times$3, 5$\times$5, 7$\times$7, and 11$\times$11 pixels. Large kernels in semantic segmentation, according to [1,] provide better per pixel approximation with an effective receptive field. It is true that using large kernels frequently in all blocks raises the overall cost of the network. As a result, we effectively utilized the benefits of a large kernel while not significantly increasing the number of trainable parameters.

This contrasts with other segmentation networks, which are based on a set of small, single-size kernels \cite{Badrinarayanan2017,Ronneberger2015}. After the input convolutional block, six more convolutional blocks are used. All the two-scale multi-kernel blocks employ 3$\times$3 and 5$\times$5 filters. Altogether, the number of convolutional layers is small and, as a result, the network is shallow, with a low computational complexity and small number of trainable parameters. While MKIS-Net is a shallow network, it adds in an element-wise fashion the receptive fields produced by each of the multi-kernel blocks. This has the effect of creating a multiscale feature map comprised by the receptive field of kernels of different sizes. Notice that our architecture is devoid of pooling layers. Therefore, to control the feature map size, we employ two strided convolutions, which yield a feature map size reduction by half. At the output stage, transposed convolutions and a softmax layer are used. As dropout helps to reduce the issues of over-fitting arising from shallow networks \cite{JMLR:v15:srivastava14a,9175266}, we have used dropout in the output block with a probability of 0.4. In our network we employ a classification layer to deliver a categorical label for each image pixel.

\section{Experimental Setup}\label{experimentalResults}
\subsection{Medical Image Datasets}

Our network was evaluated using three different medical image segmentation applications and four different datasets that are publicly available. These are retinal vessel segmentation, skin lesion segmentation, and chest X-ray segmentation. All considered datasets are equipped with manual annotations from experts to serve as the gold standard.

Regarding retinal vessel segmentation, we have evaluated our network utilizing two different publicly available datasets: DRIVE \cite{Staal2004}\footnote{https://drive.grand-challenge.org/} and CHASE \cite{Fraz2012c}\footnote{https://blogs.kingston.ac.uk/retinal/chasedb1/}. The DRIVE dataset was created by a Netherlands-based screening diabetic retinopathy program. It comprises 40 color images, 20 color images for training and 20 for testing, each having a resolution of 584$\times$565 pixels. Only 7 of these 40 images show little signs of mild early retinopathy. There is a binary field-of-view (FOV) mask associated with each individual image.

The CHASE dataset includes 28 color images from 14 British schoolchildren. Using a FOV of $30^\circ$ centered on optical disks, each image has a resolution of $999\times960$ pixels. The gold standard consists of two different manual segmentation maps. For our experiments, we used the first expert annotation. No dedicated training or test sets are available in the CHASE dataset. Thus, we chose to use the first 20 images as training images and the last 8 test images for testing. This is consistent with the approach taken by others \cite{M.Khan2020}.

For the skin lesion segmentation experiments, we used the ISBI 2016 Skin Lesion Challenge dataset \cite{8363547}\footnote{https://challenge.kitware.com/\#phase/566744dccad3a56fac786787} and the PH2 dataset \cite{Mendonca2013}\footnote{https://www.fc.up.pt/addi/ph2\%20database.html}. The ISBI 2016 dataset is part of a competition called ``Skin Lesion Analysis Towards Melanoma Detection''. This dataset contains 900 images of varying dimensions, all in a variety of formats. The PH2 dataset contains 200 dermoscopic images of $768\times560$ pixels each, acquired at the Hospital Petro Hispano, Matosinhos, Portugal.

Finally, for the chest X-ray segmentation task, we used the Montgomery County chest X-ray dataset (MC) \cite{Jaeger2014}\footnote{https://lhncbc.nlm.nih.gov/LHC-downloads/downloads.html}. This is a lung segmentation dataset consisting of 138 frontal chest X-ray images. The dataset was acquired by the Montgomery County tuberculosis program in Maryland, USA. It has 58 tuberculosis and 80 normal cases covering a wide range of abnormalities. A Eureka stationary X-ray machine was used to acquire the images, yielding high-resolution imagery of either $4,020\times4,892$ or $4,892\times4,020$ pixels.

\subsection{Augmentation and Training}
As the datasets used in the retinal vessel segmentation experiment are relatively small, we applied data augmentation. Specifically, each training image was rotated by one degree, and the brightness was adjusted in a random manner, both higher and lower, resulting in 7,600 images for the DRIVE and CHASE datasets. For the skin segmentation experiment, we used a protocol \cite{7942129} where the 900 images of the ISBI 2016 dataset are used for training, and the 200 images of the PH2 dataset for testing. For the chest X-ray segmentation experiment, we followed others \cite{Souza2019} and used 80 images for training purposes and the remaining images for testing.

For all our experiments we used a weighted cross-entropy loss and trained MKIS-Net using adaptive moment estimation (Adam) optimization \cite{kingma2014adam} with an initial learning rate of 0.001 and an exponential decay rate of 0.9. The maximum number of iterations during the training phase was set to 10. Note that different methods can be used for assigning loss weights. The method of median frequency balance was utilized here for the calculation of class association weights \cite{Badrinarayanan2017}.

\subsection{Evaluation Criteria}
The manually annotated gold-standard segmentation maps are binary, with each pixel corresponding to either an object of interest (retinal vessels, skin lesions, abnormalities) or the background. Thus, for each pixel in an output image, there are four possible outcomes: the pixel belongs to an object of interest and has been correctly predicted as such (true positive, TP), it belongs to the background and has been correctly predicted as such (true negative, TN), or it has been incorrectly predicted as an object pixel (false positive, FP) or as a background pixel (false negative, FN). Using these four categories of pixels, we computed common measures of performance \cite{Taha-2015, Reinke-2022}, depending on the experiment: sensitivity (Se), specificity (Sp), accuracy (Acc), Dice/F1-score (F1). For the skin lesion segmentation experiment, we used the F1 and the Jaccard coefficient (Jacc) so as to allow direct comparison with methods in literature\cite{Fan2020}. In the retinal vessel segmentation experiment, we also made use of the area under the receiver operating characteristic curve (AUC), as the datasets have an unbalanced distribution of positive and negative classes and AUC has been found to be a good measure to gauge how the model can separate those classes in segmentation problems \cite{Li2016}.

\begin{figure}[!htbp]
    \centering
    \resizebox{1\textwidth}{!}{%
    \begin{tabular}{@{}c@{\ }c@{\ }c@{\ }c@{\ }c@{}}
        \includegraphics[width=0.2\textwidth]{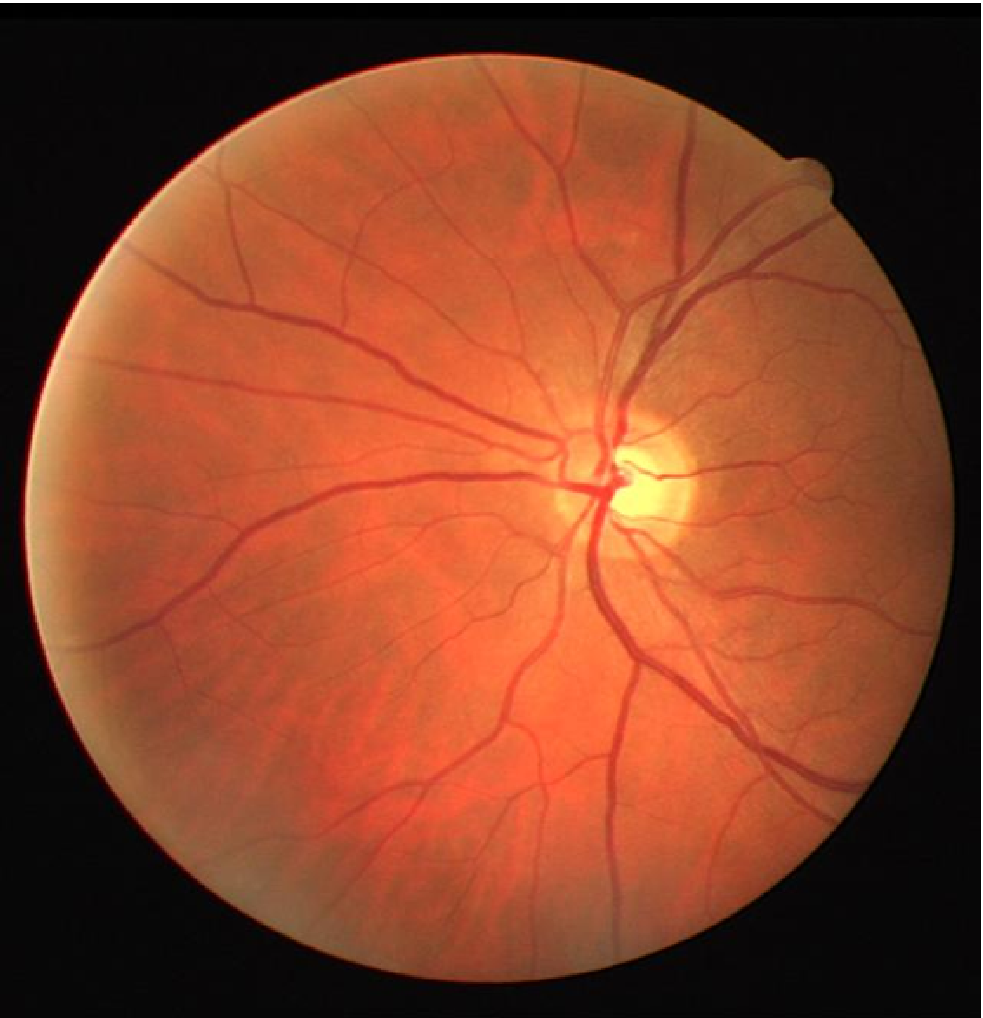} &
        \includegraphics[width=0.2\textwidth]{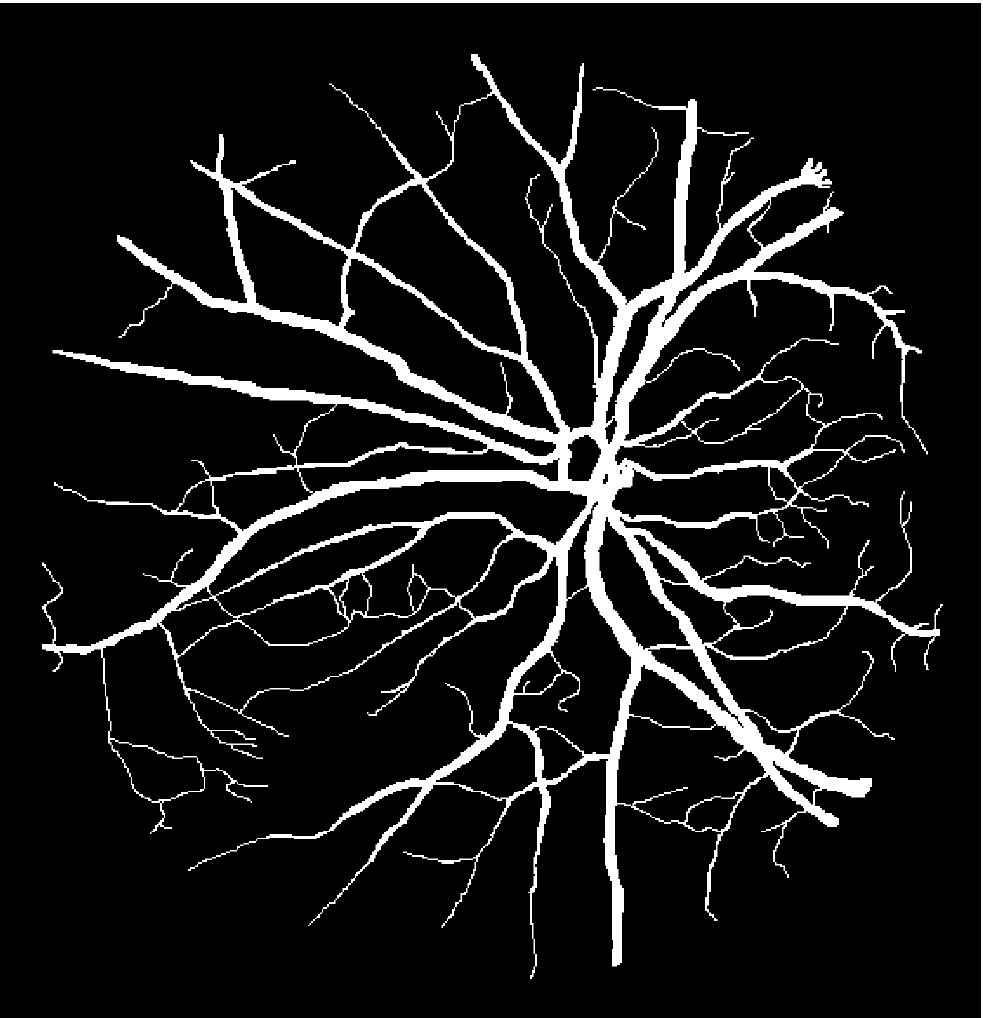} &
        \includegraphics[width=0.2\textwidth]{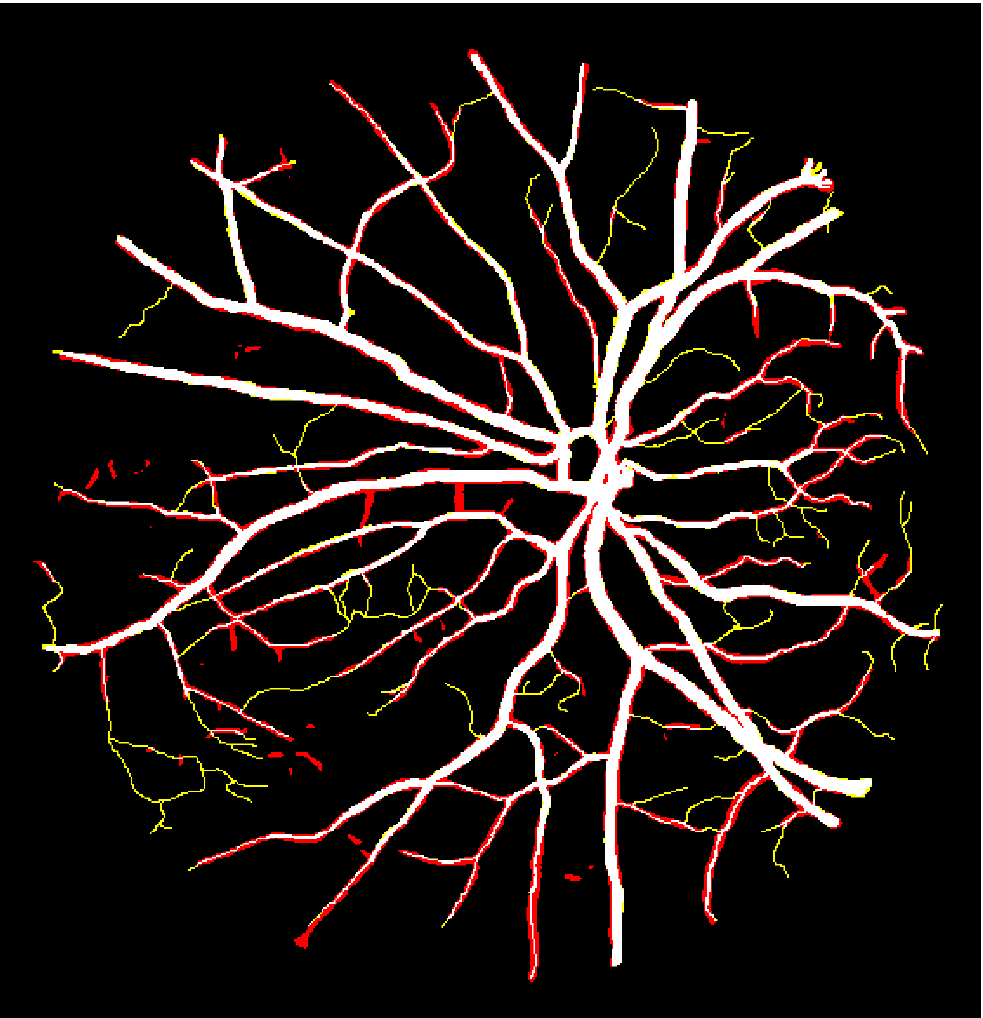} &
        \includegraphics[width=0.2\textwidth]{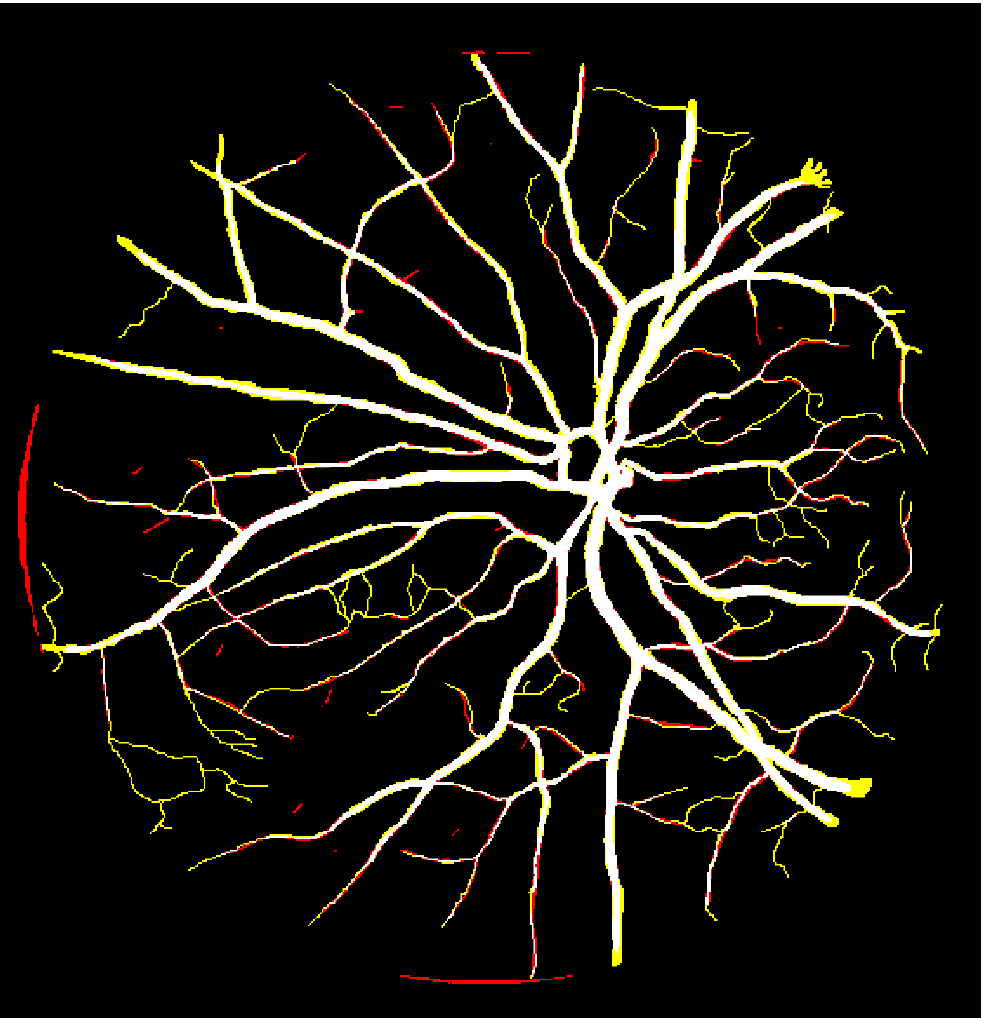} &
        \includegraphics[width=0.2\textwidth]{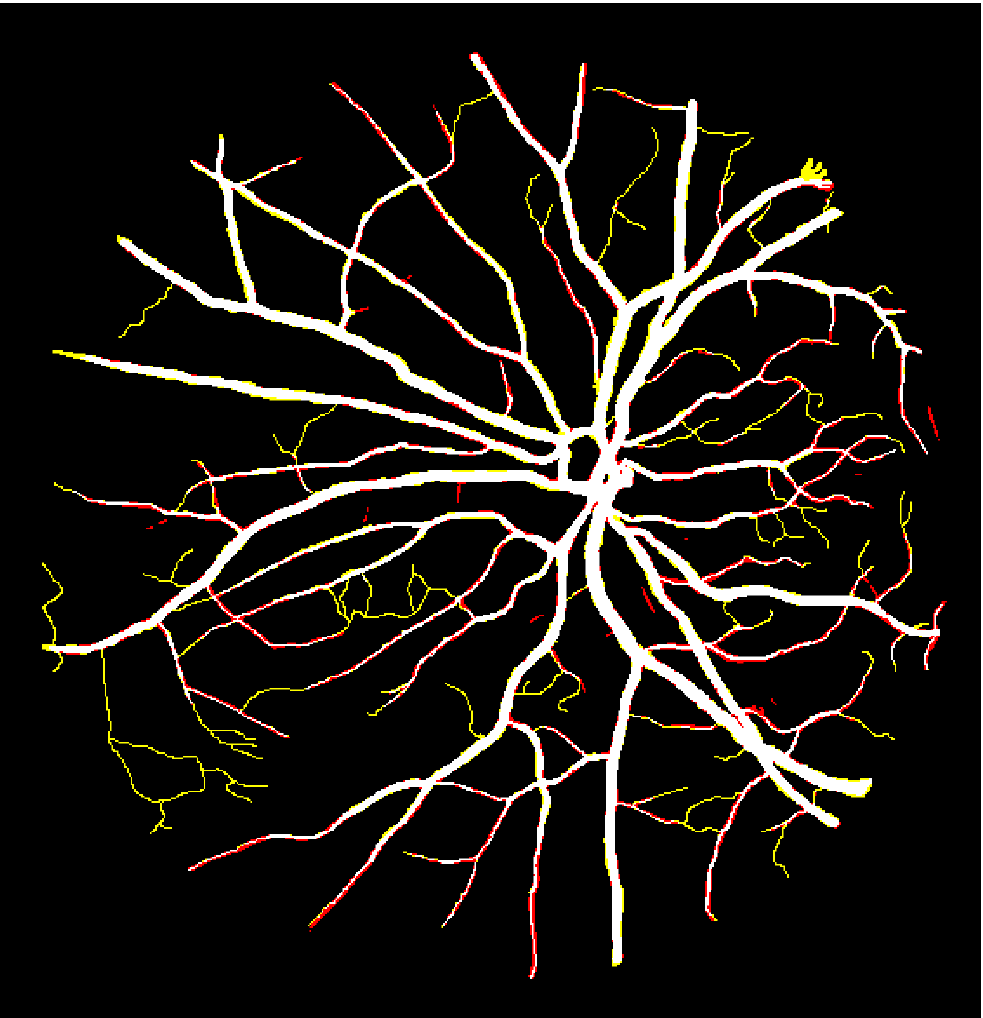} \\
        \includegraphics[width=0.2\textwidth]{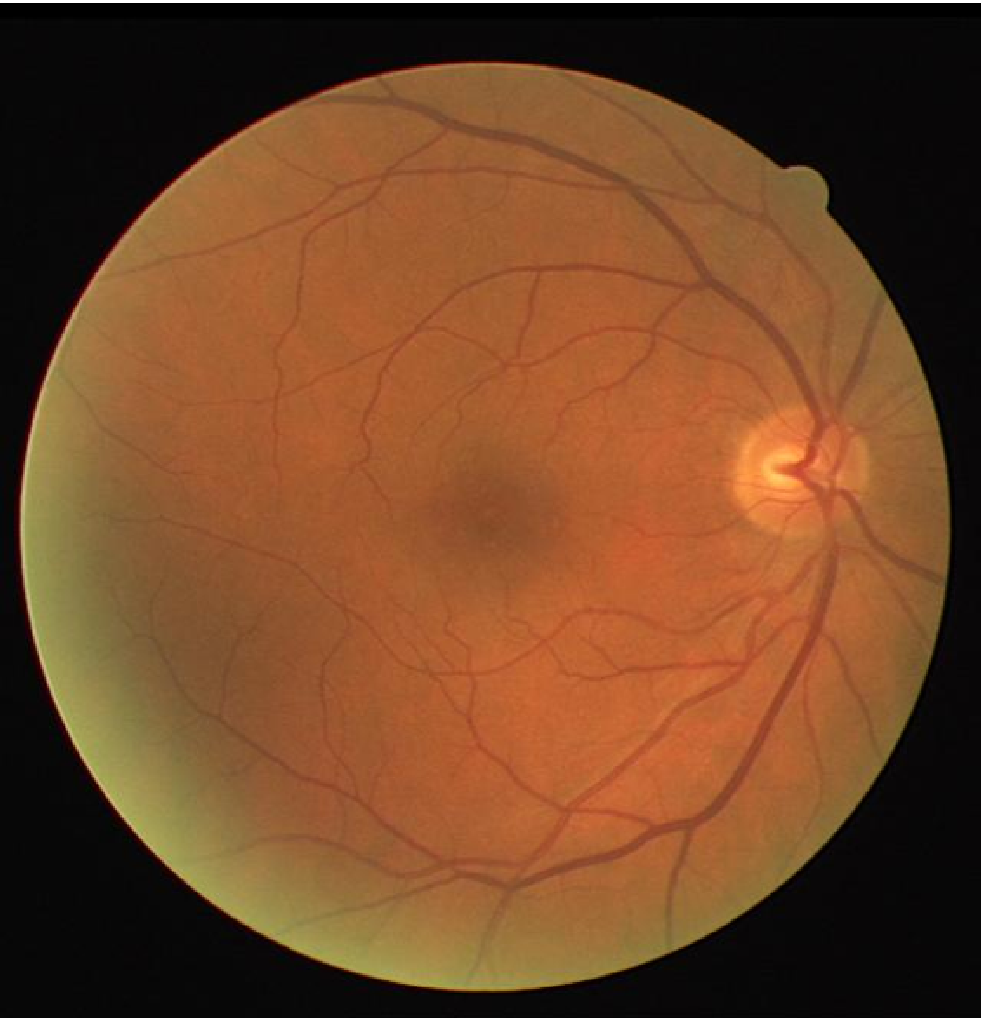} &
        \includegraphics[width=0.2\textwidth]{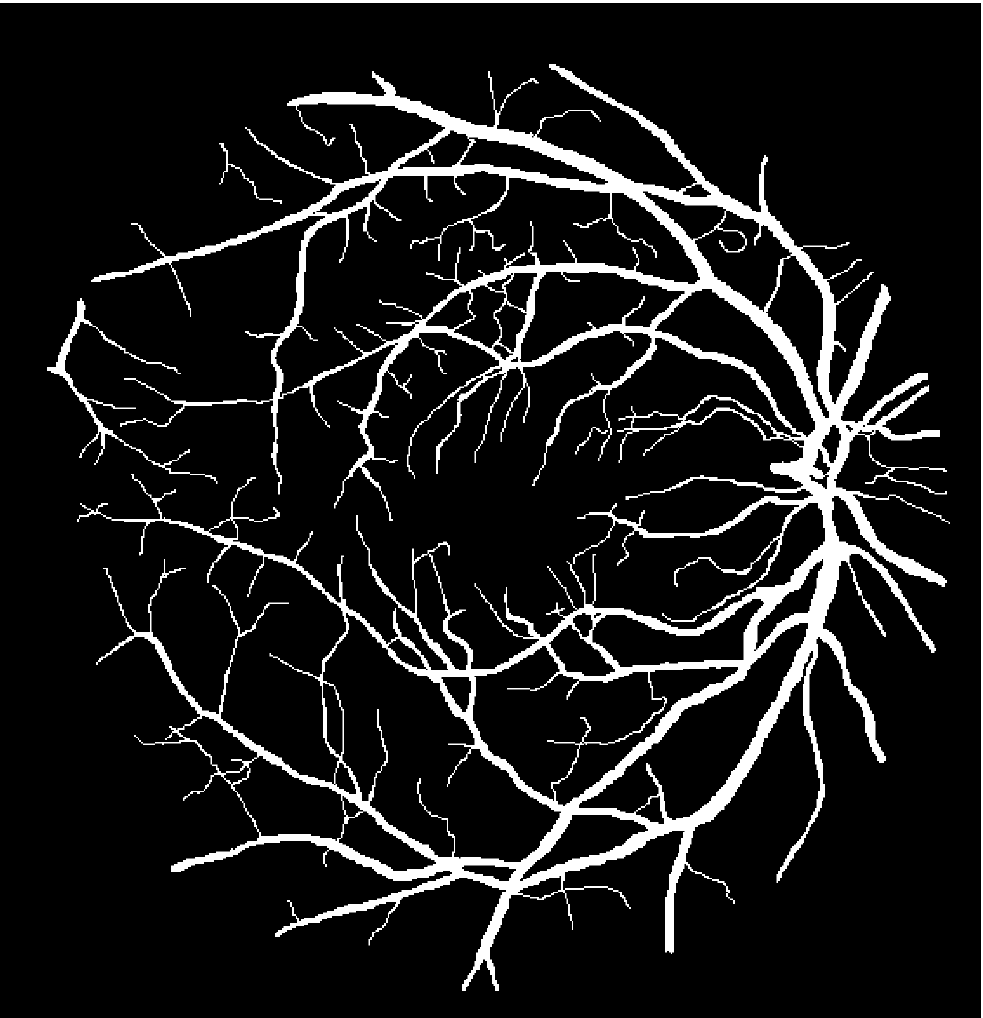} &
        \includegraphics[width=0.2\textwidth]{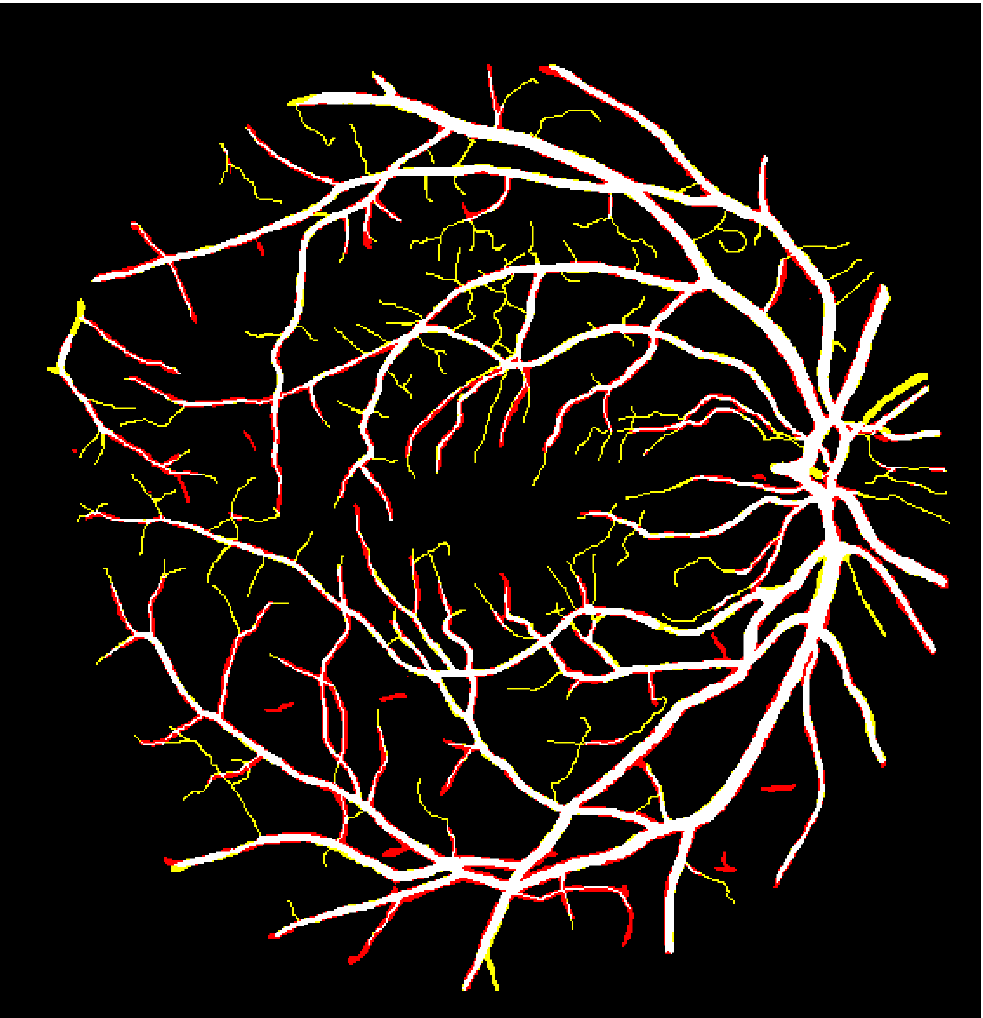} &
        \includegraphics[width=0.2\textwidth]{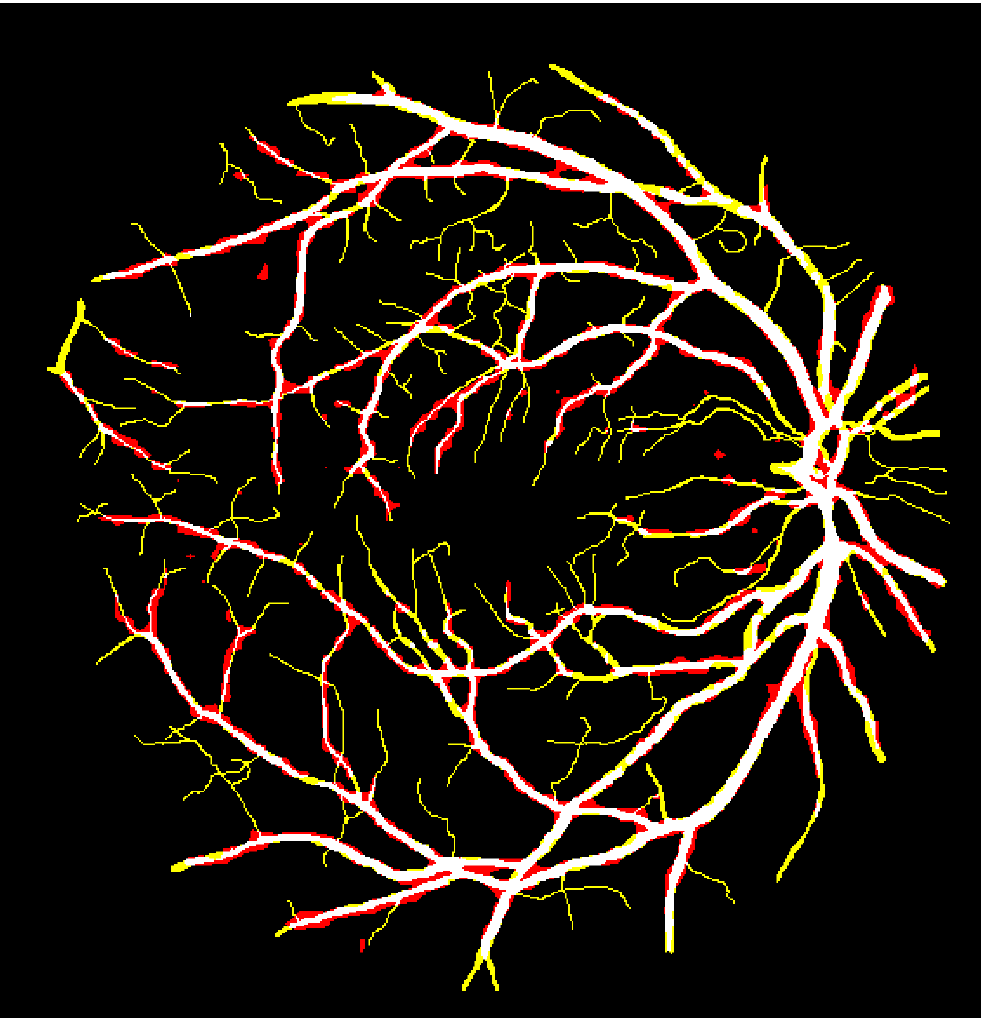} &
        \includegraphics[width=0.2\textwidth]{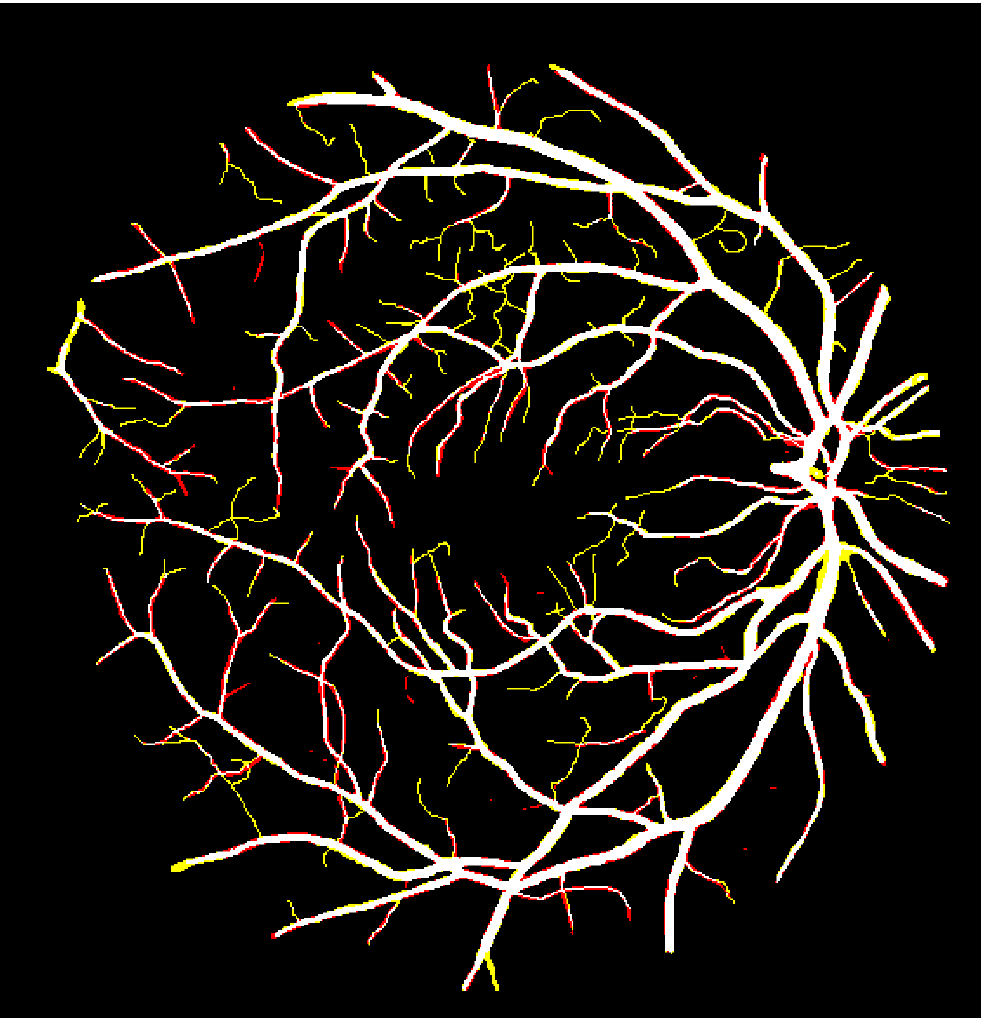} \\
        \includegraphics[width=0.2\textwidth]{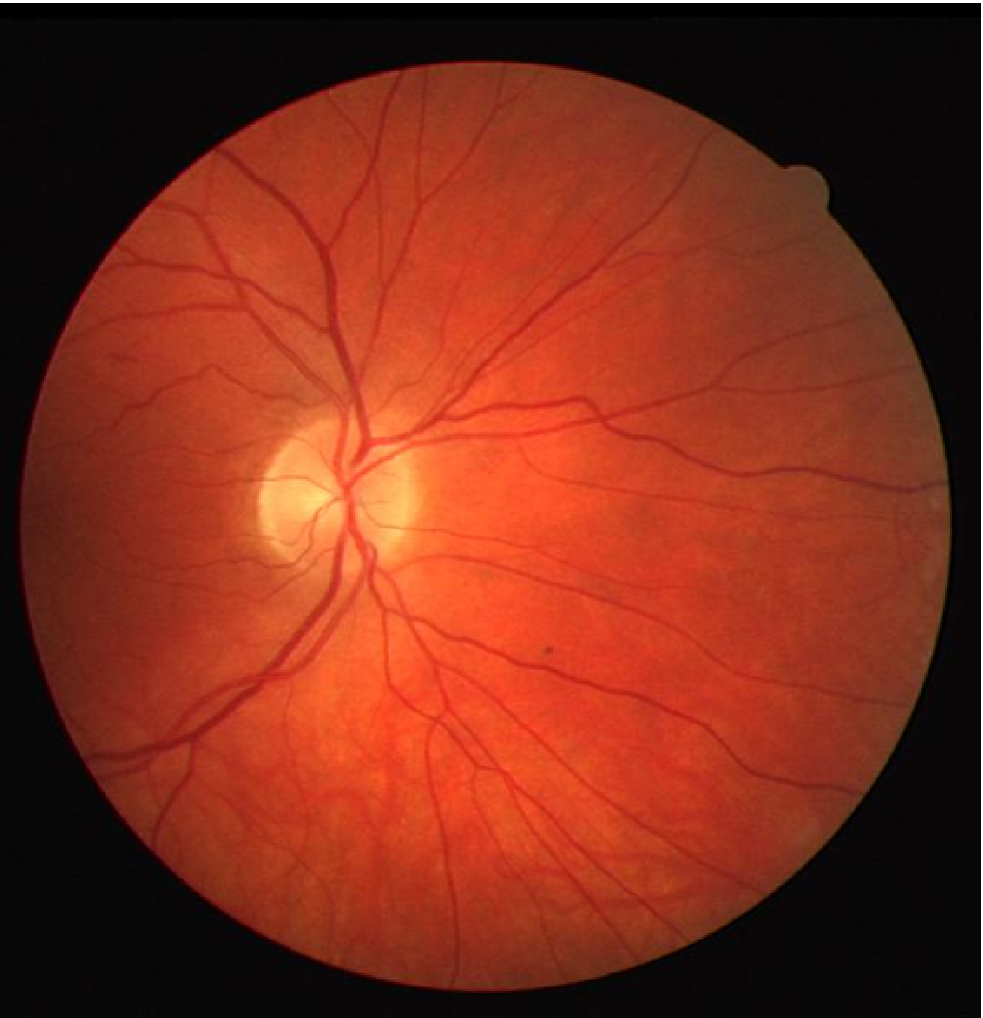} &
        \includegraphics[width=0.2\textwidth]{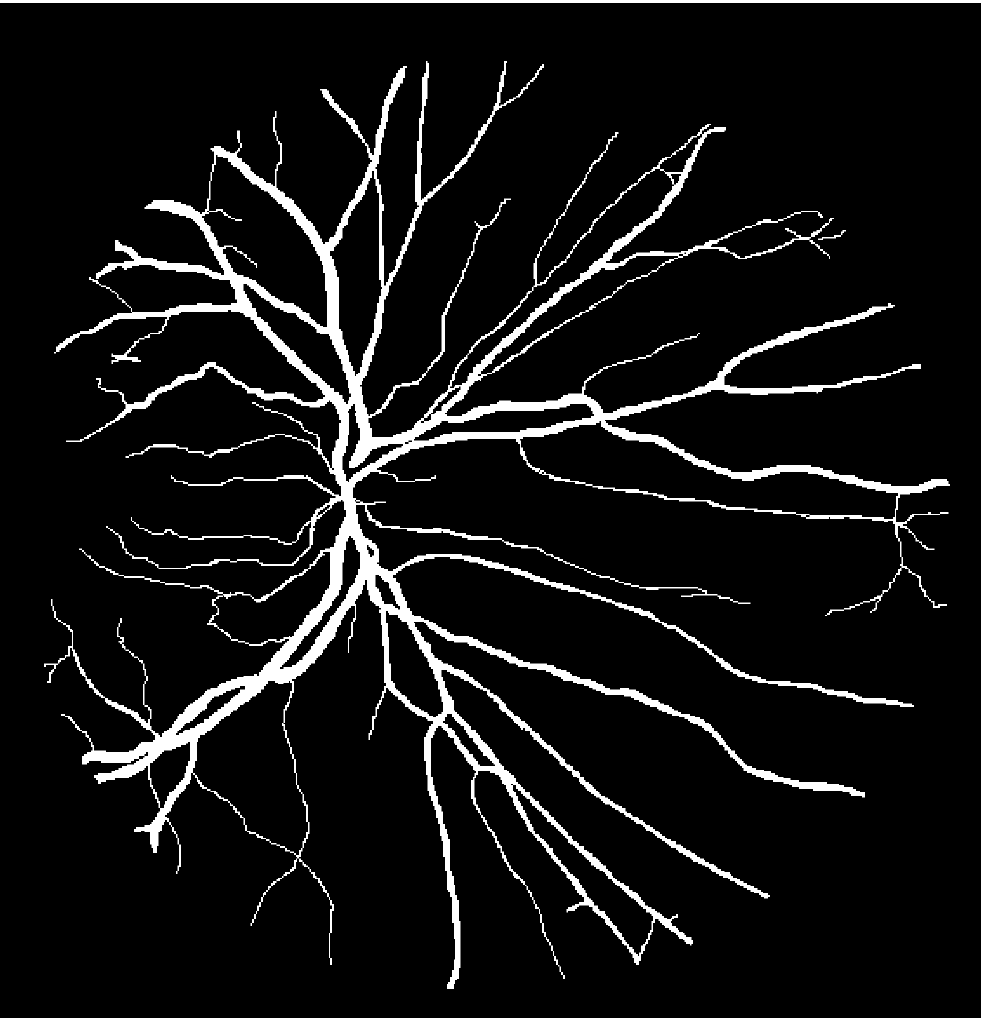} &
        \includegraphics[width=0.2\textwidth]{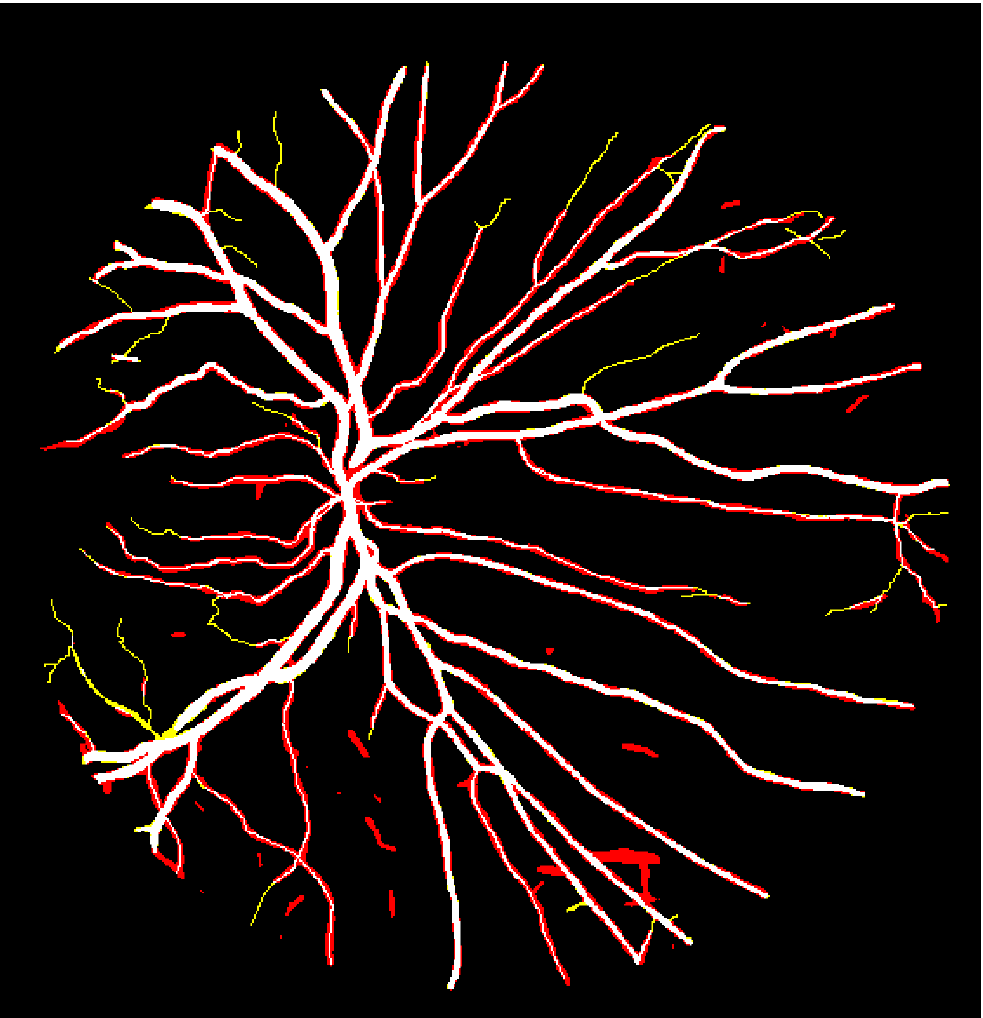} &
        \includegraphics[width=0.2\textwidth]{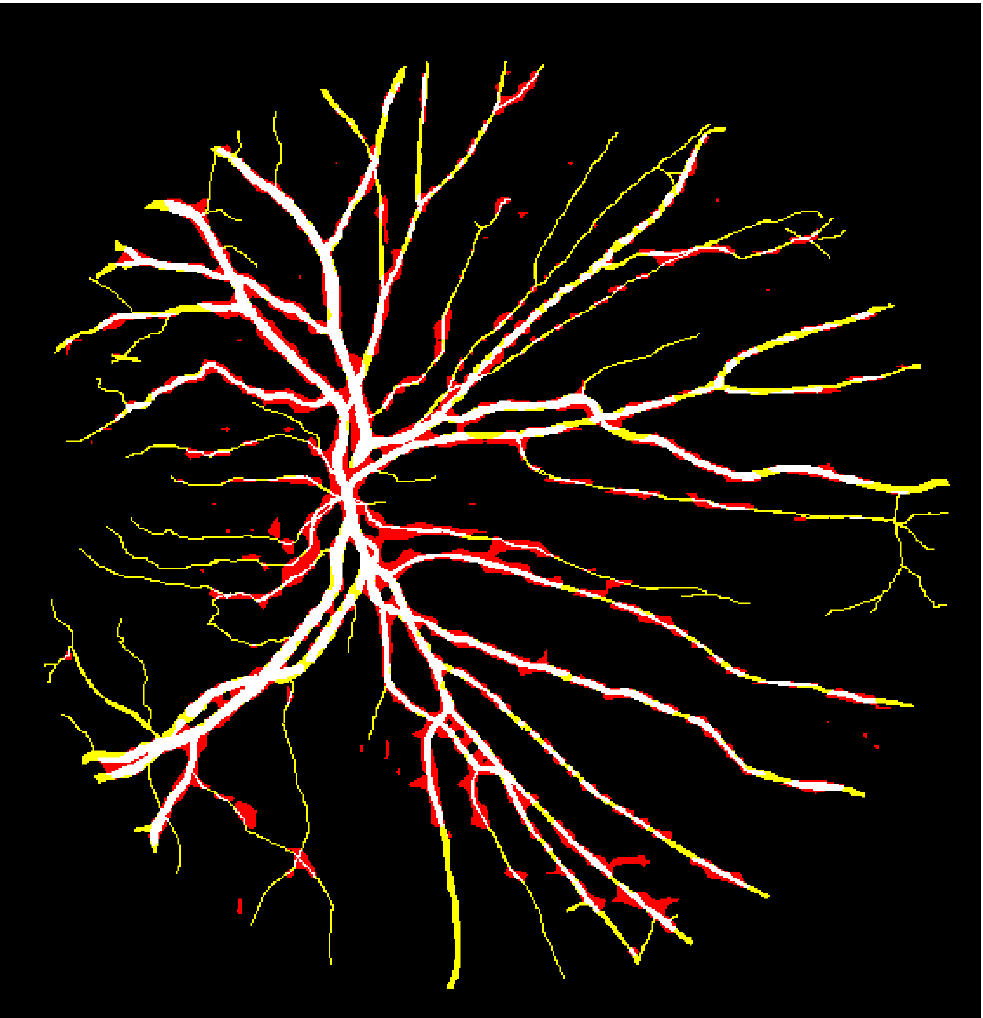} &
        \includegraphics[width=0.2\textwidth]{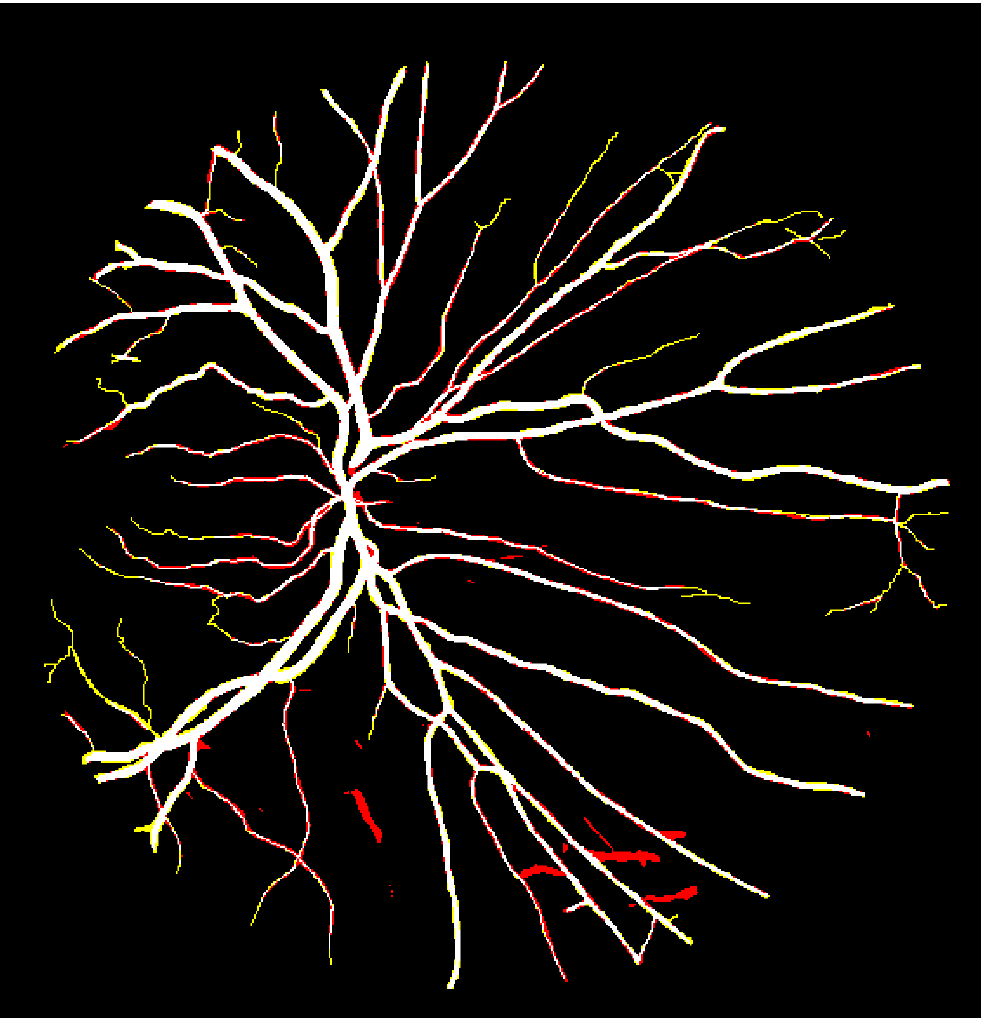} \\
    \end{tabular}}
    \caption{Sample visual results on the DRIVE dataset. From left to right: the input images, the manually annotated gold standard vessel maps, and the results generated by SegNet, U-Net, and the proposed MKIS-Net.}
    \label{visualDRIVE}
\end{figure}

\begin{figure}[!htbp]
    \centering
    \resizebox{1\textwidth}{!}{%
    \begin{tabular}{@{}c@{\ }c@{\ }c@{\ }c@{\ }c@{}}
        \includegraphics[width=0.2\textwidth]{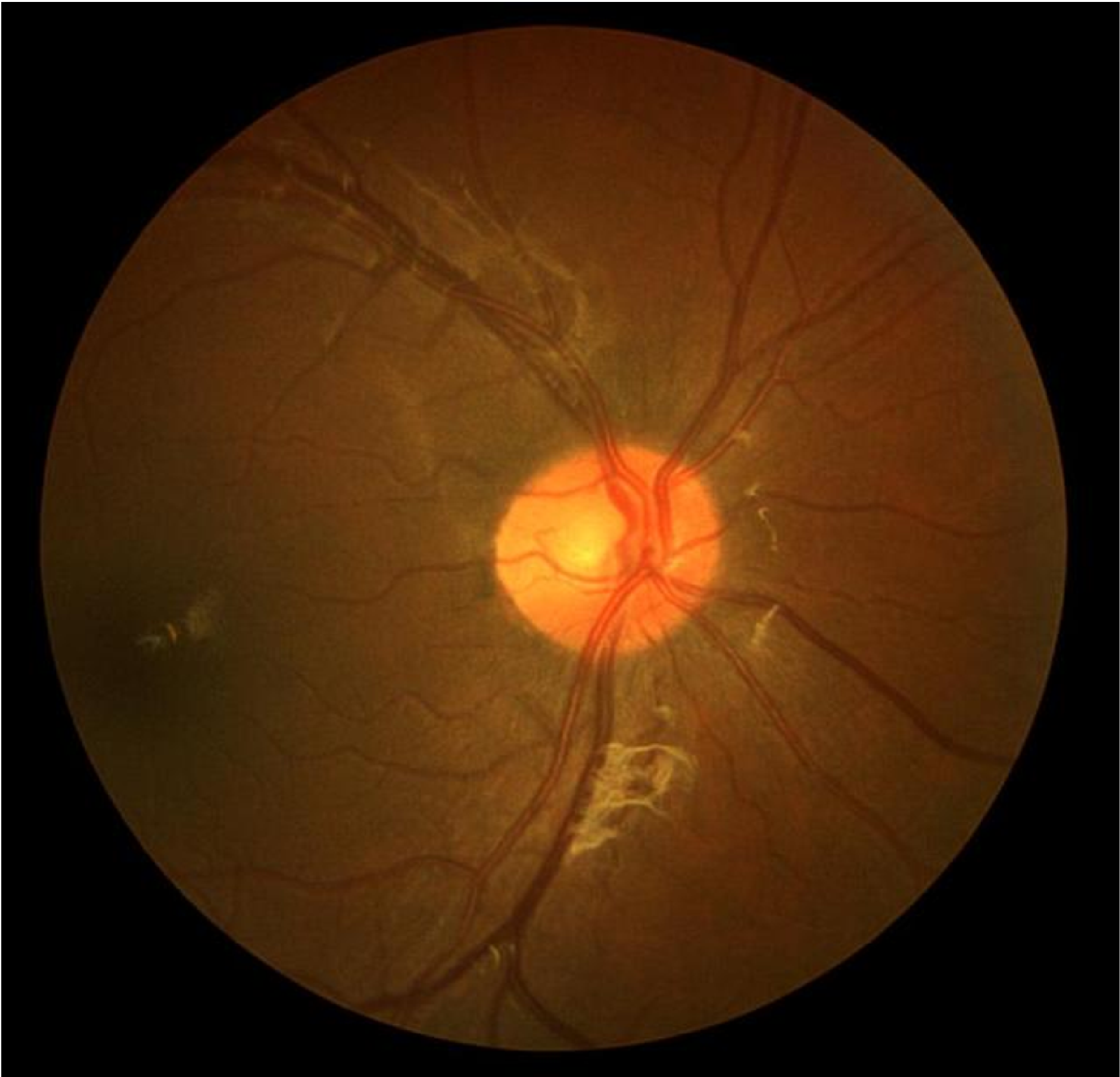} &
        \includegraphics[width=0.2\textwidth]{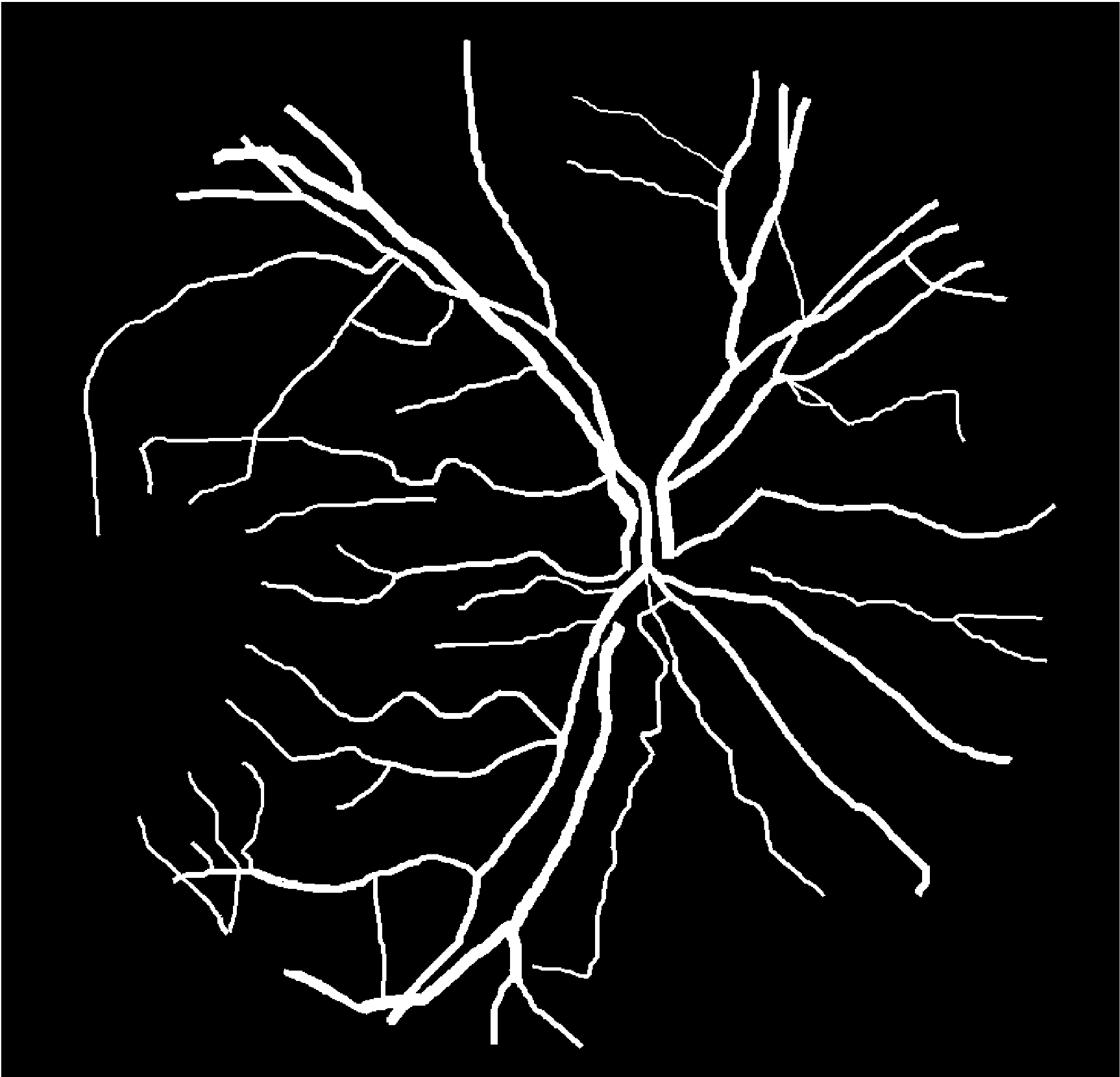} &
        \includegraphics[width=0.2\textwidth]{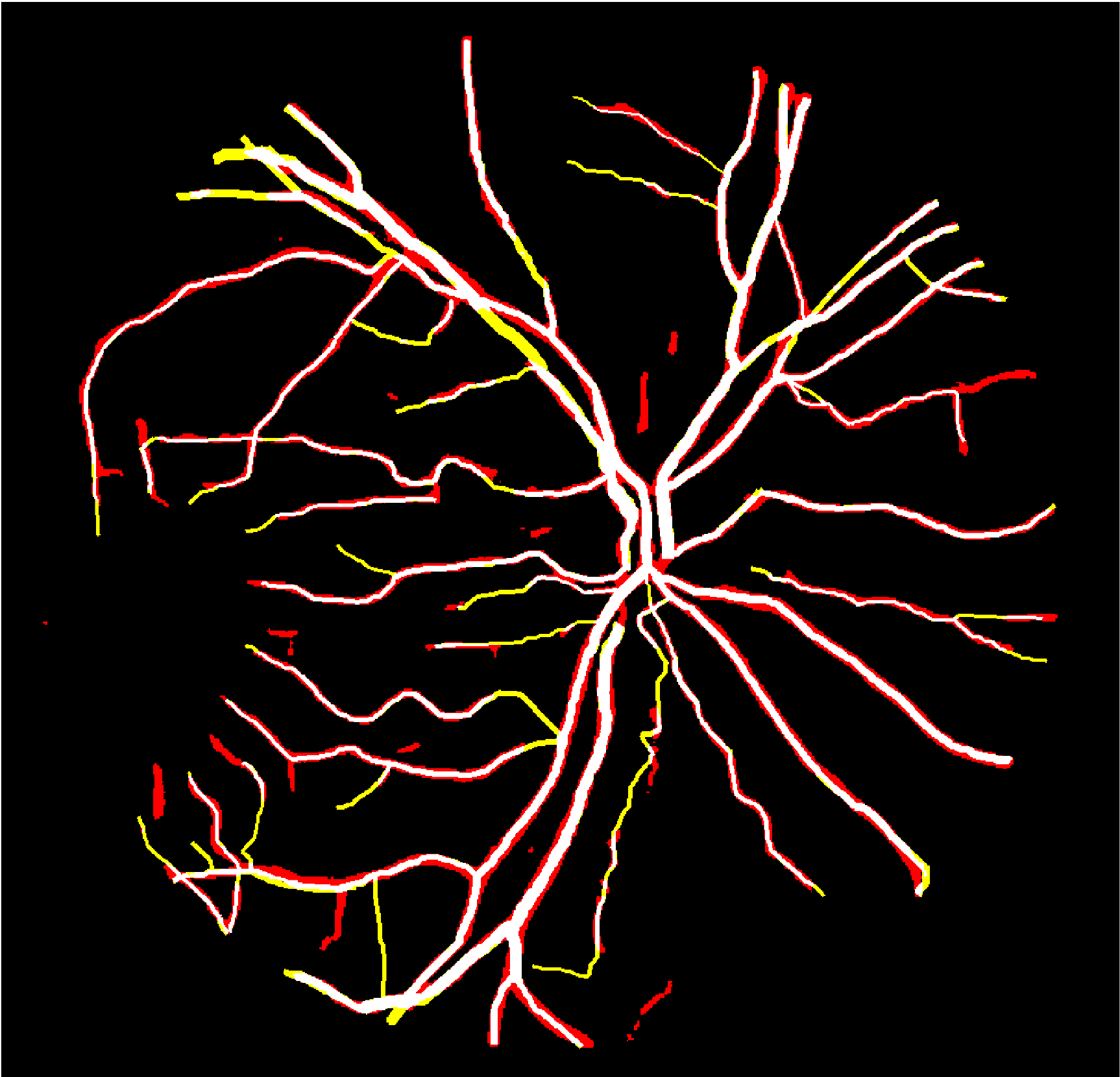} &
        \includegraphics[width=0.2\textwidth]{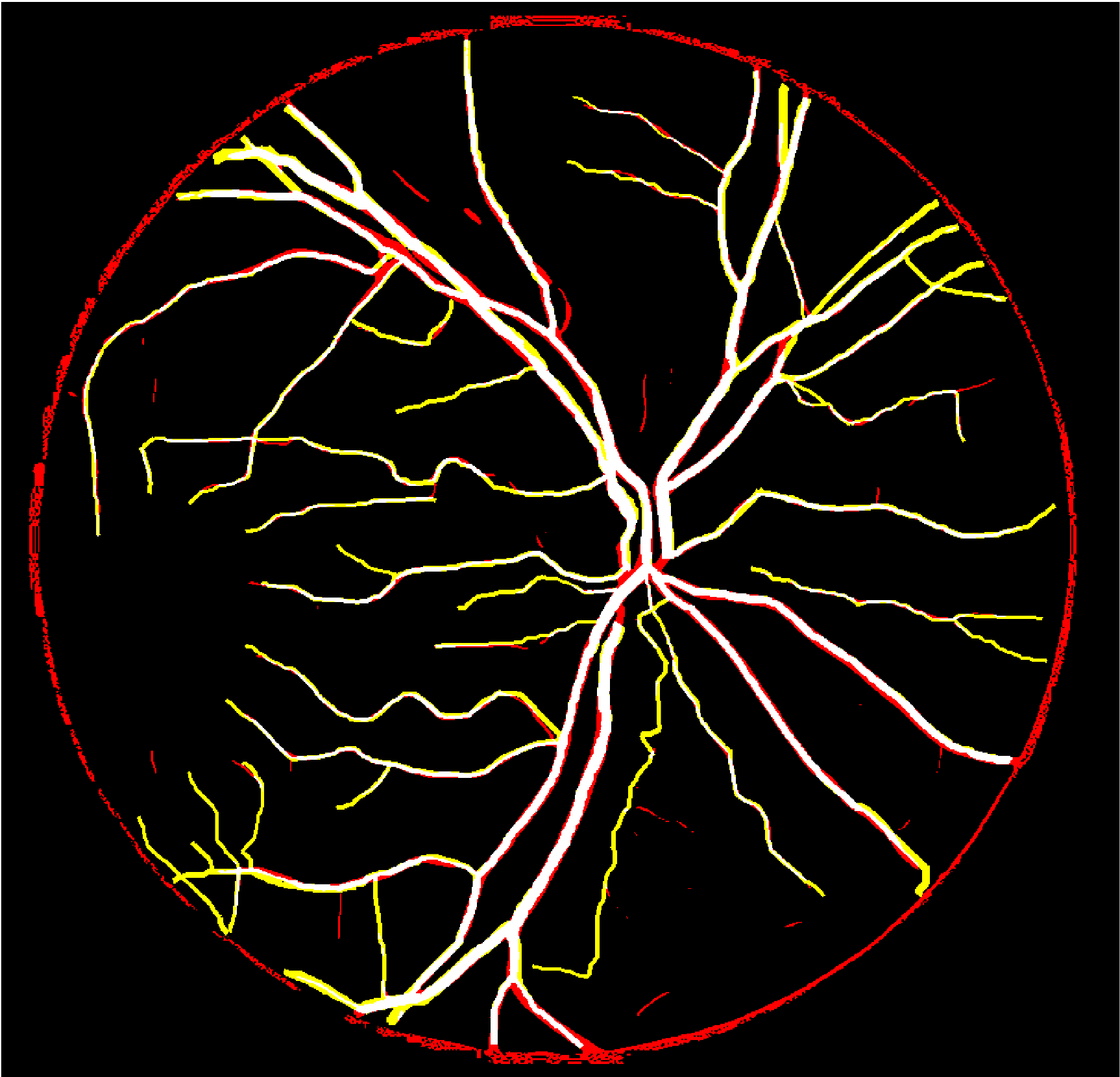} &
        \includegraphics[width=0.2\textwidth]{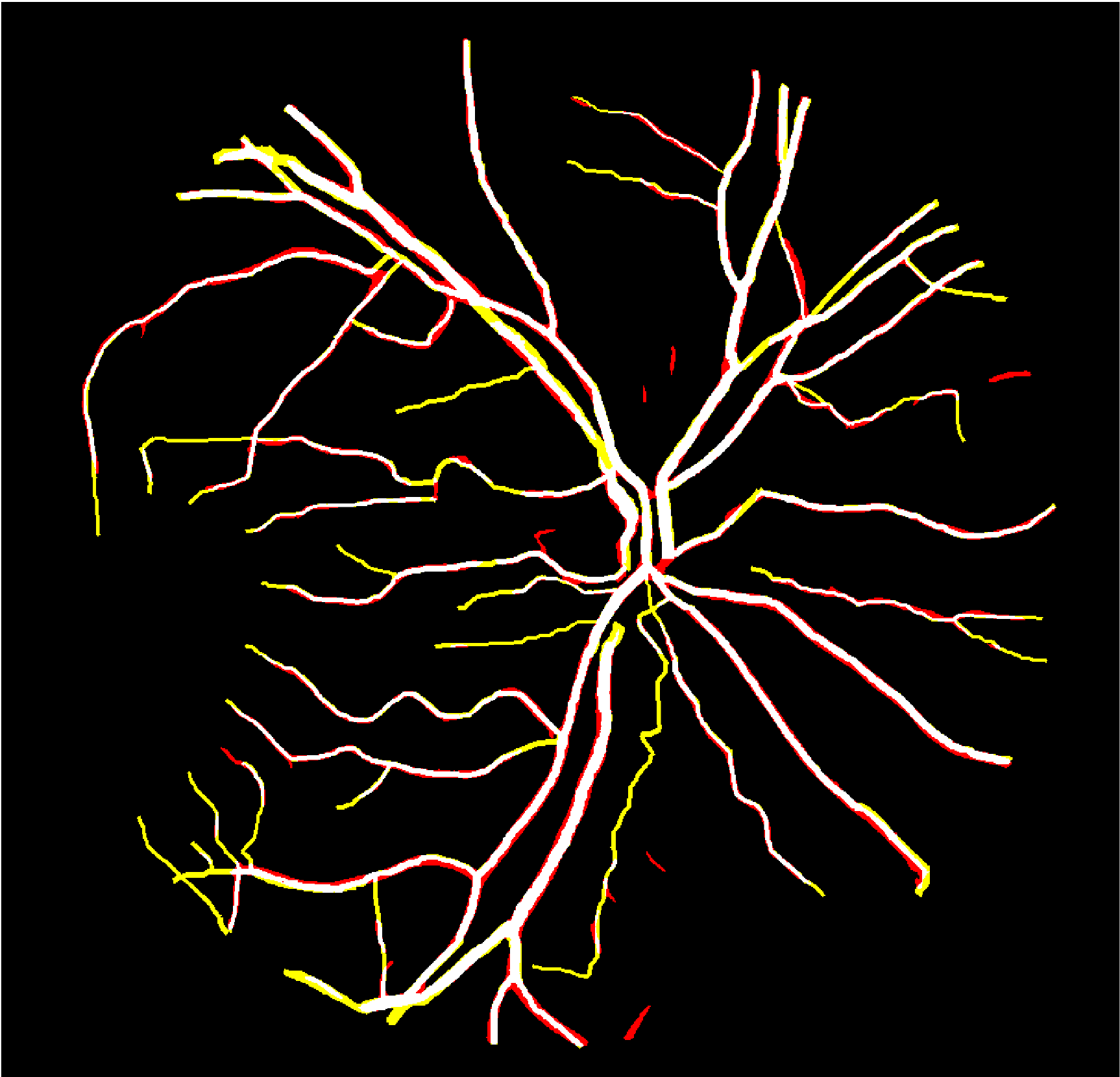} \\
        \includegraphics[width=0.2\textwidth]{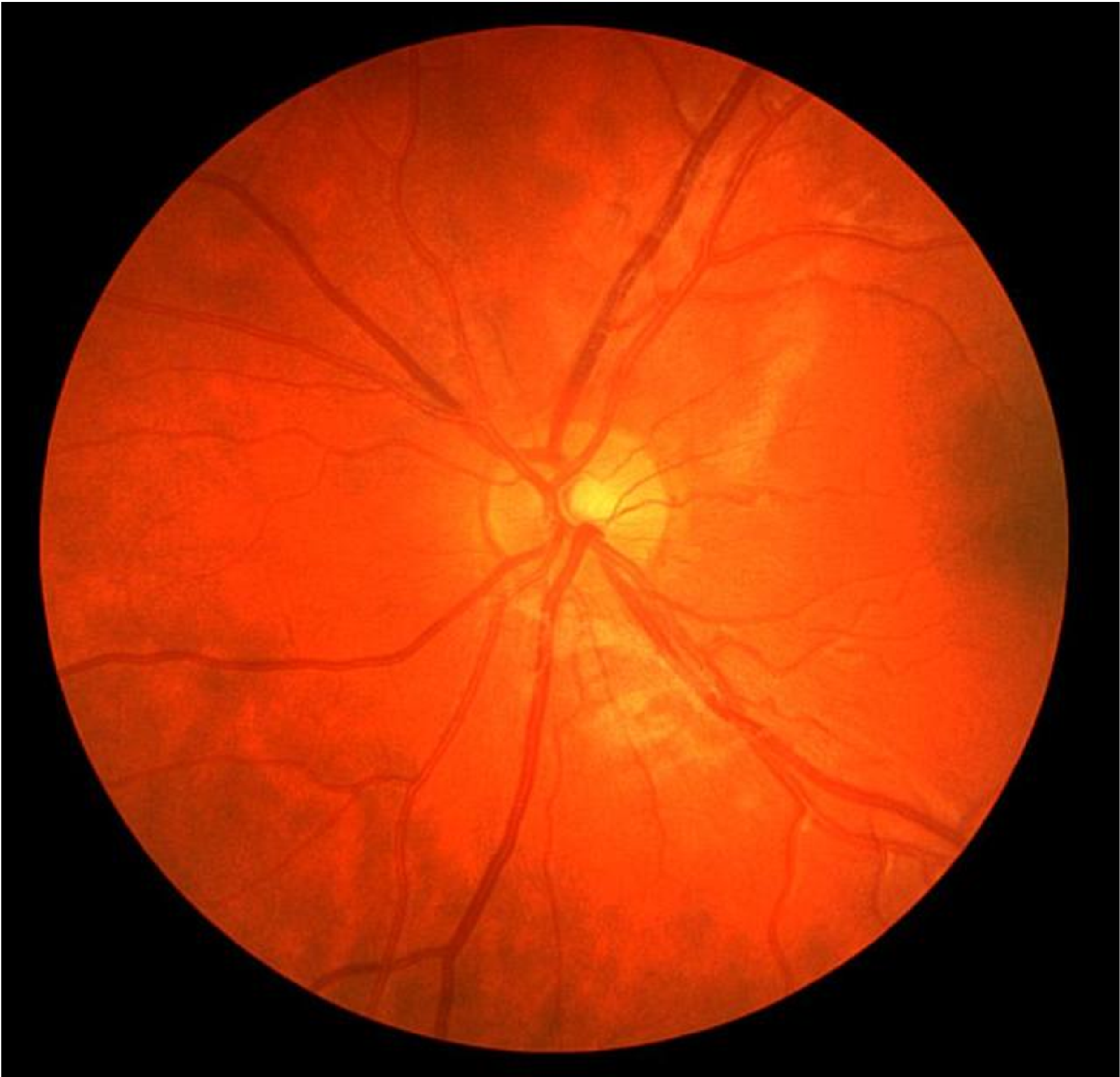} &
        \includegraphics[width=0.2\textwidth]{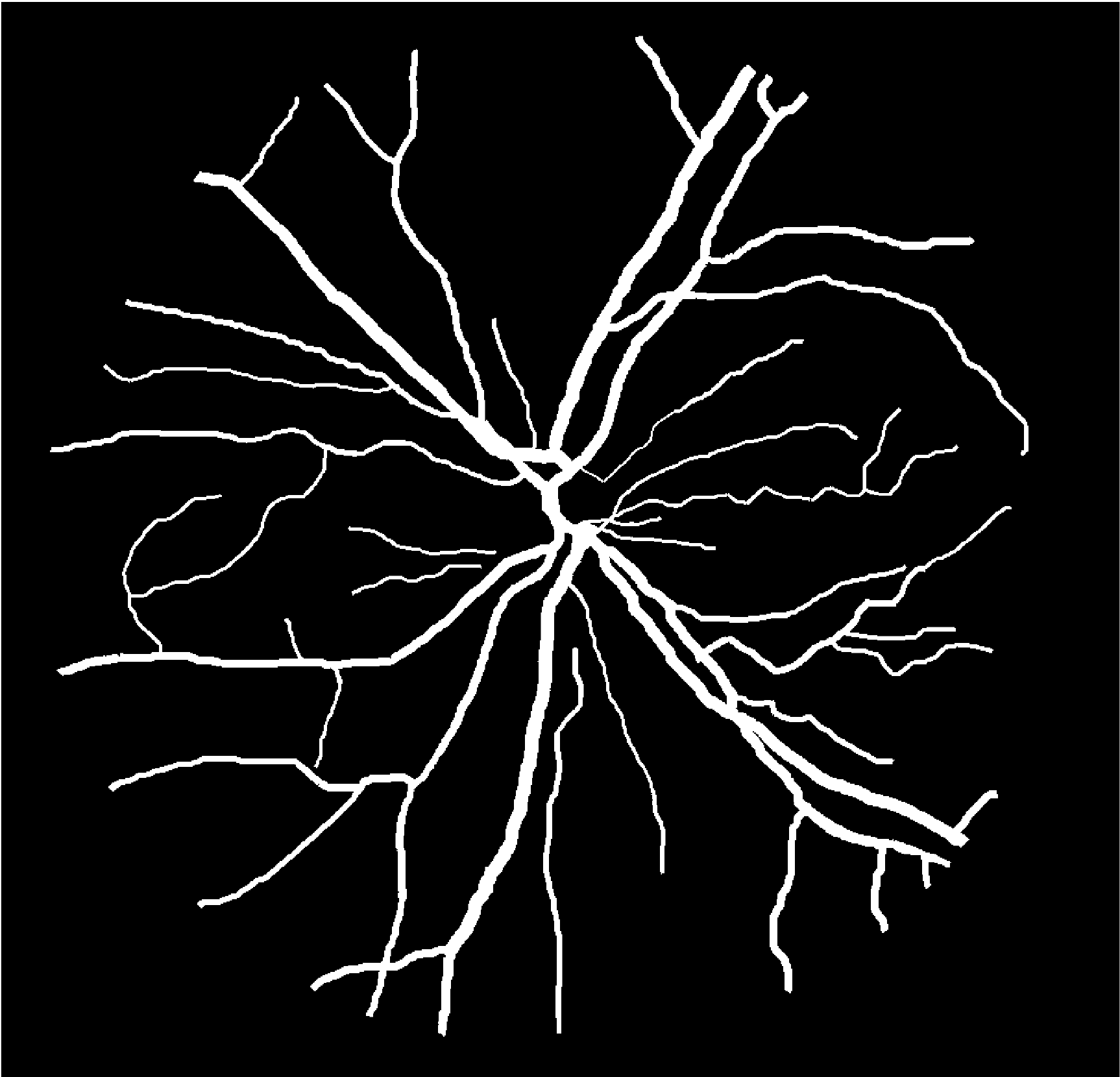} &
        \includegraphics[width=0.2\textwidth]{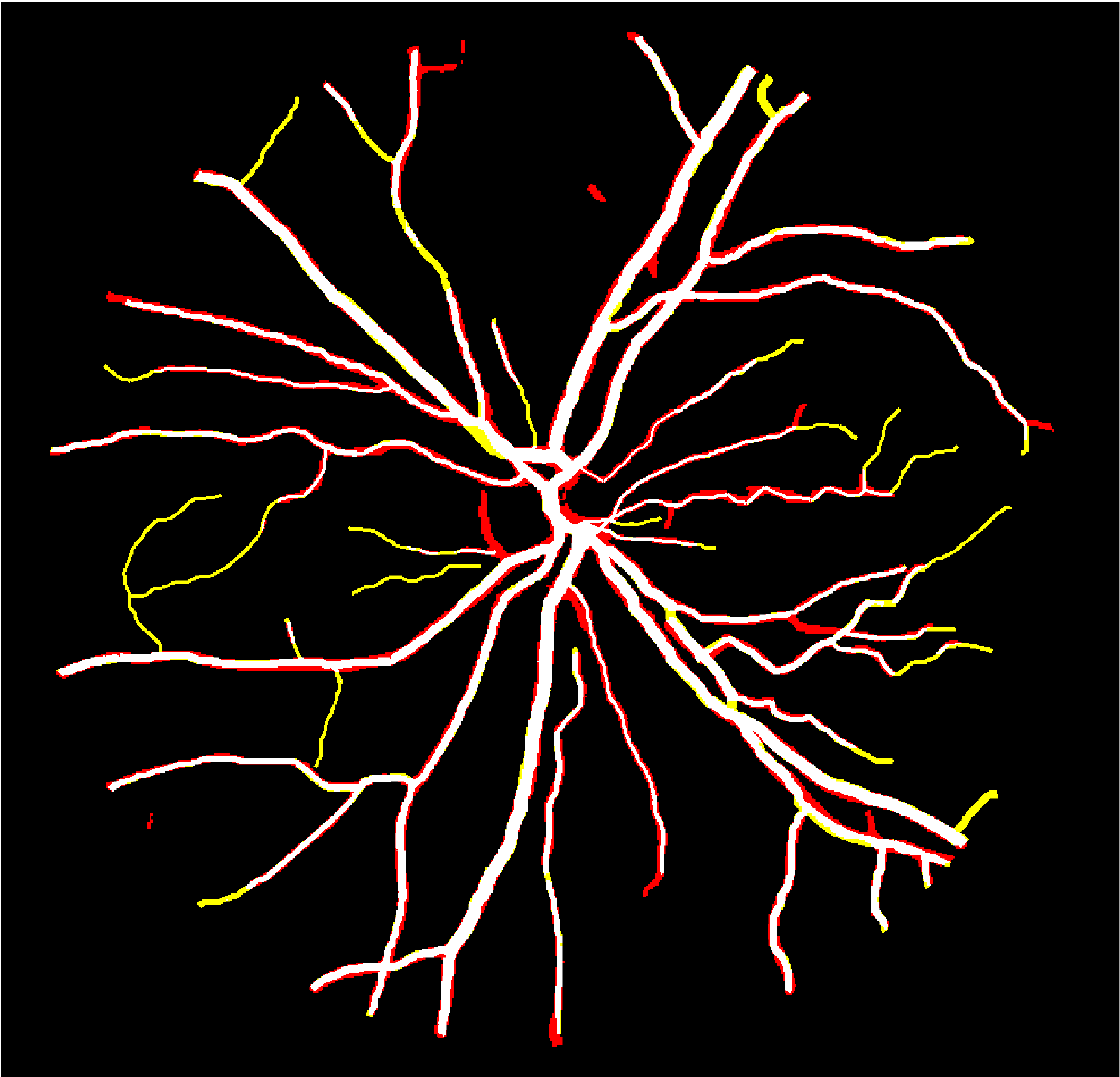} &
        \includegraphics[width=0.2\textwidth]{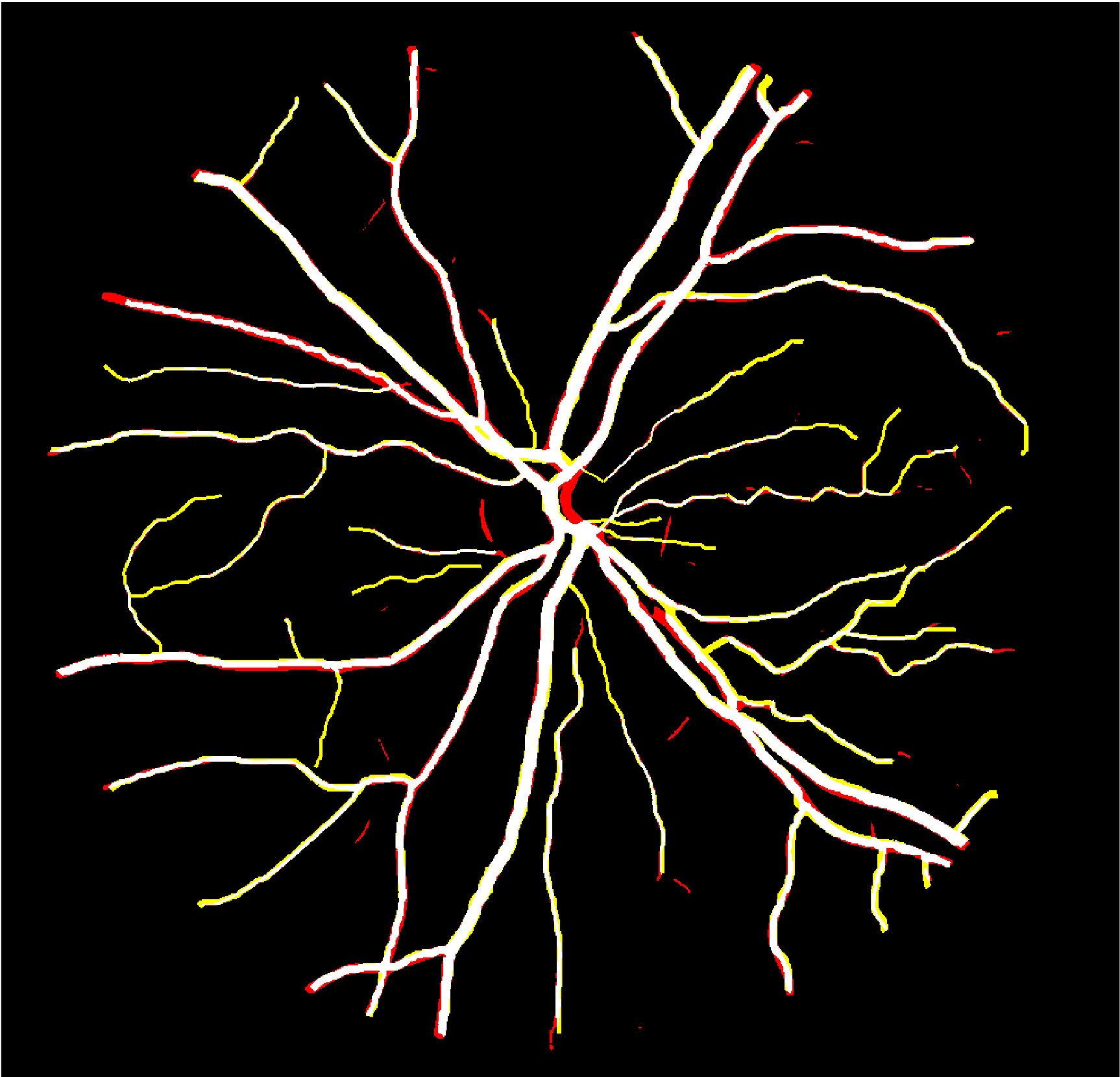} &
        \includegraphics[width=0.2\textwidth]{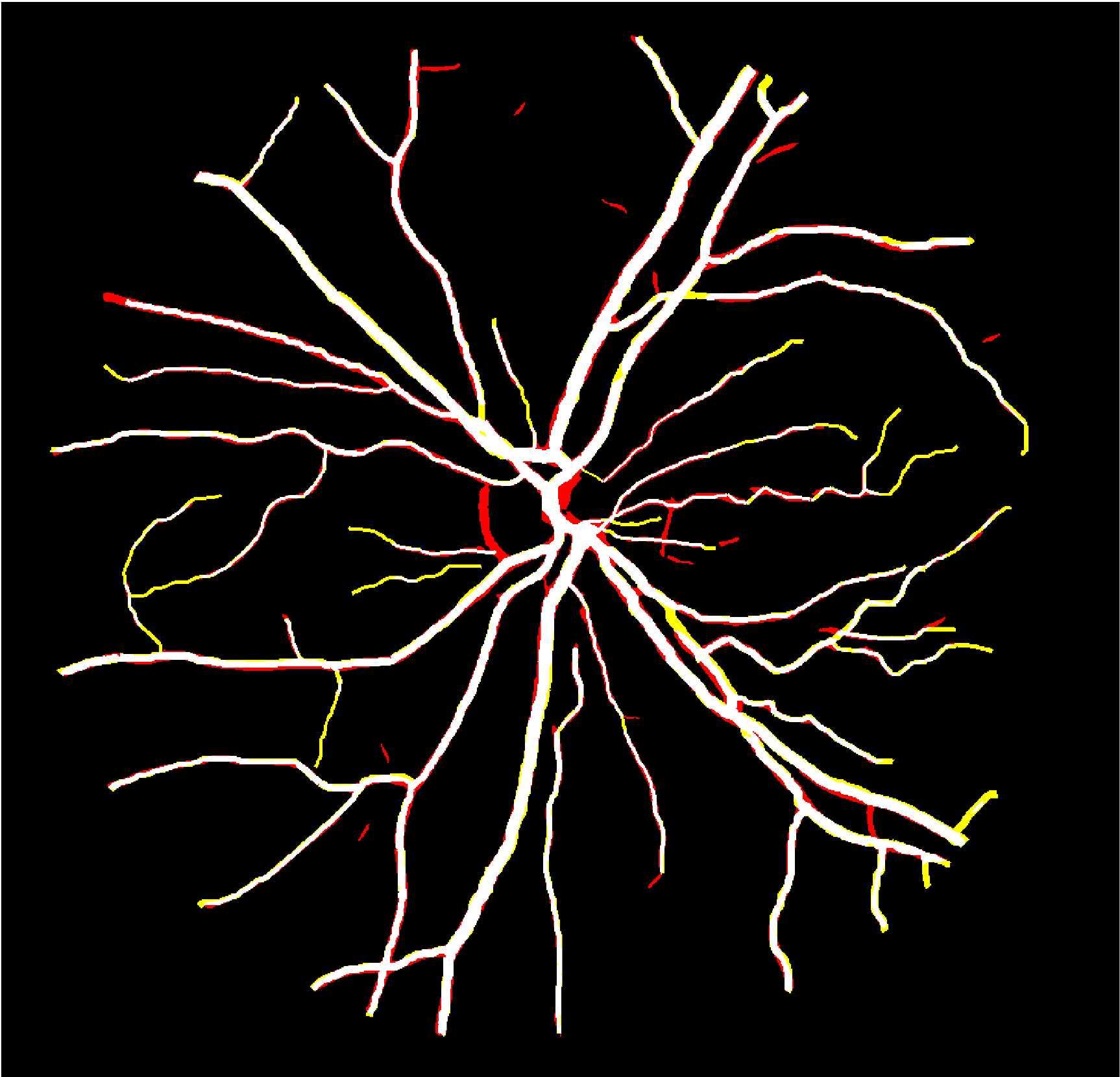} \\
        \includegraphics[width=0.2\textwidth]{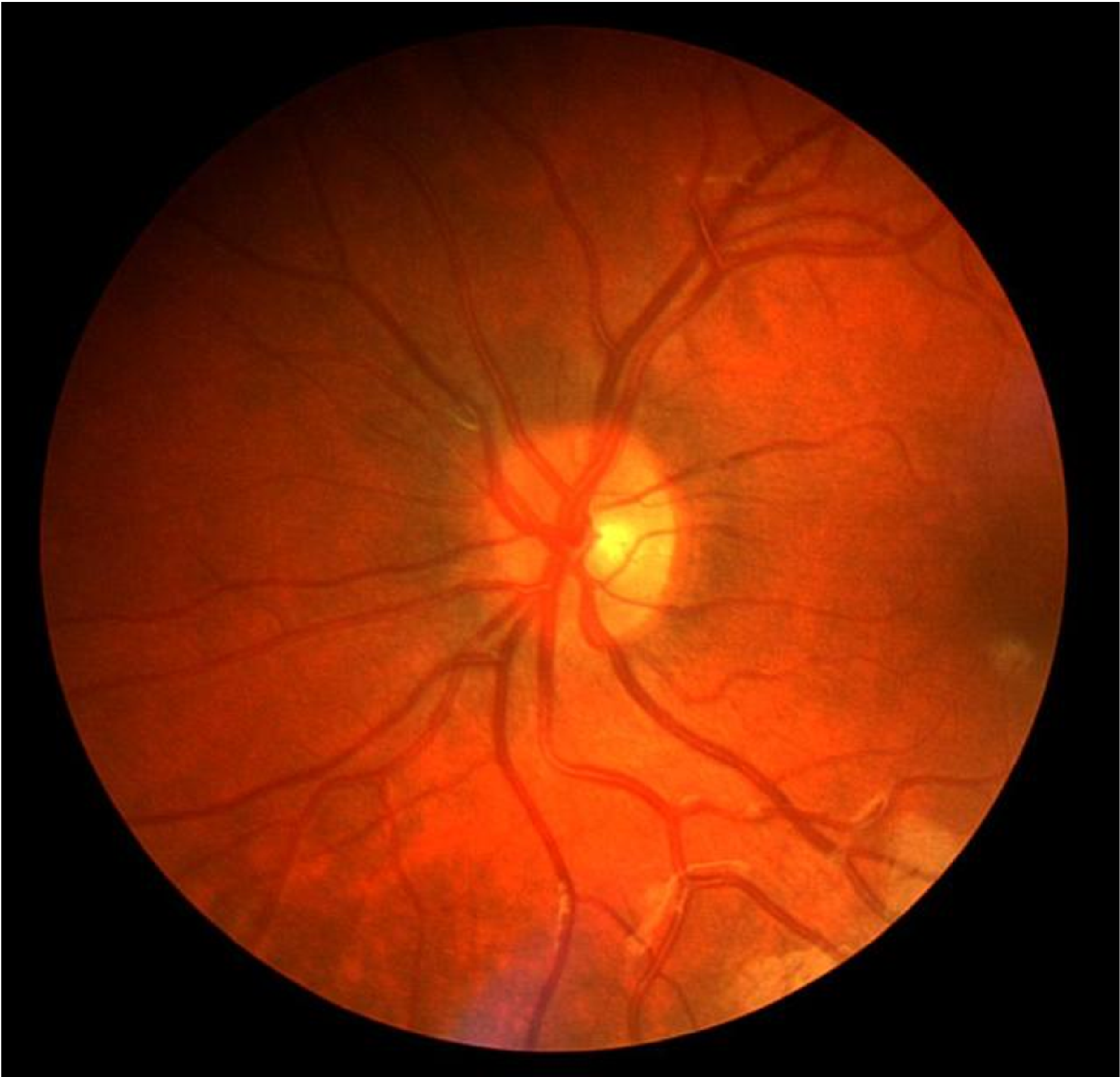} &
        \includegraphics[width=0.2\textwidth]{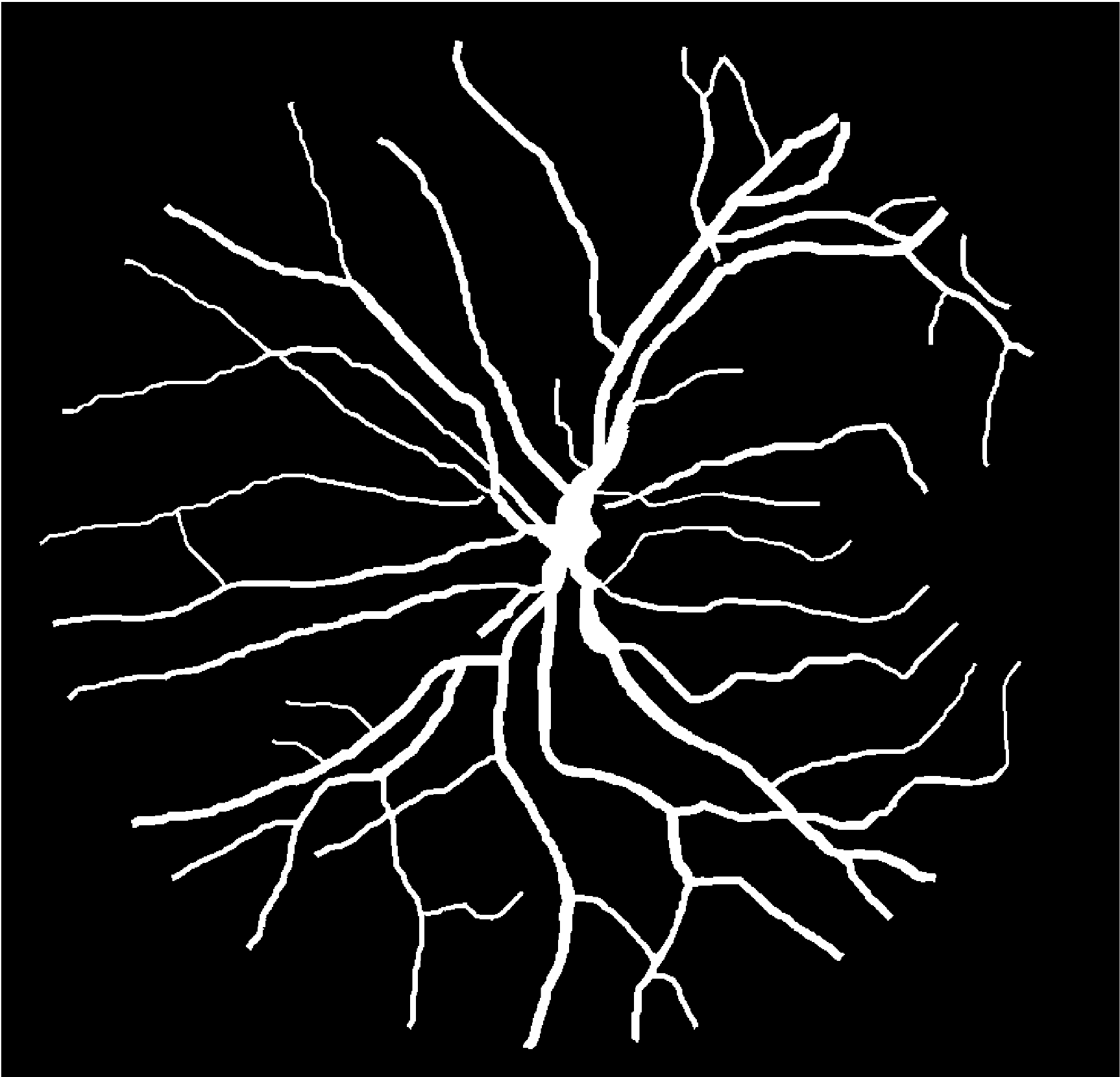} &
        \includegraphics[width=0.2\textwidth]{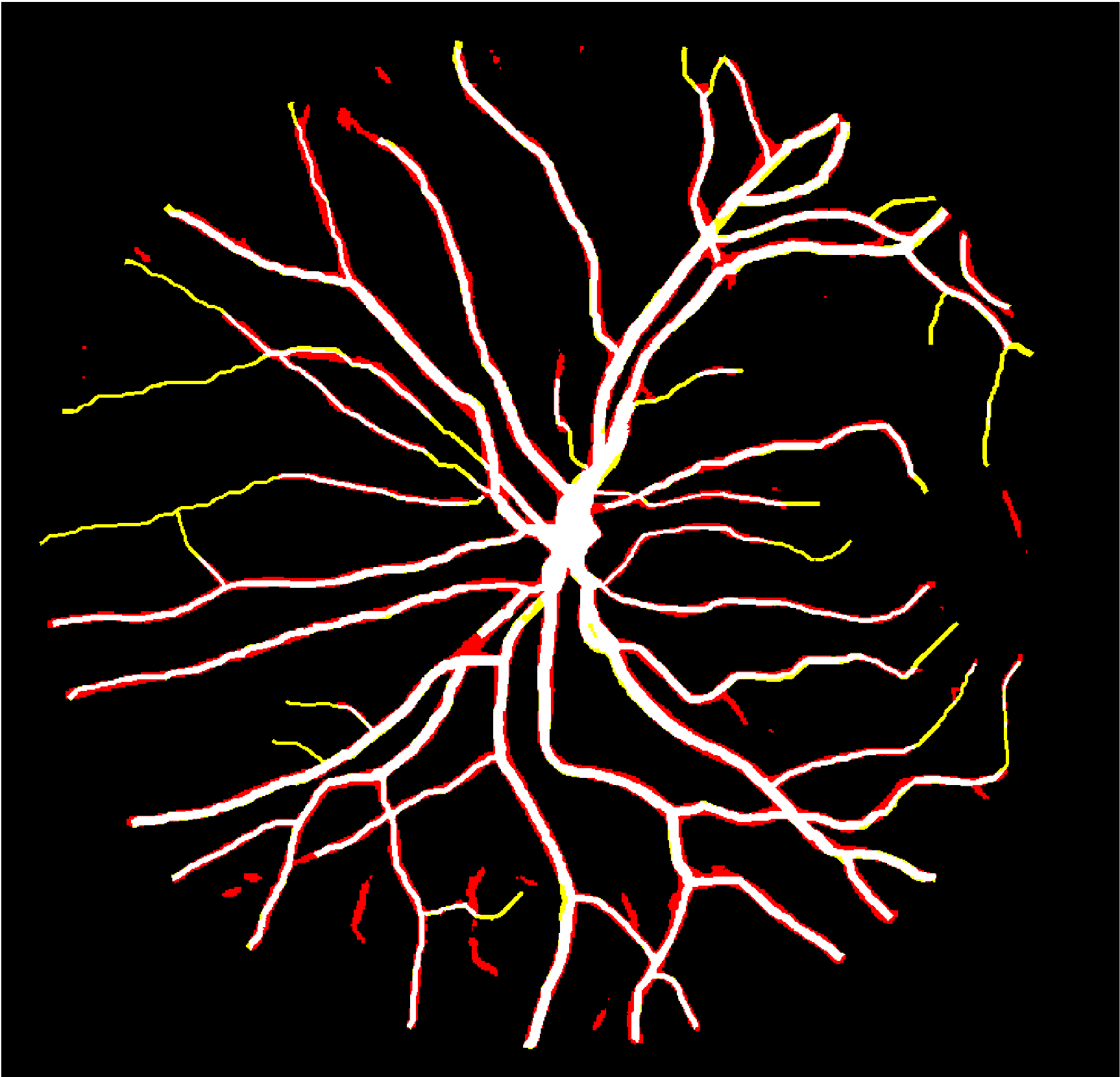} &
        \includegraphics[width=0.2\textwidth]{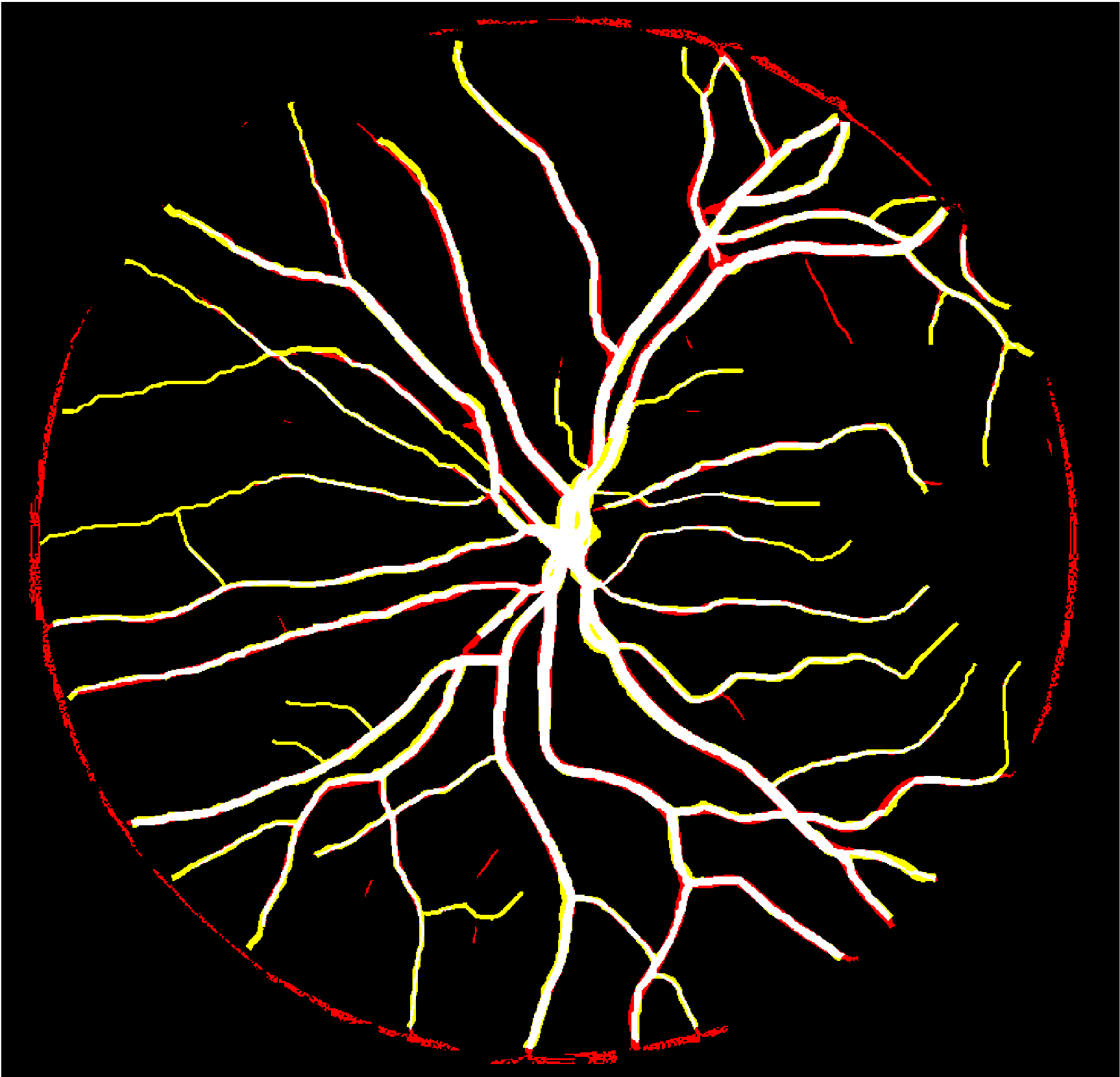} &
        \includegraphics[width=0.2\textwidth]{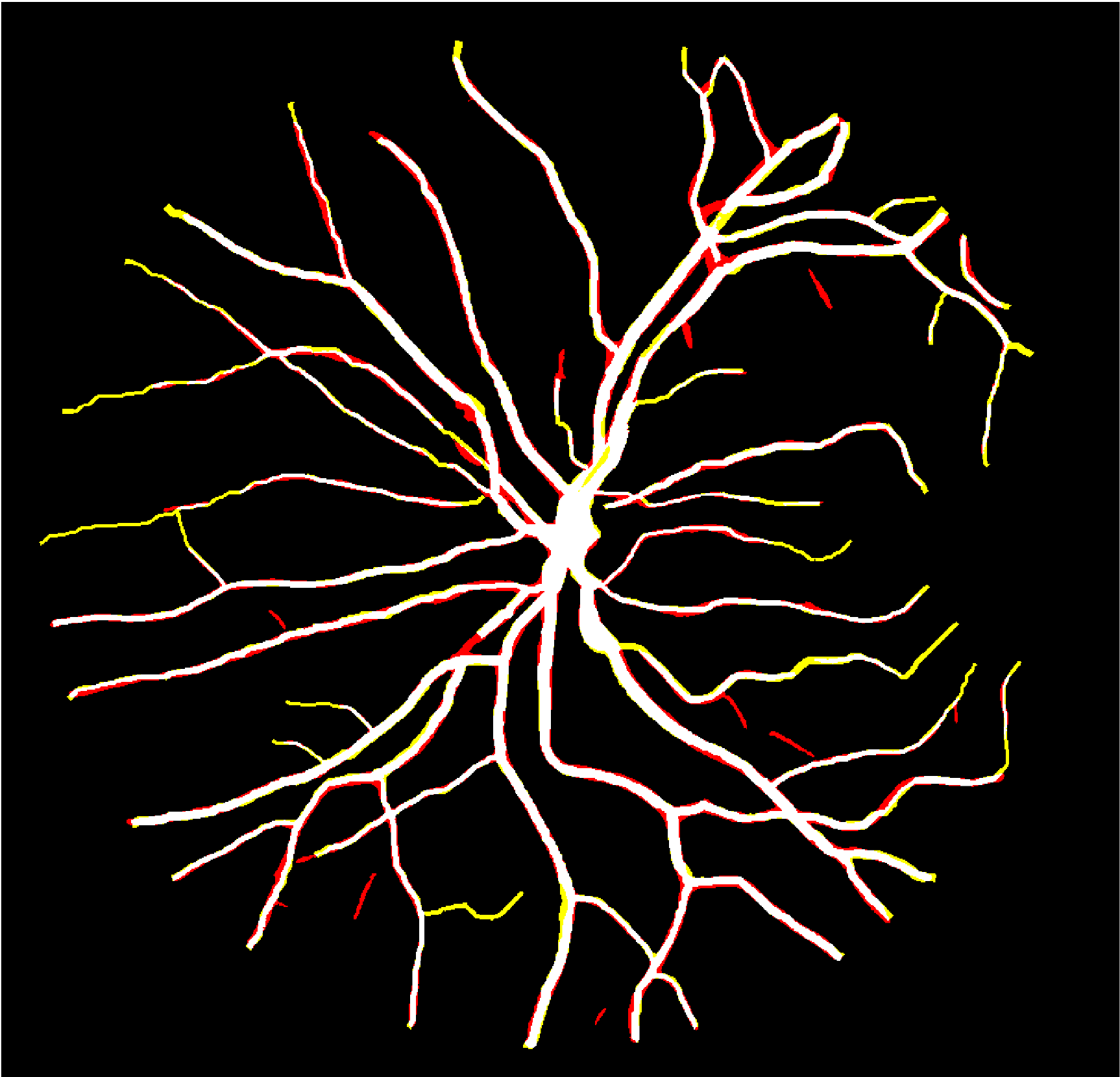} \\
    \end{tabular}}
    \caption{Sample visual results on the CHASE dataset. From left to right: the input images, the manually annotated gold standard vessel maps, and the results generated by SegNet, U-Net, and the proposed MKIS-Net.}
    \vspace{-2.5cm}
    \label{visualCHASE}
\end{figure}

\section{Results and Comparison}
\label{Results}

\subsection{Comparison with State-of-the-Art Methods}
\label{Comparison}
We compared our method both qualitatively and quantitatively with existing state-of-the-art methods in the medical image segmentation literature. First, we evaluated MKIS-Net and alternative nets such as SegNet \cite{M.Khan2020} and U-Net \cite{guo2020dpn} on the DRIVE and CHASE datasets for retinal vessel segmentation. From the qualitative results (Figs.\ \ref{visualDRIVE} and \ref{visualCHASE}) and quantitative results (Tables \ref{DRIVE} and \ref{CHASEDB1}) we see that our method is quite competitive, despite having a comparatively small number of learnable parameters (Table \ref{DRIVE}). In fact, it outperforms alternatives in the DRIVE dataset, and is the best performer in terms of accuracy and specificity for the CHASE dataset. U-Net \cite{guo2020dpn} outperforms MKIS-Net's sensitivity on the CHASE dataset, with our network coming second.

\begin{figure}[!t]
    \centering
    \resizebox{1\textwidth}{!}{%
    \begin{tabular}{@{}c@{\ }c@{\ }c@{\ }c@{}}
        \includegraphics[width=0.25\textwidth]{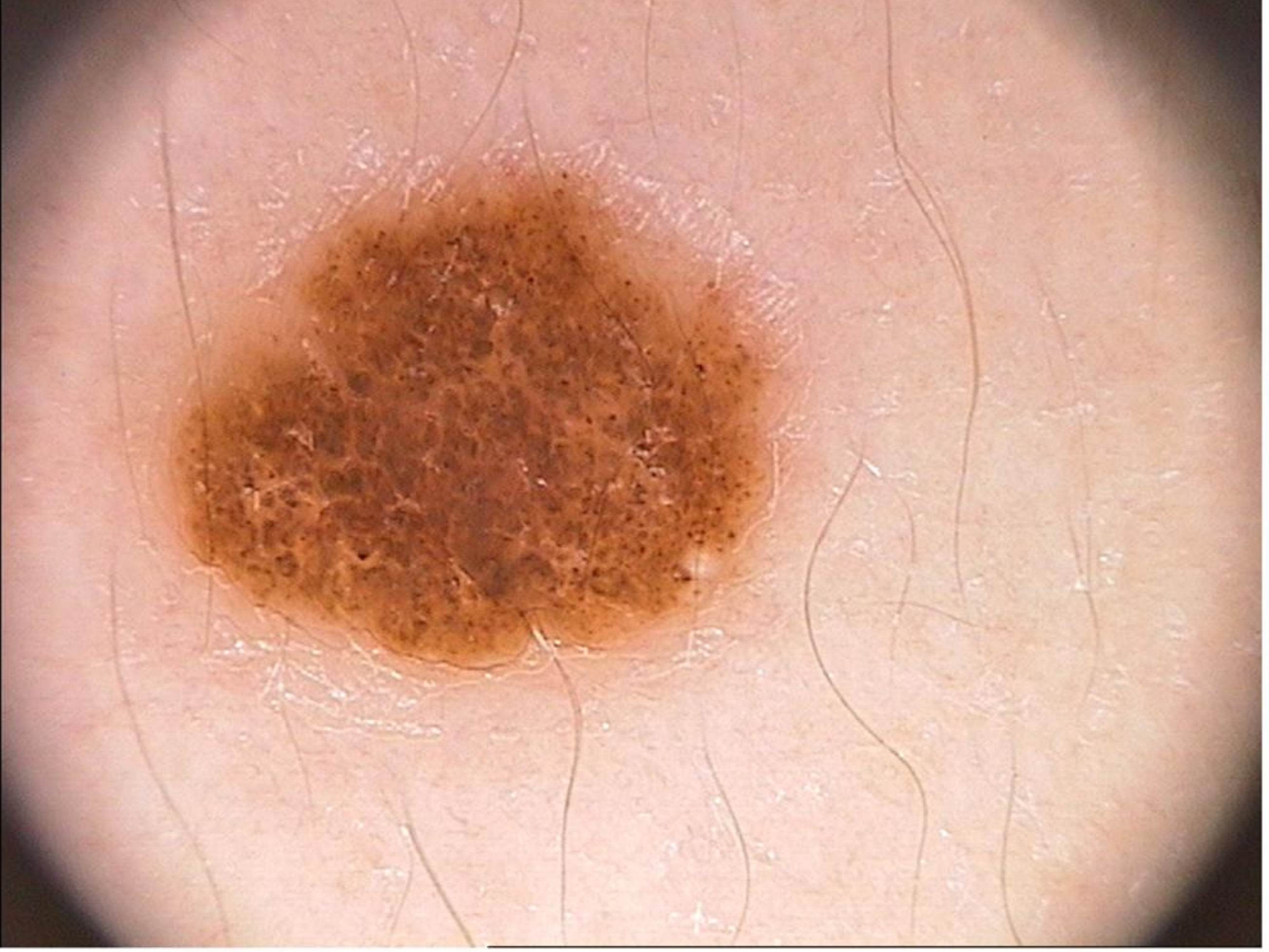} &
        \includegraphics[width=0.25\textwidth]{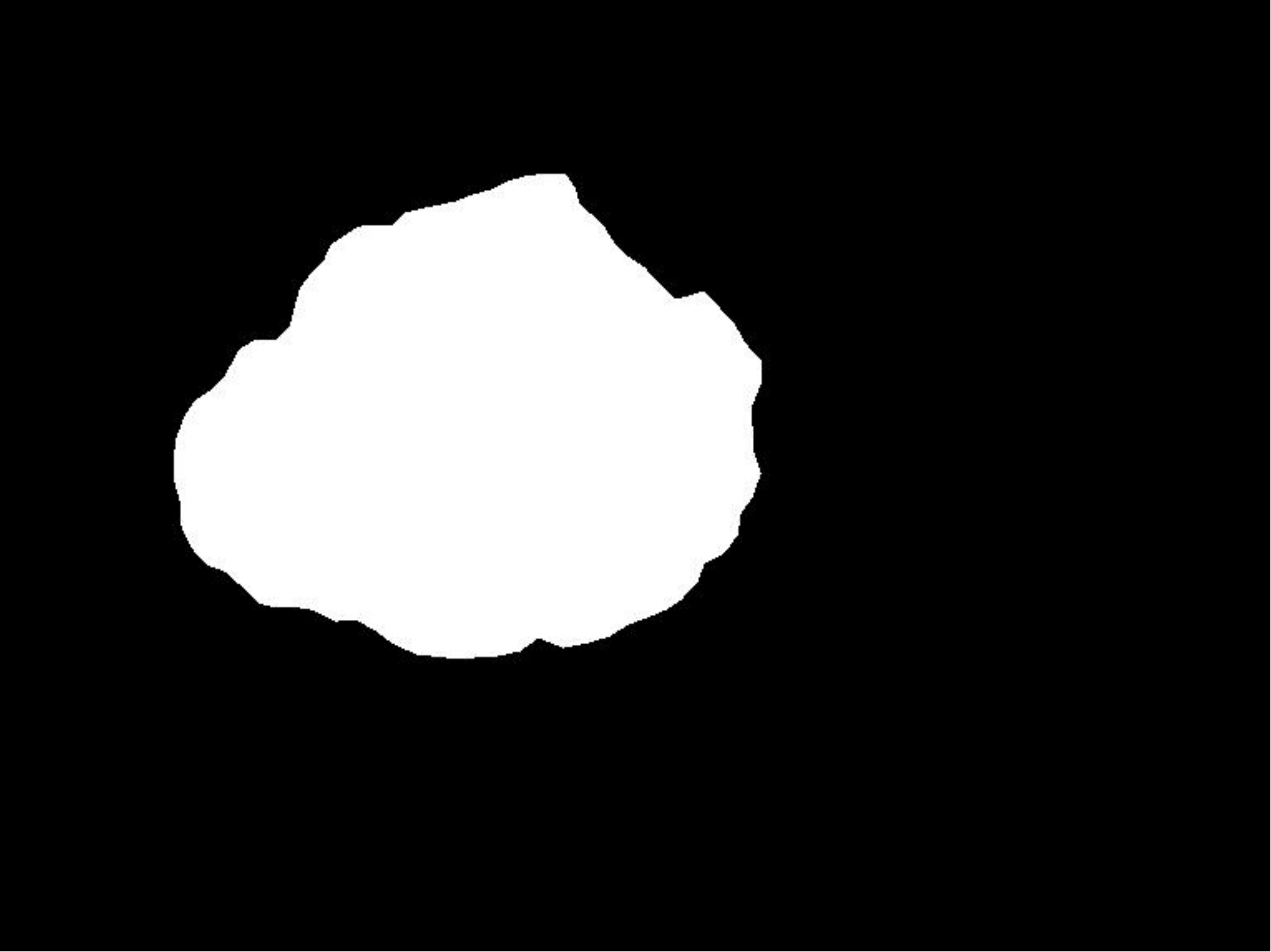} &
        \includegraphics[width=0.25\textwidth]{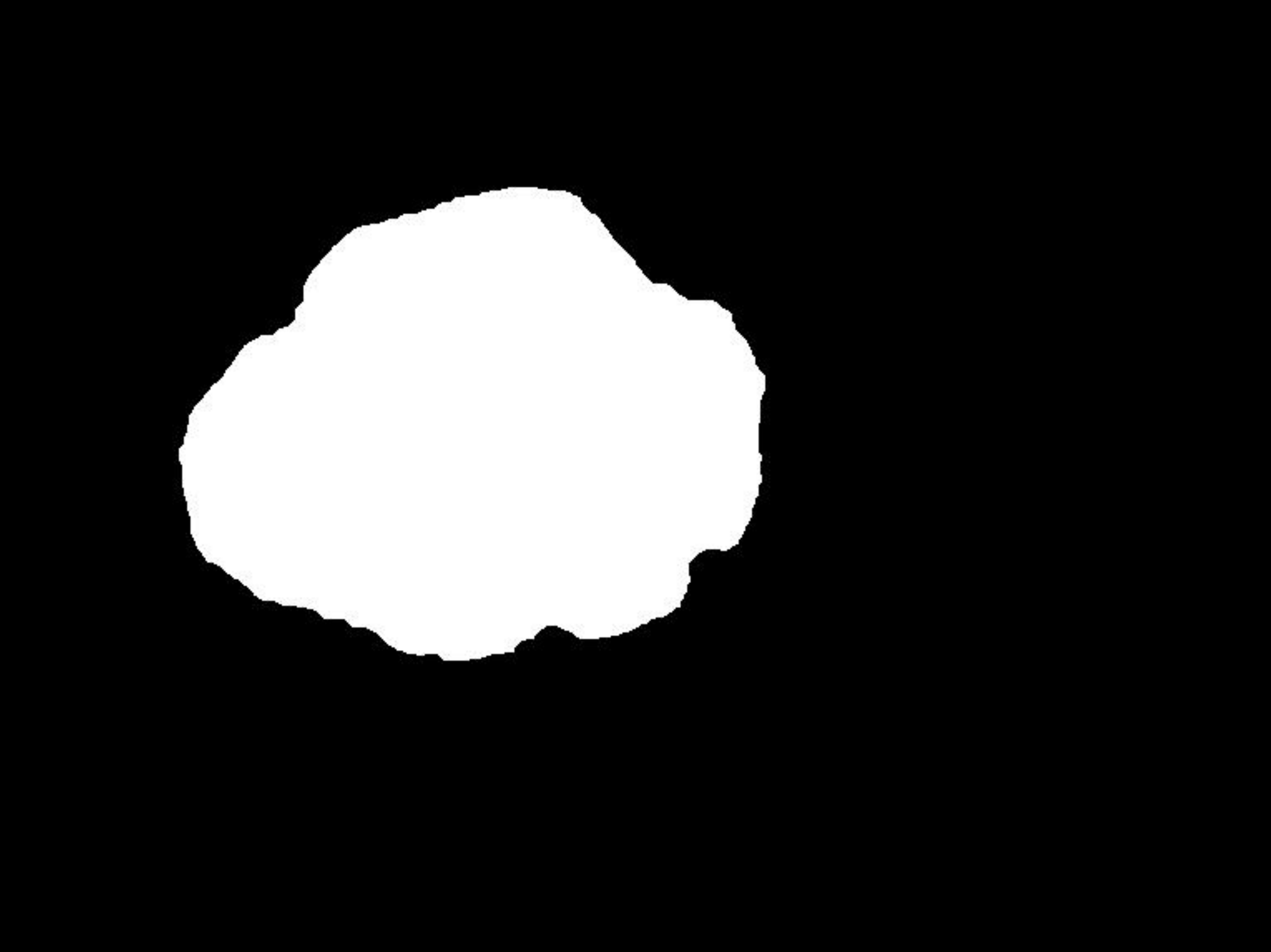} &
        \includegraphics[width=0.25\textwidth]{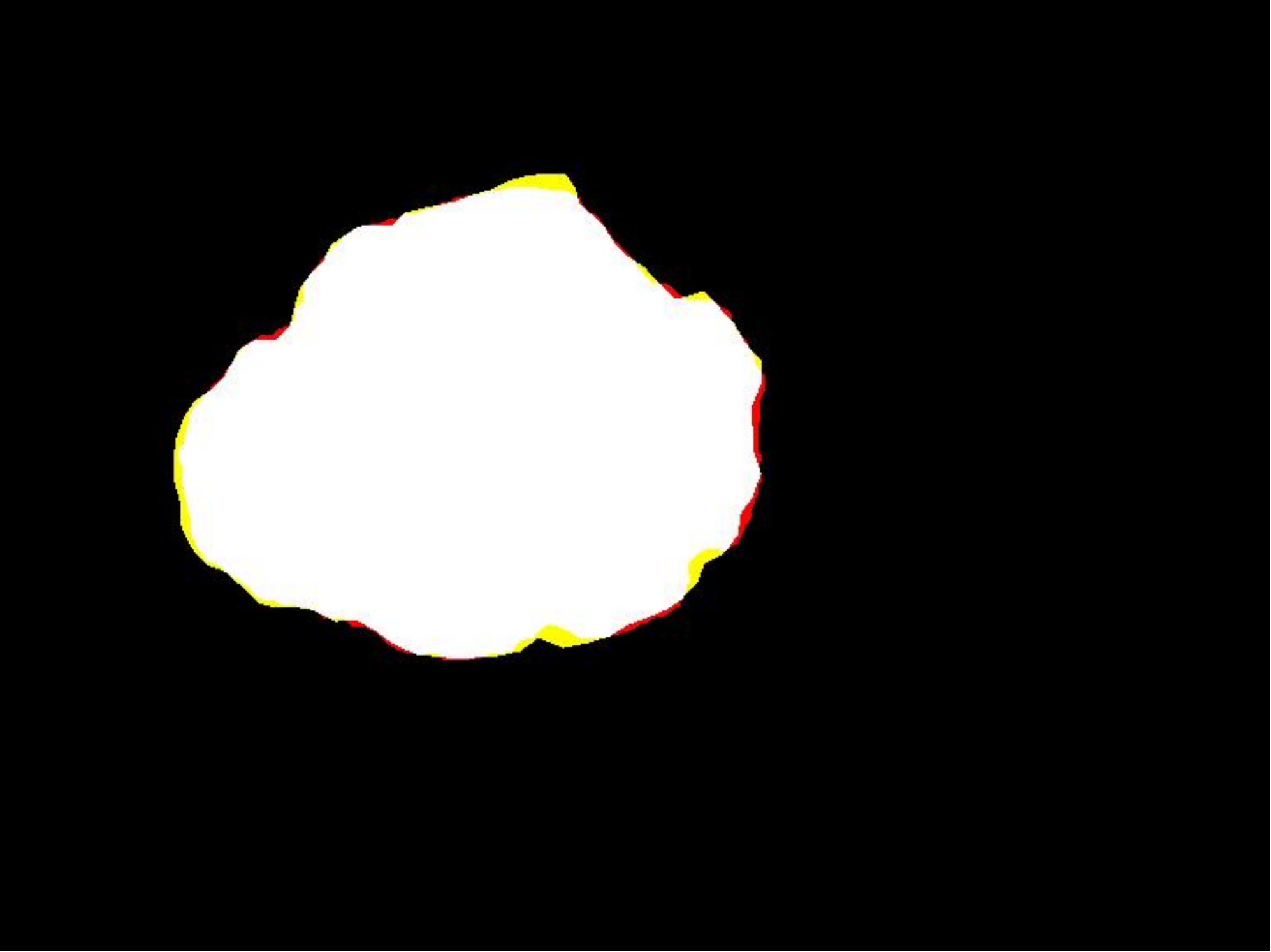} \\
        \includegraphics[width=0.25\textwidth]{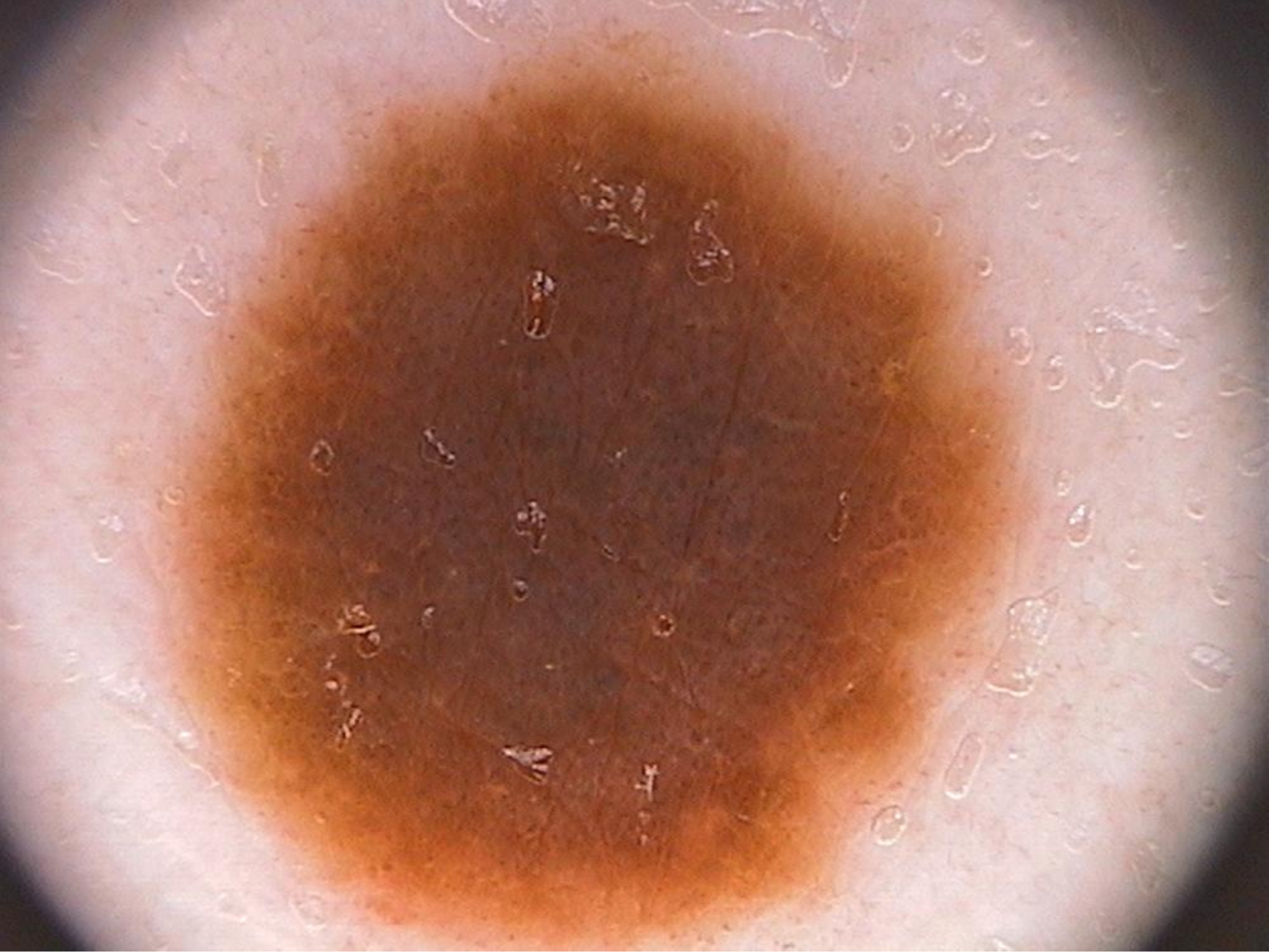} &
        \includegraphics[width=0.25\textwidth]{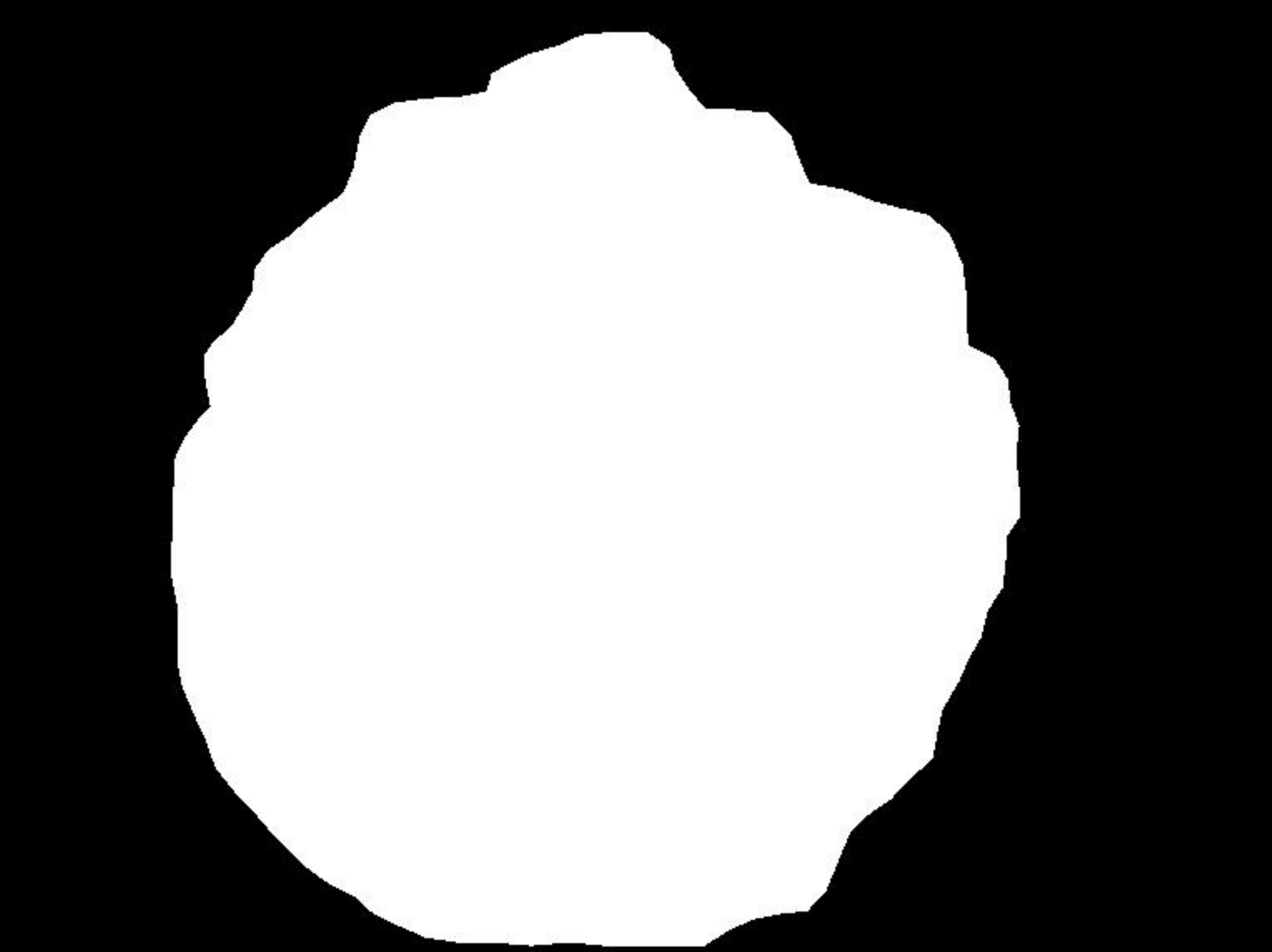} &
        \includegraphics[width=0.25\textwidth]{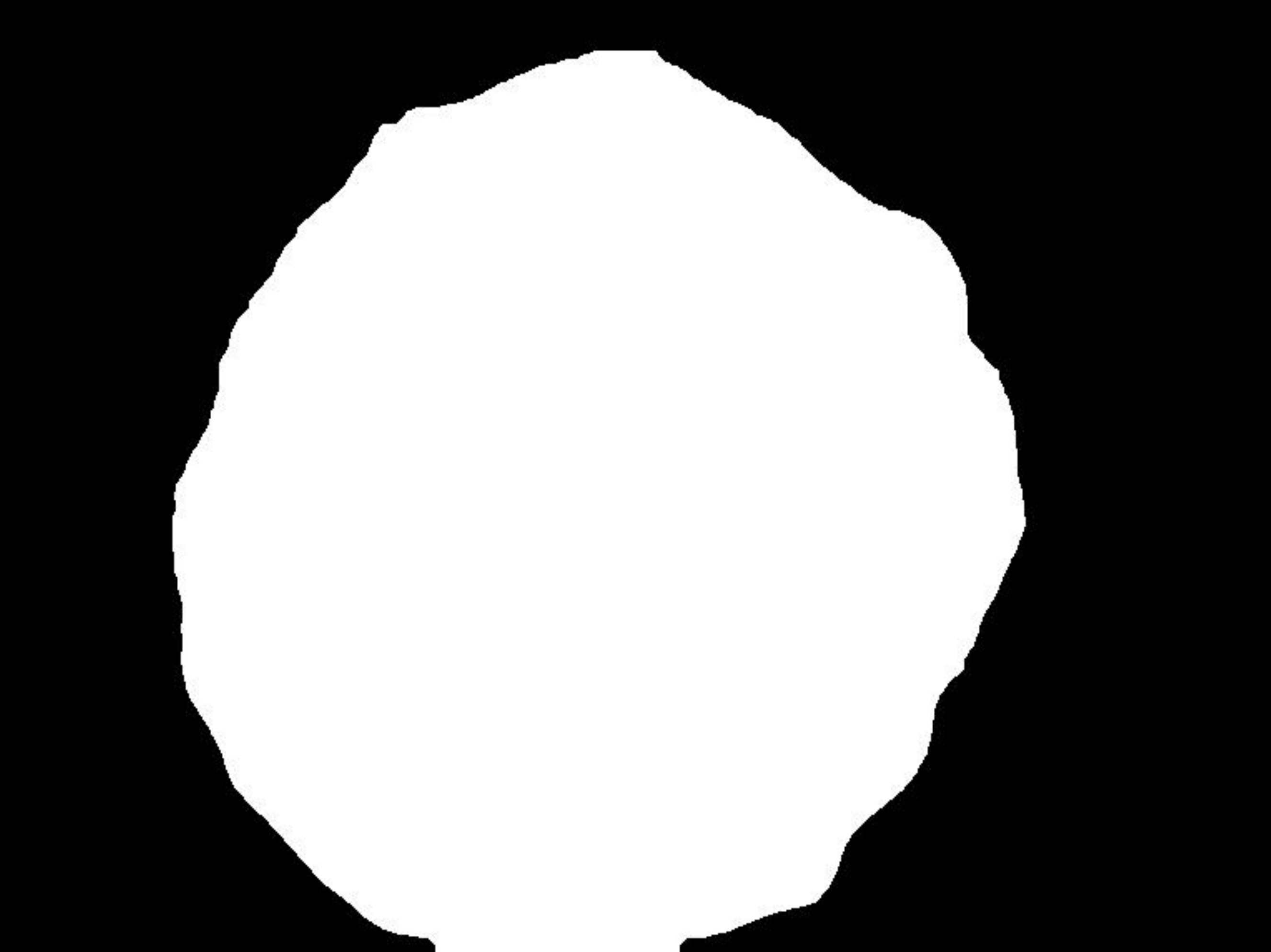} &
        \includegraphics[width=0.25\textwidth]{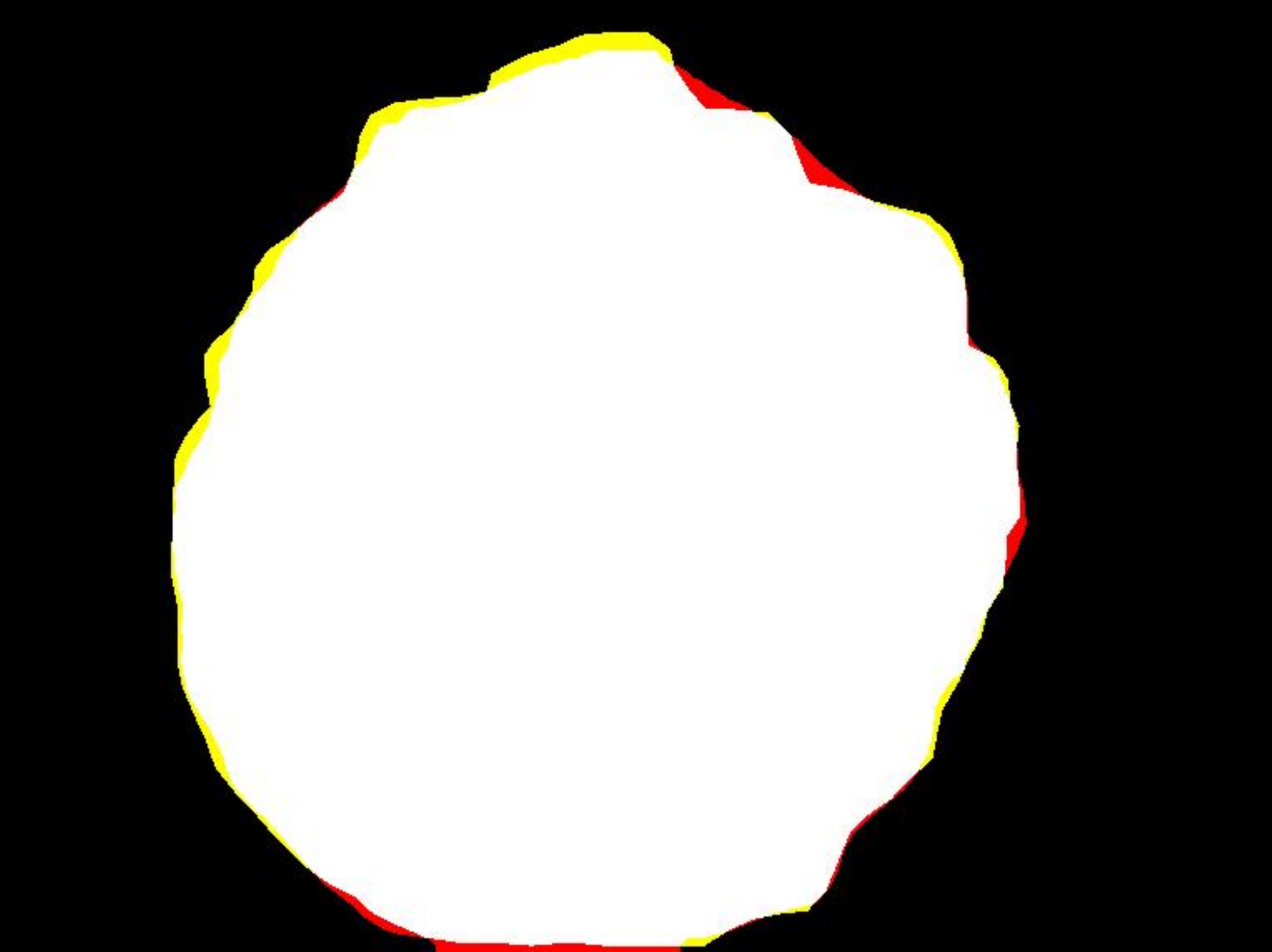} \\
        \includegraphics[width=0.25\textwidth]{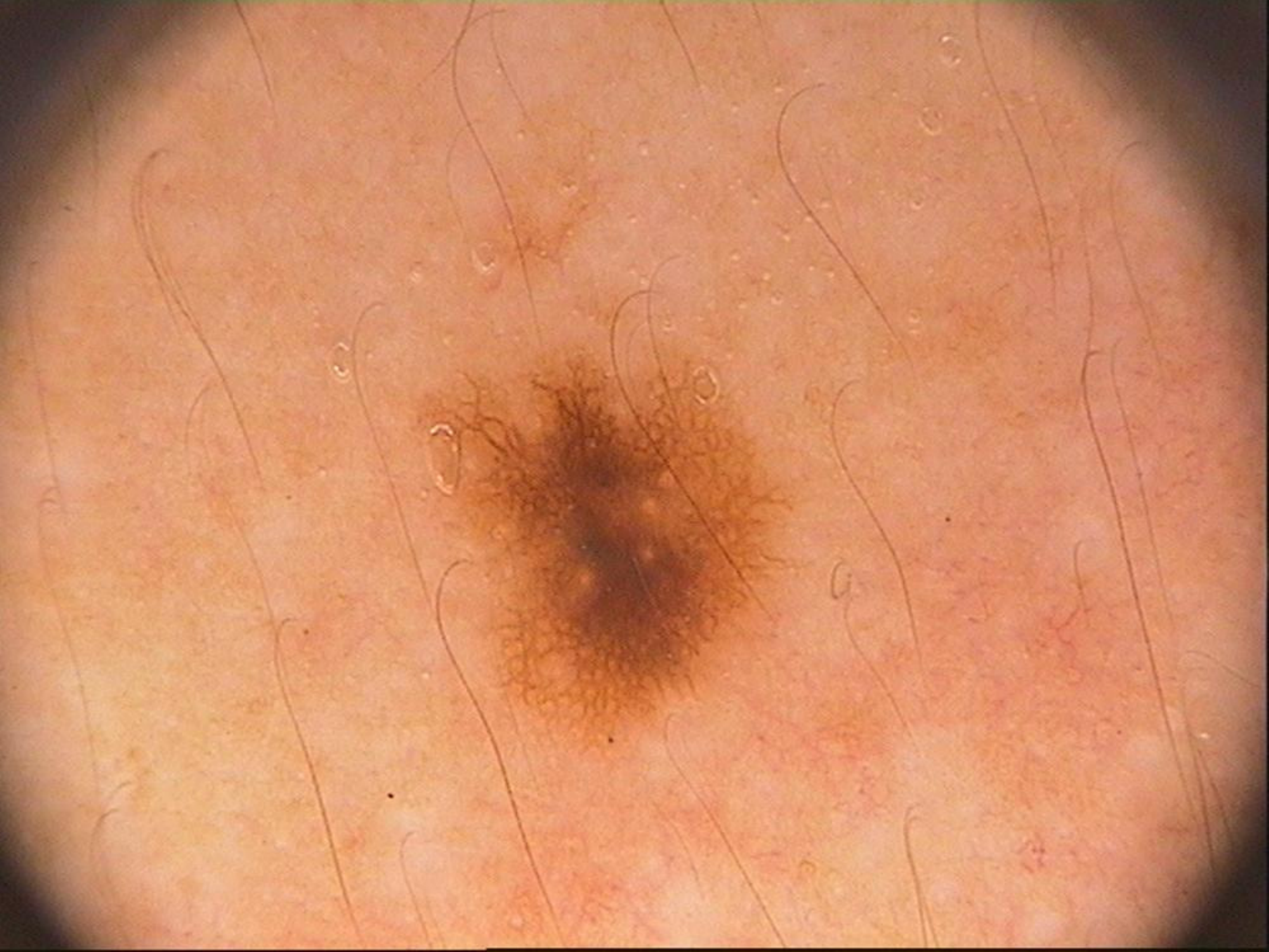} &
        \includegraphics[width=0.25\textwidth]{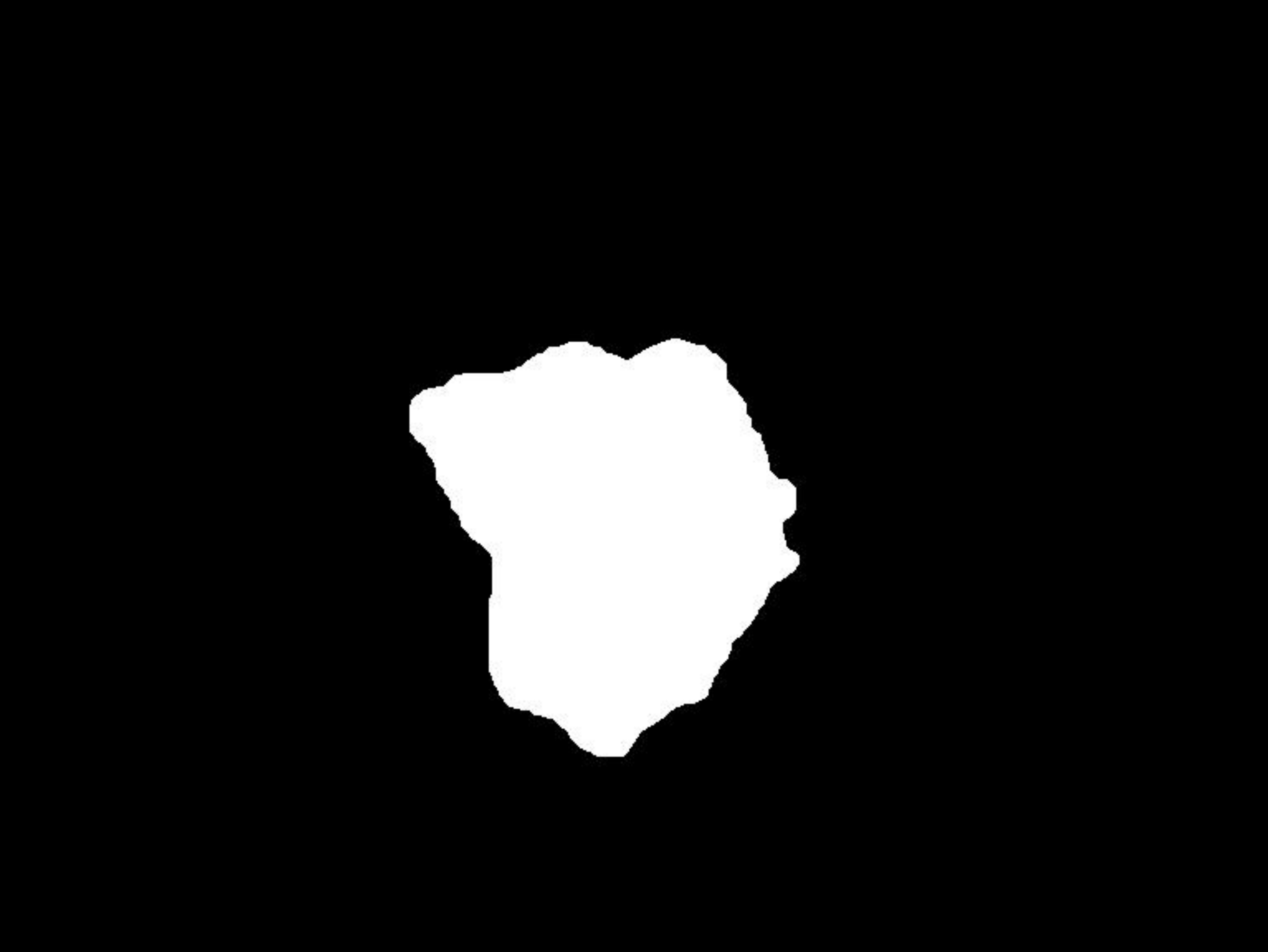} &
        \includegraphics[width=0.25\textwidth]{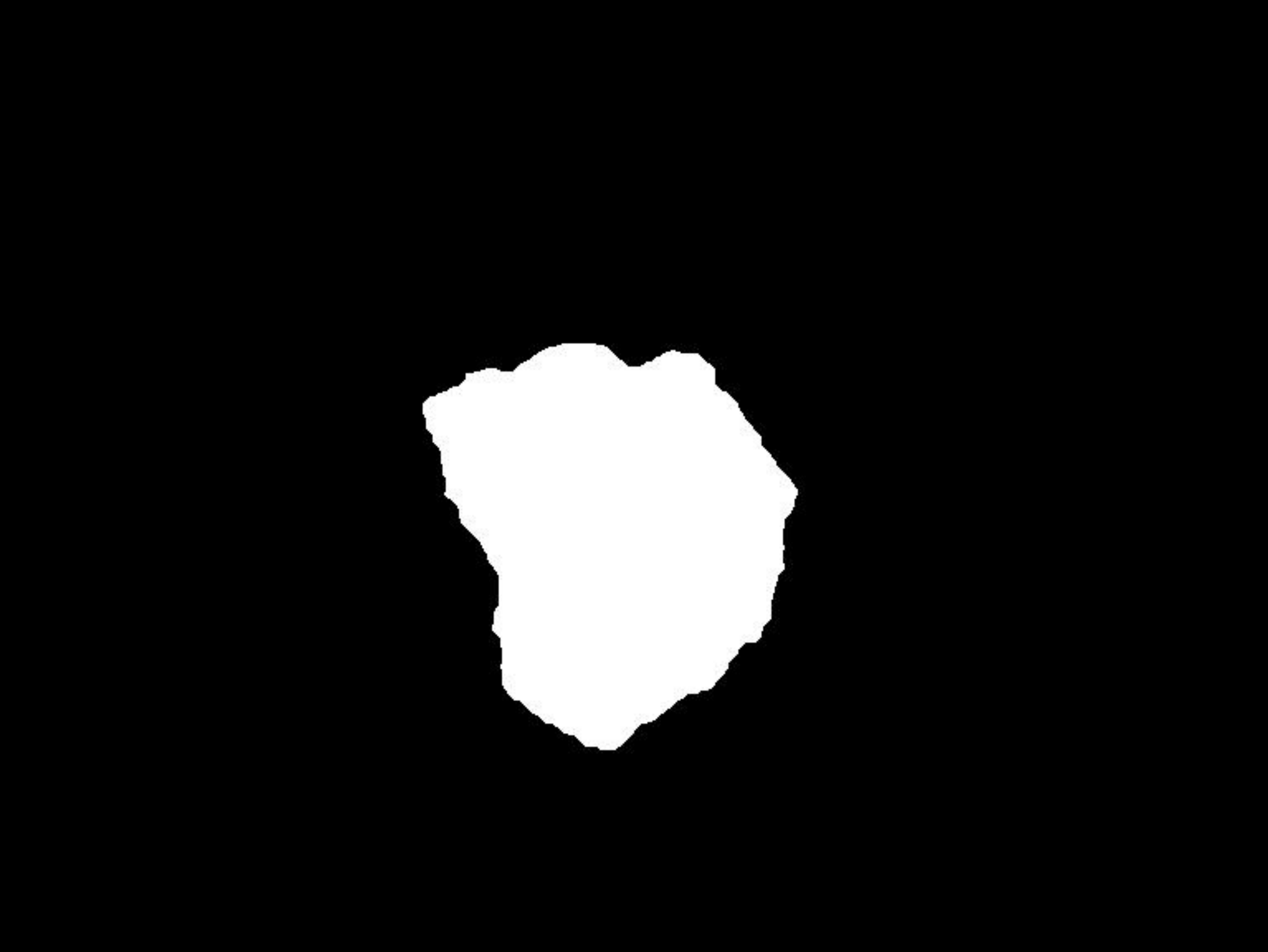} &
        \includegraphics[width=0.25\textwidth]{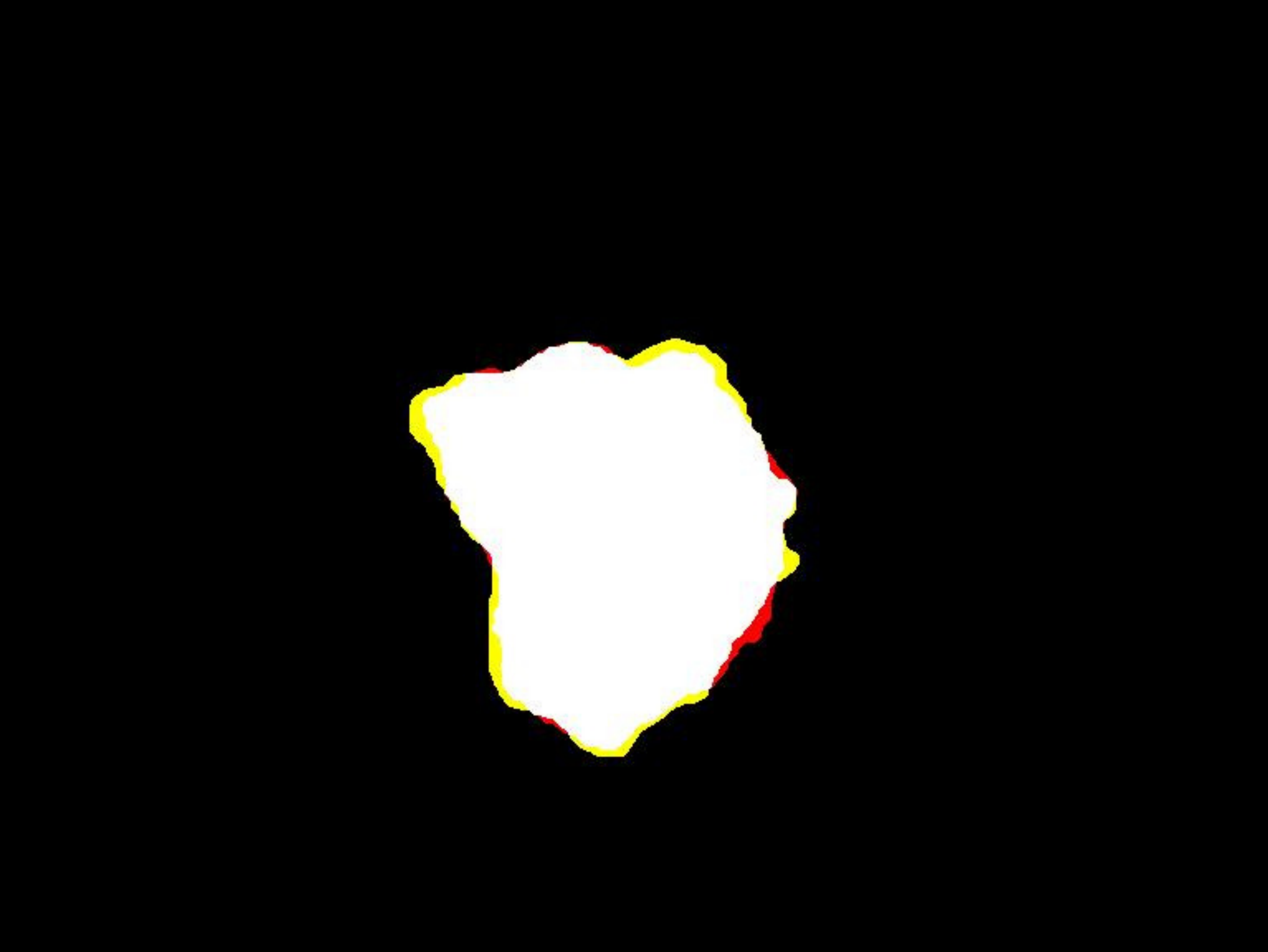} \\
    \end{tabular}}
    \caption{Sample visual results of the experiments on the PH2+ISBI 2016 challenge dataset. From left to right: the input images, the manually annotated gold standard maps, the binary segmentation output from MKIS-Net, and the visual accuracy map. The latter shows true positive and negative predictions in black and white and the false positive and negative predictions in red and yellow colors, respectively.}
  \label{visualSKIN}
\end{figure}

Next, we evaluated MKIS-Net on the PH2+ISBI 2016 skin lesion segmentation dataset. From the qualitative results (Fig.\ \ref{visualSKIN}) we see that our method works well for a wide range of lesion sizes, shapes, colours, and textures. The quantitative results (Table~\ref{PH2+ISBI}) indicate that MKIS-Net achieved the best results in terms of F1 and Jaccard compared to other state-of-the-art segmentation methods.

\begin{figure}[!t]
    \centering
    \resizebox{1\textwidth}{!}{%
    \begin{tabular}{@{}c@{\ }c@{\ }c@{\ }c@{}}
        \includegraphics[width=0.25\textwidth]{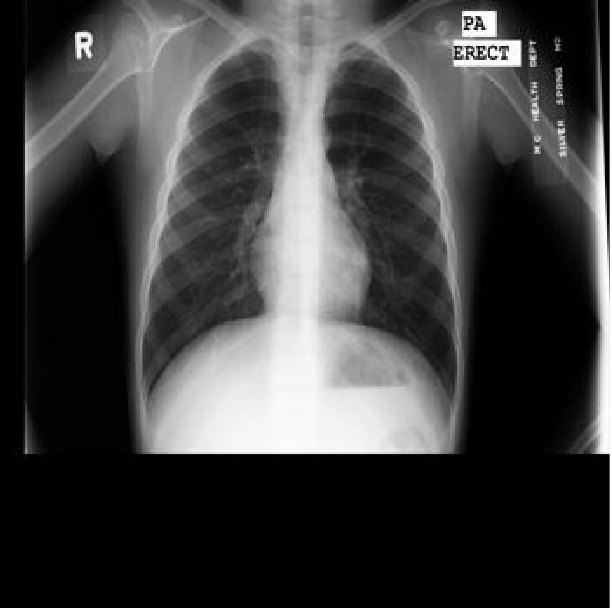} &
        \includegraphics[width=0.25\textwidth]{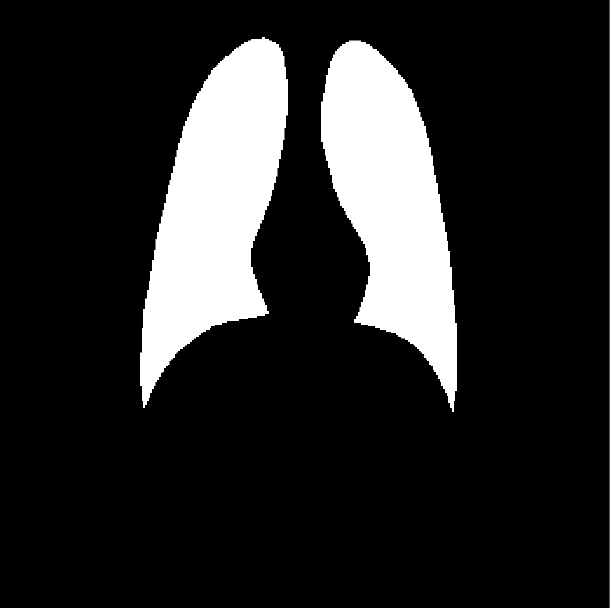} &
        \includegraphics[width=0.25\textwidth]{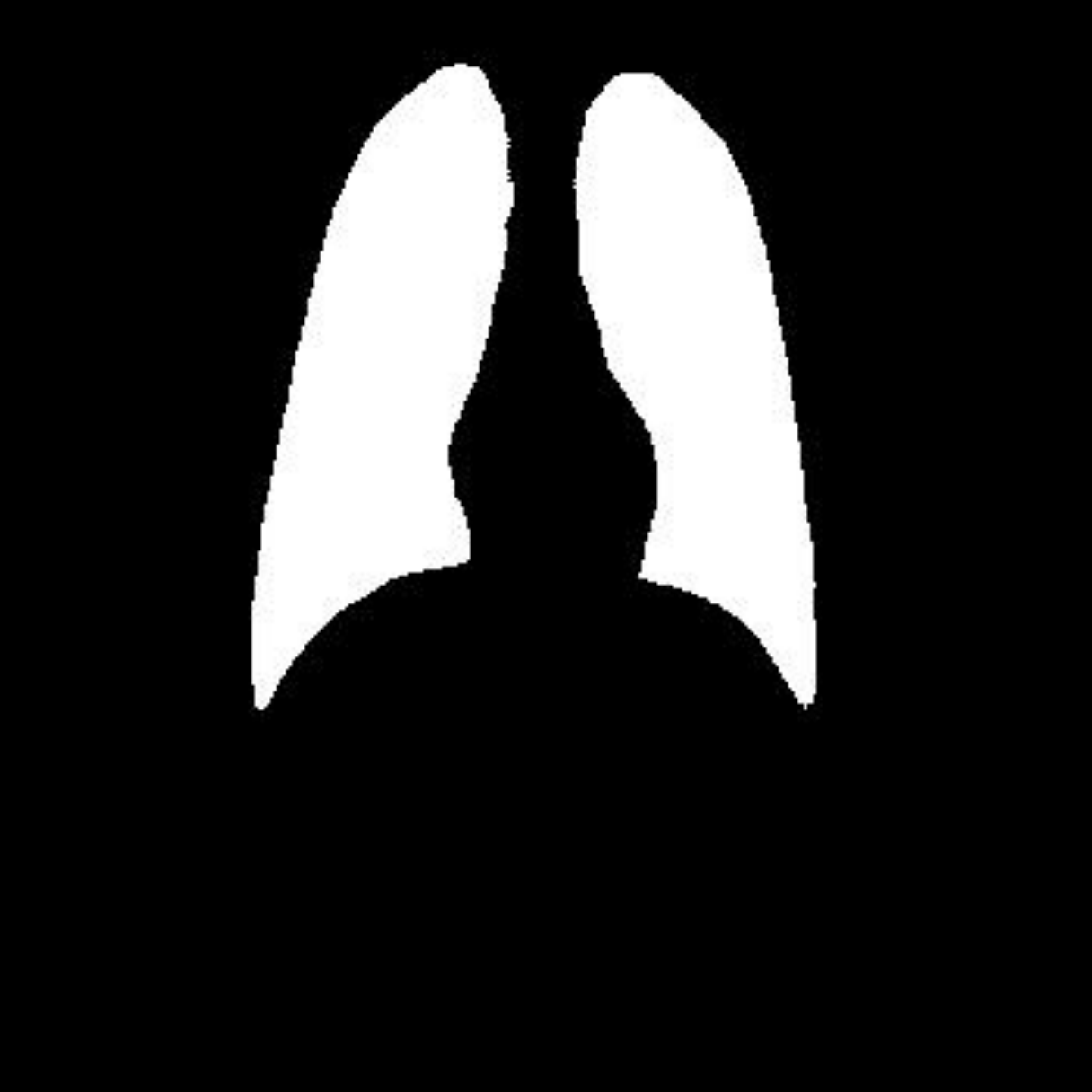} &
        \includegraphics[width=0.25\textwidth]{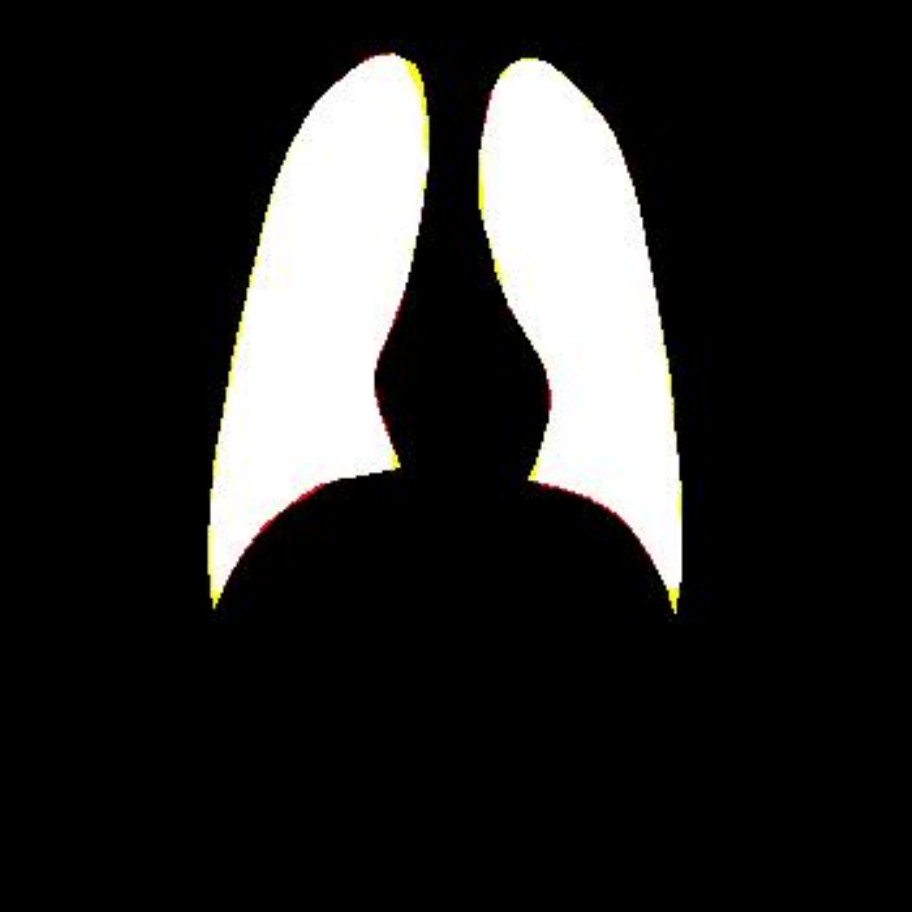} \\
        \includegraphics[width=0.25\textwidth]{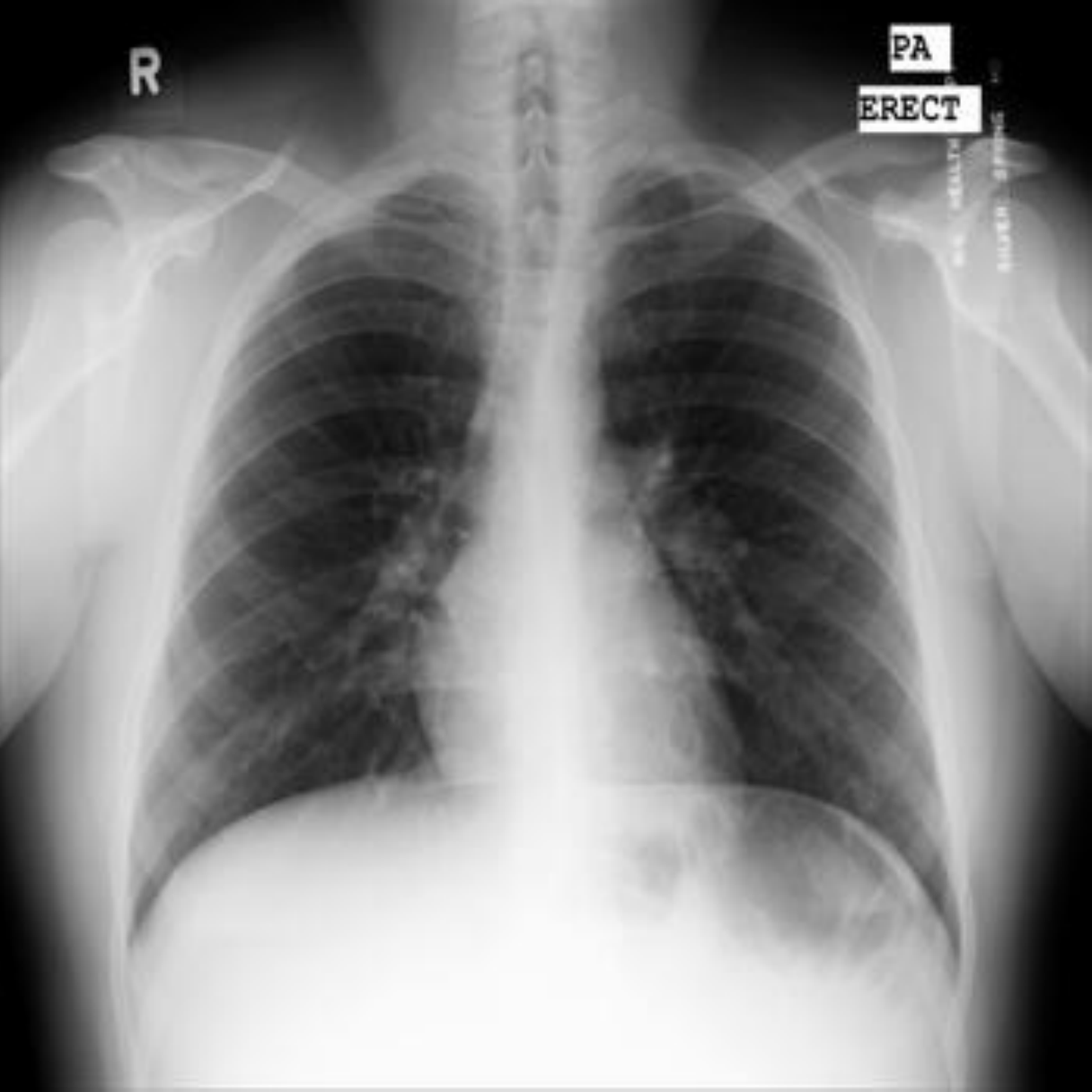} &
        \includegraphics[width=0.25\textwidth]{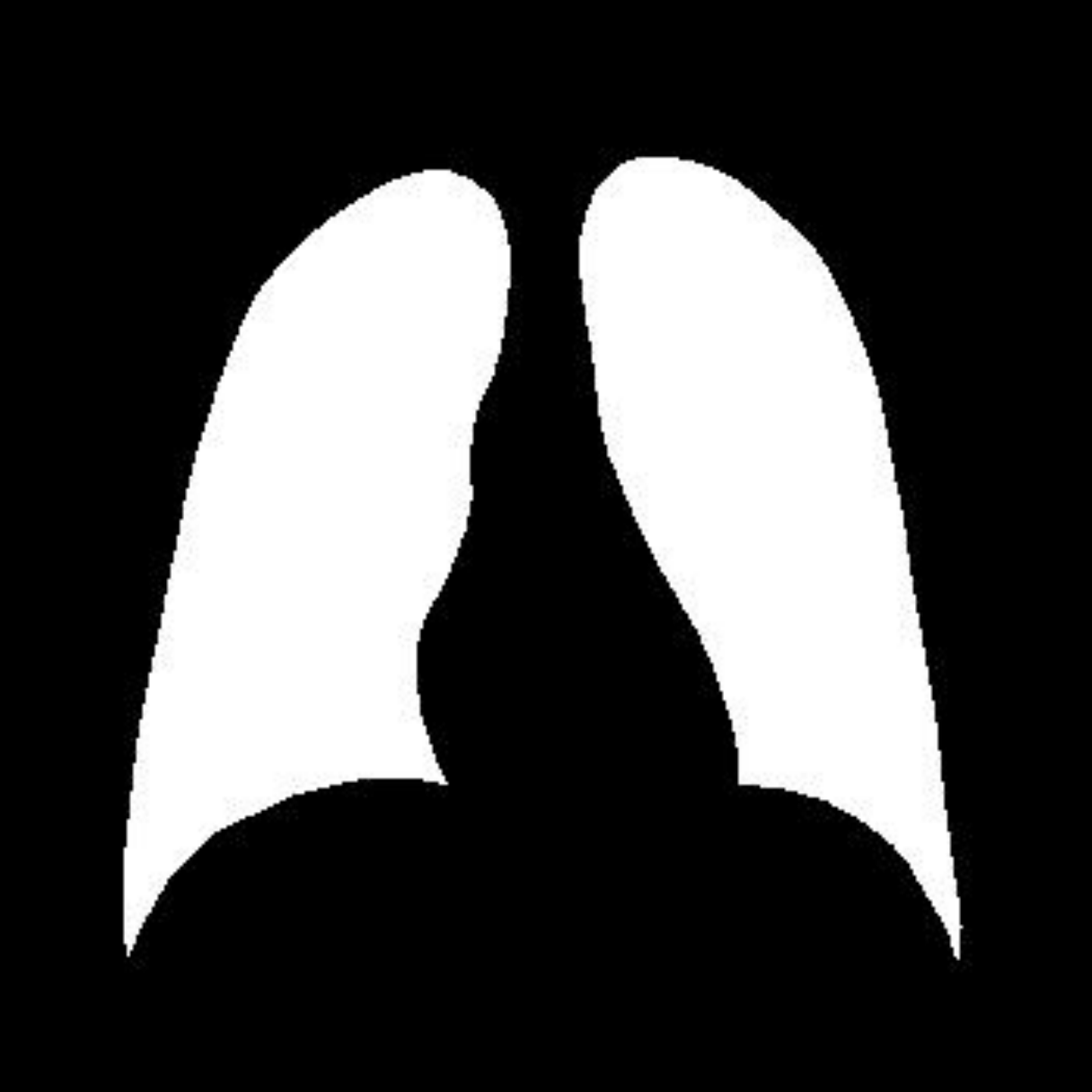} &
        \includegraphics[width=0.25\textwidth]{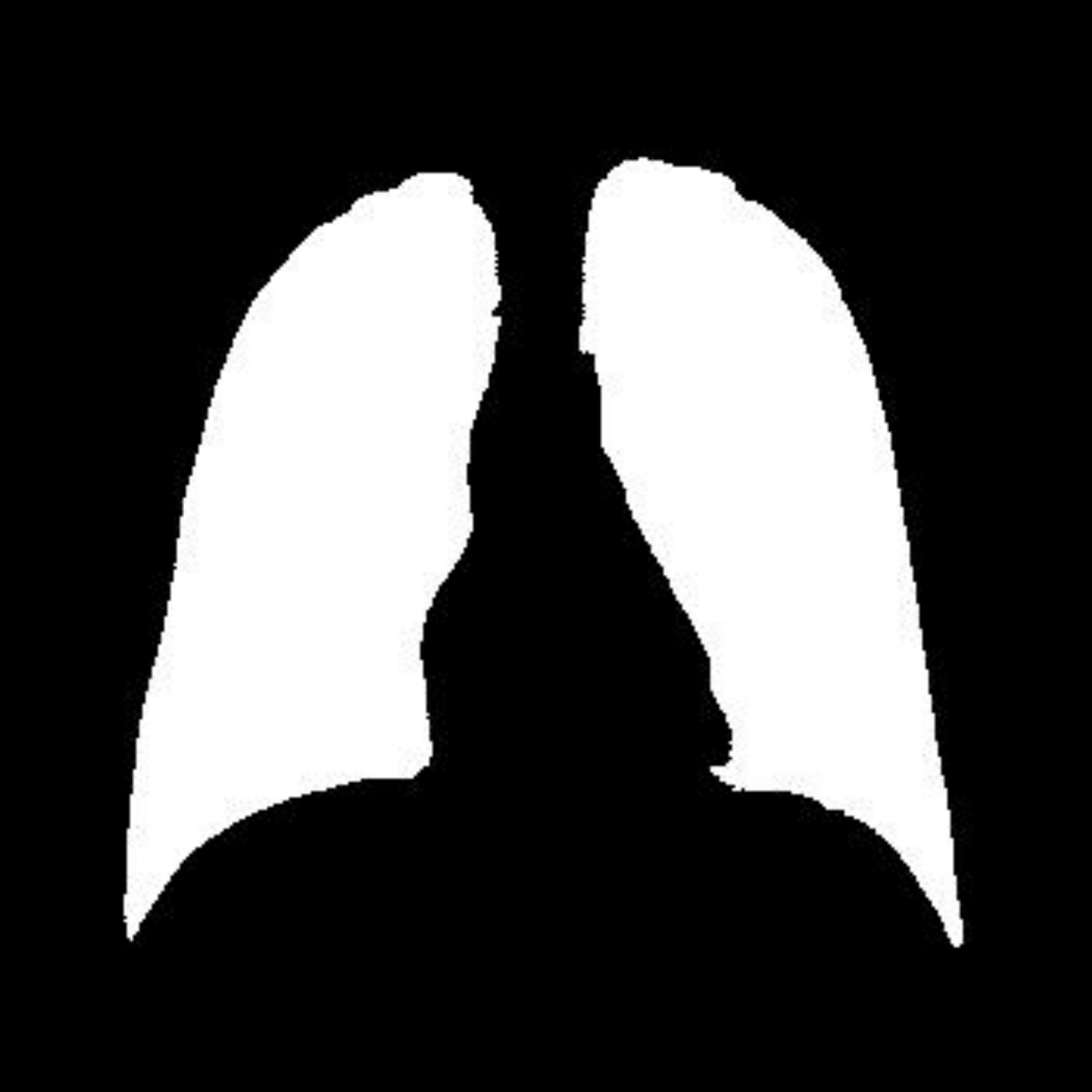} &
        \includegraphics[width=0.25\textwidth]{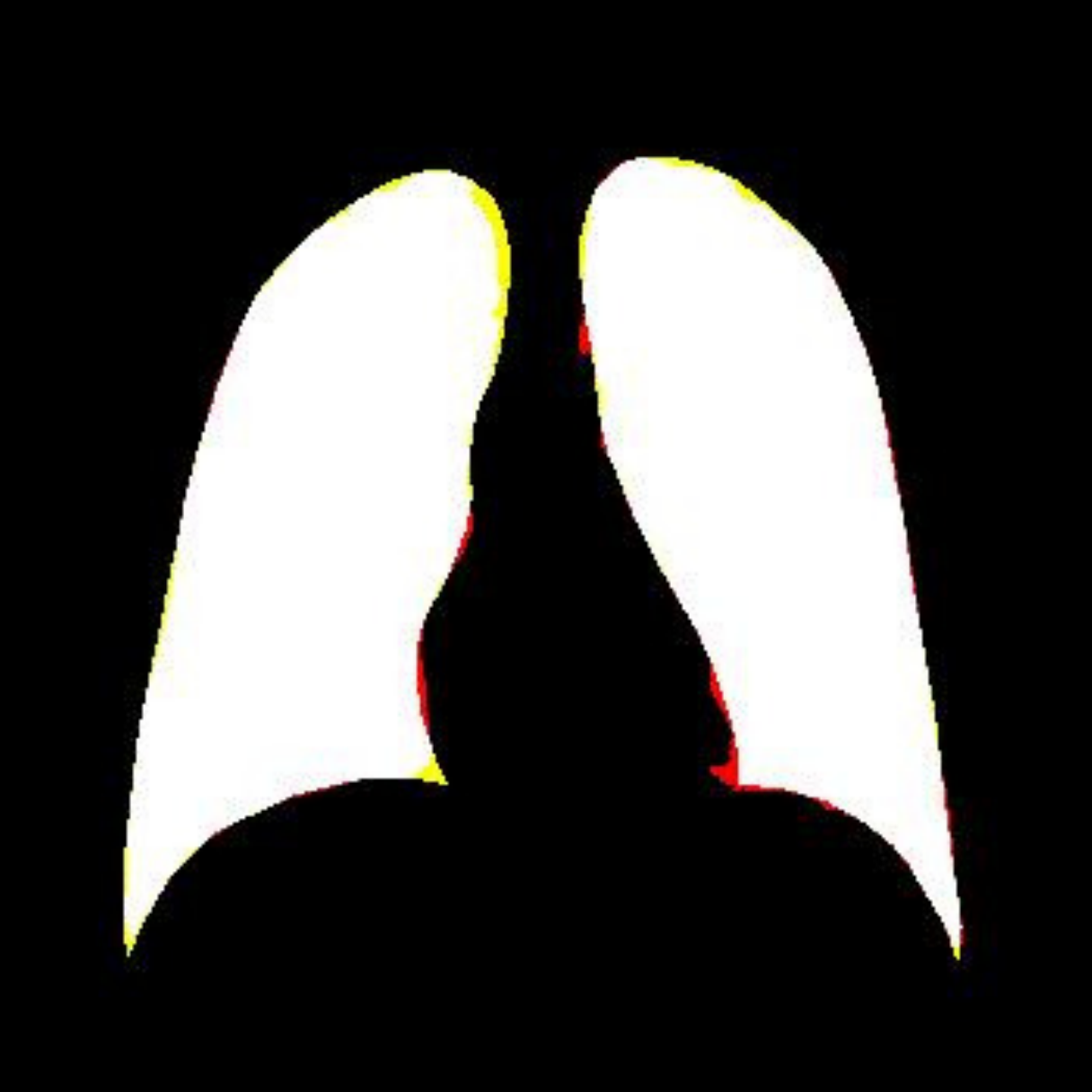} \\
        \includegraphics[width=0.25\textwidth]{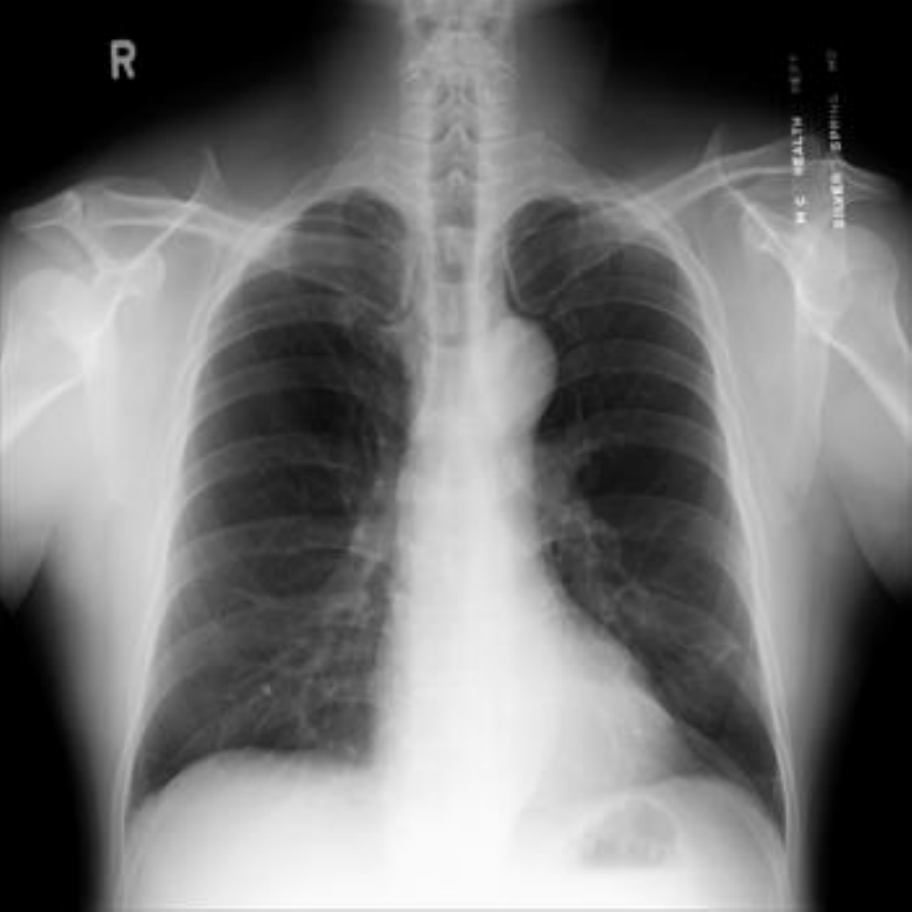} &
        \includegraphics[width=0.25\textwidth]{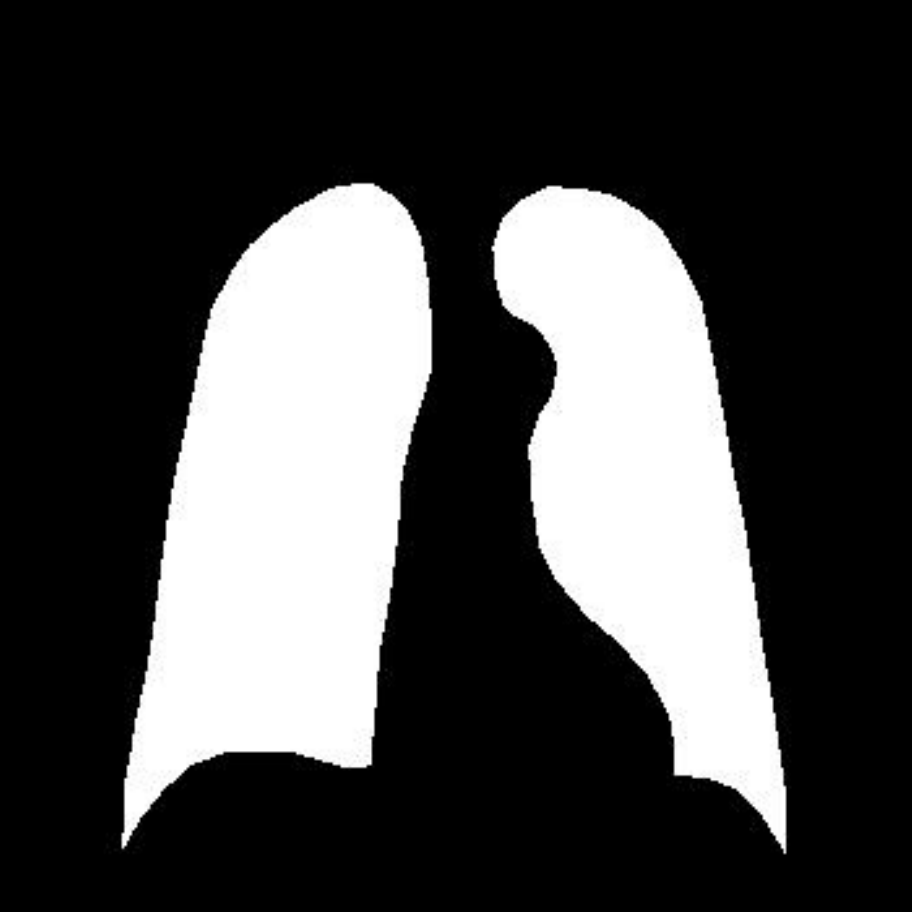} &
        \includegraphics[width=0.25\textwidth]{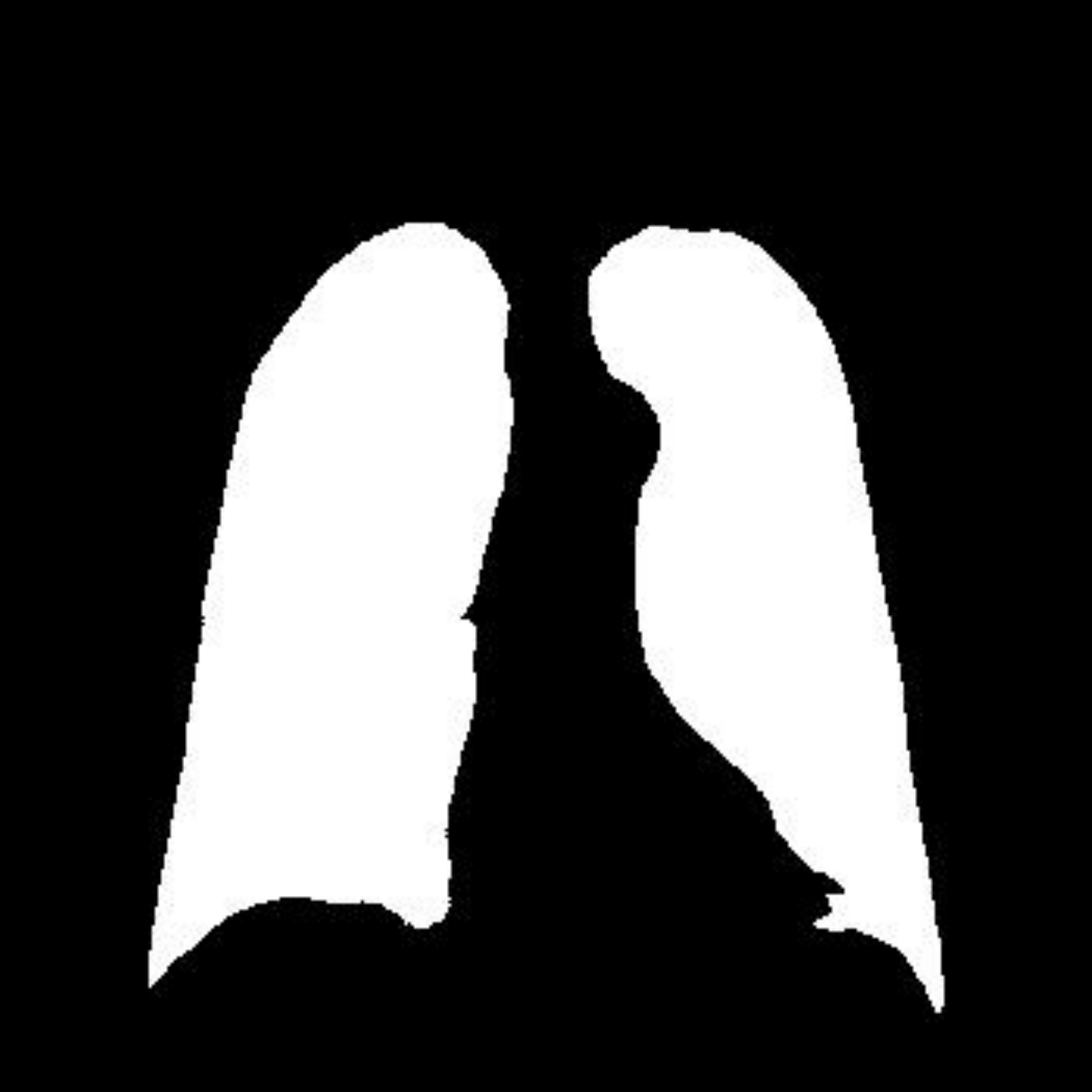} &
        \includegraphics[width=0.25\textwidth]{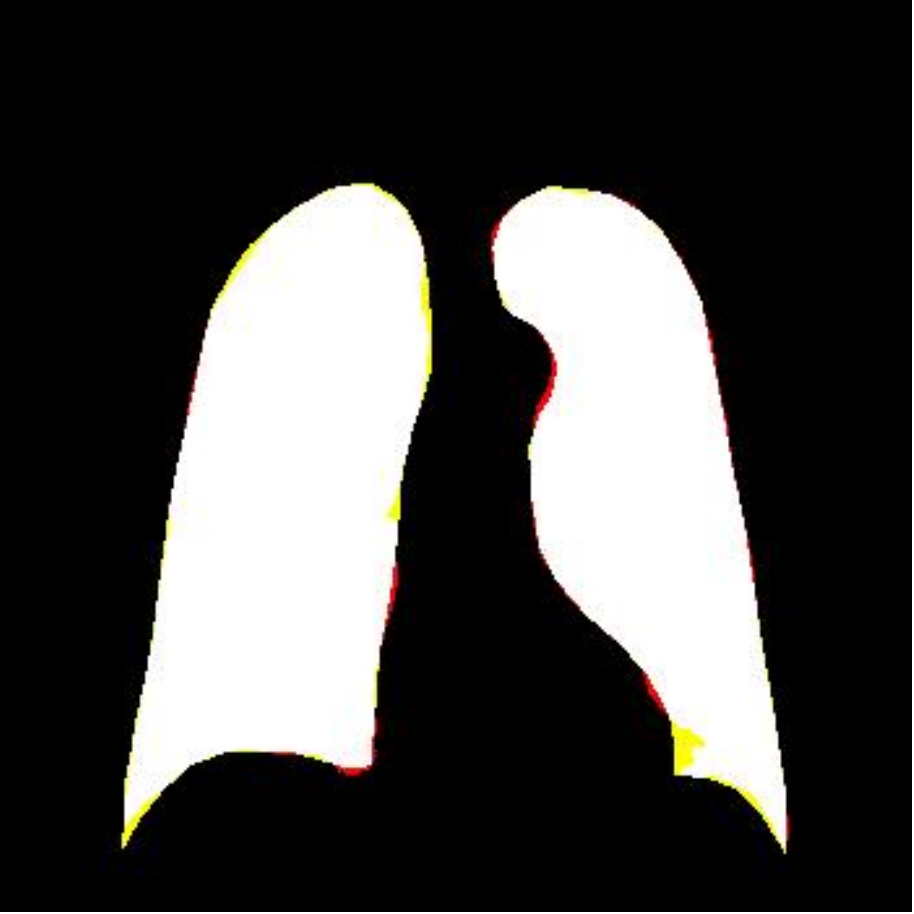} \\
    \end{tabular}}
    \caption{Sample visual results of the experiments on the MC dataset. From left to right: the input images, the manually annotated gold standard maps, the binary segmentation output from MKIS-Net, and the visual accuracy map. The latter shows true positive and negative predictions in black and white and the false positive and negative predictions in red and yellow colors, respectively.}
  \label{visualCHEST}
\end{figure}

Finally, we evaluated our method on the MC chest X-ray dataset. The qualitative results (Fig.\ \ref{visualCHEST}) show that MKIS-Net also performs well for these high-resolution images where the task is to segment large regions of interest. From the quantitative results (Table~\ref{CHEST}) we see that our network achieved comparable results to X-RayNet-1 \cite{jcm9030871} and outperformed other state-of-the-art networks despite being far smaller in size ($\approx60\times$ fewer trainable parameters).

\subsection{Comparison with Light-Weight Architectures}
\begin{figure}[!t]
    \centering
    \resizebox{1\textwidth}{!}{%
    \begin{tabular}{@{}c@{\ }c@{\ }c@{\ }c@{\ }c@{}}
        \includegraphics[width=0.2\textwidth]{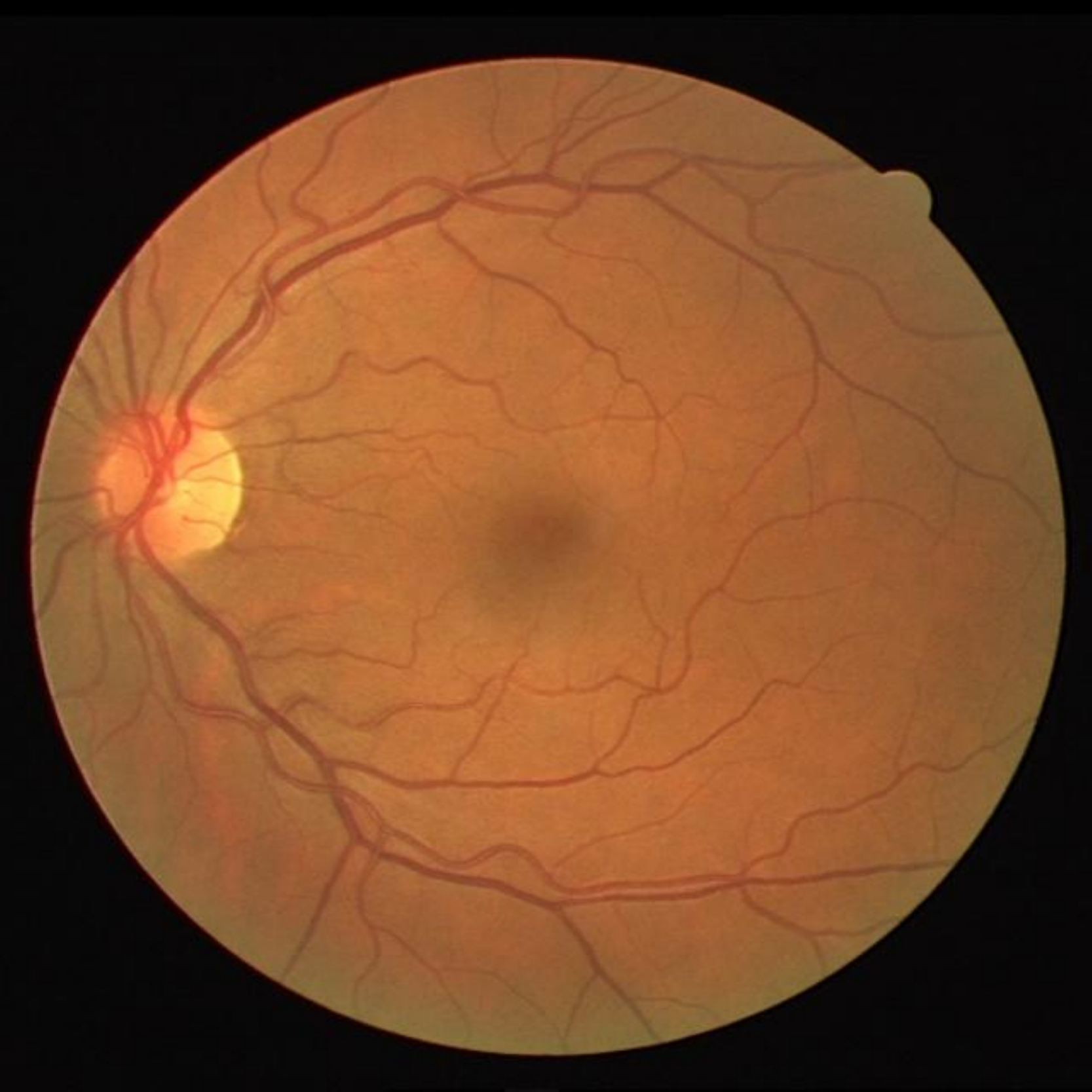} &
        \includegraphics[width=0.2\textwidth]{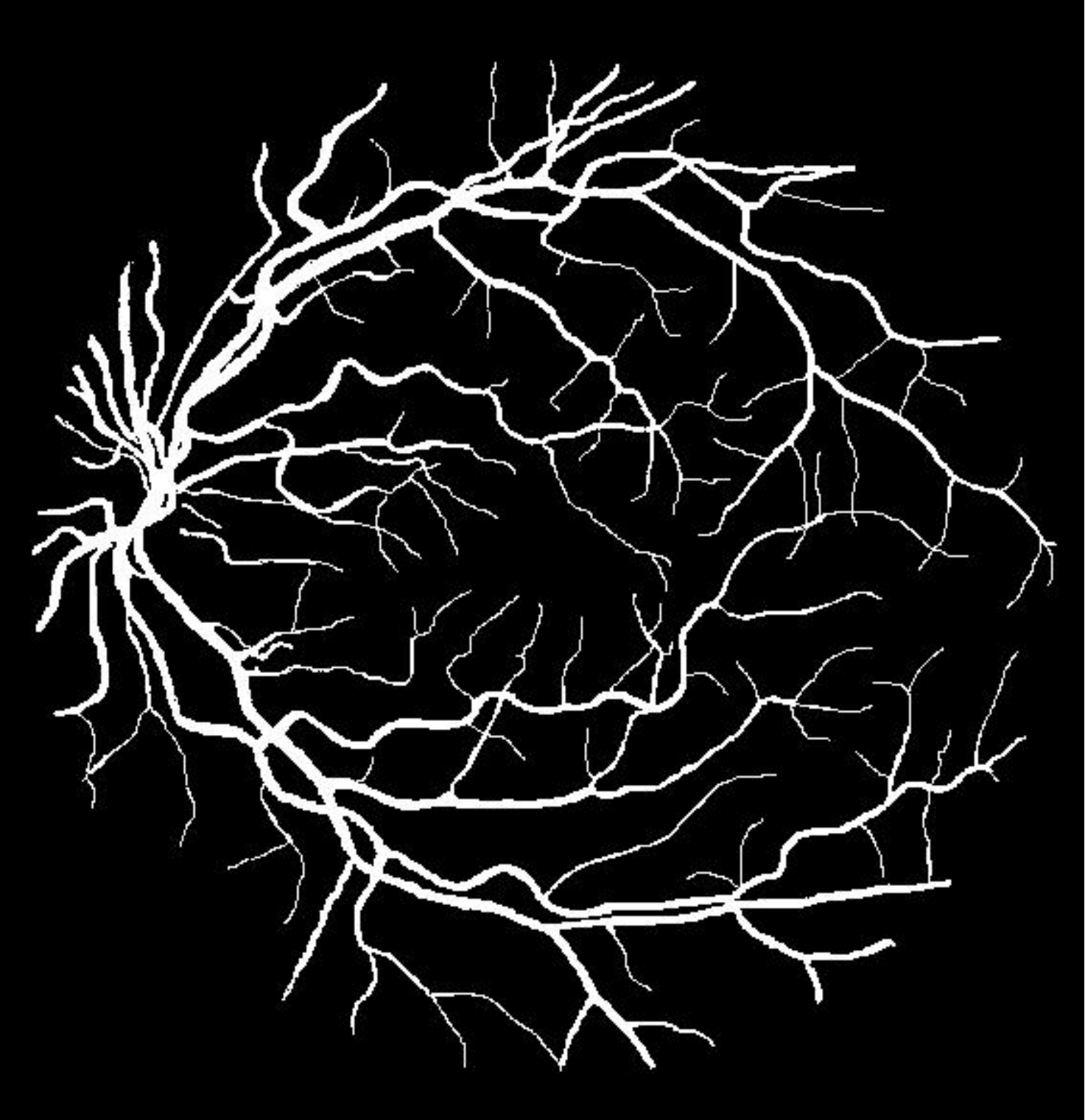} &
        \includegraphics[width=0.2\textwidth]{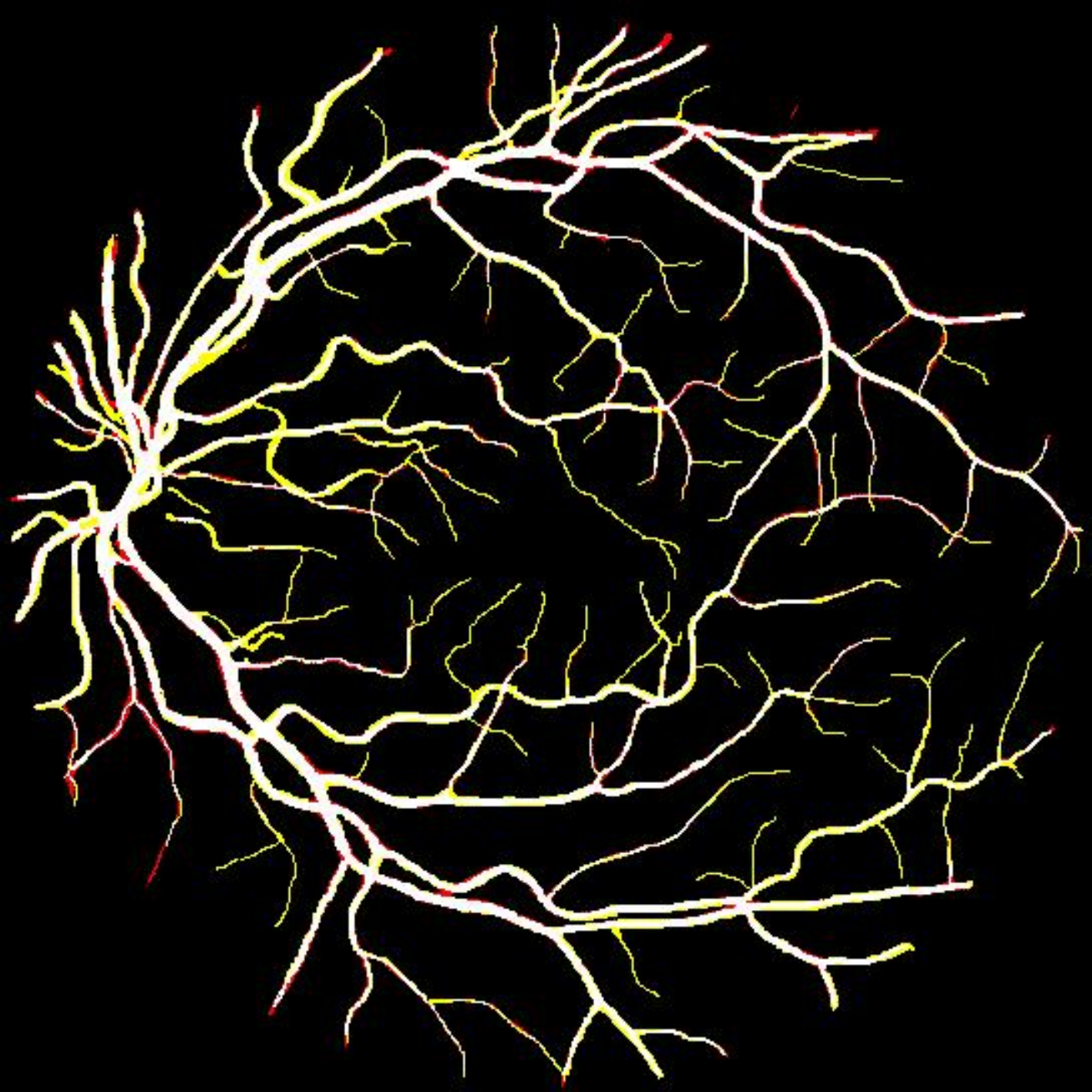} &
        \includegraphics[width=0.2\textwidth]{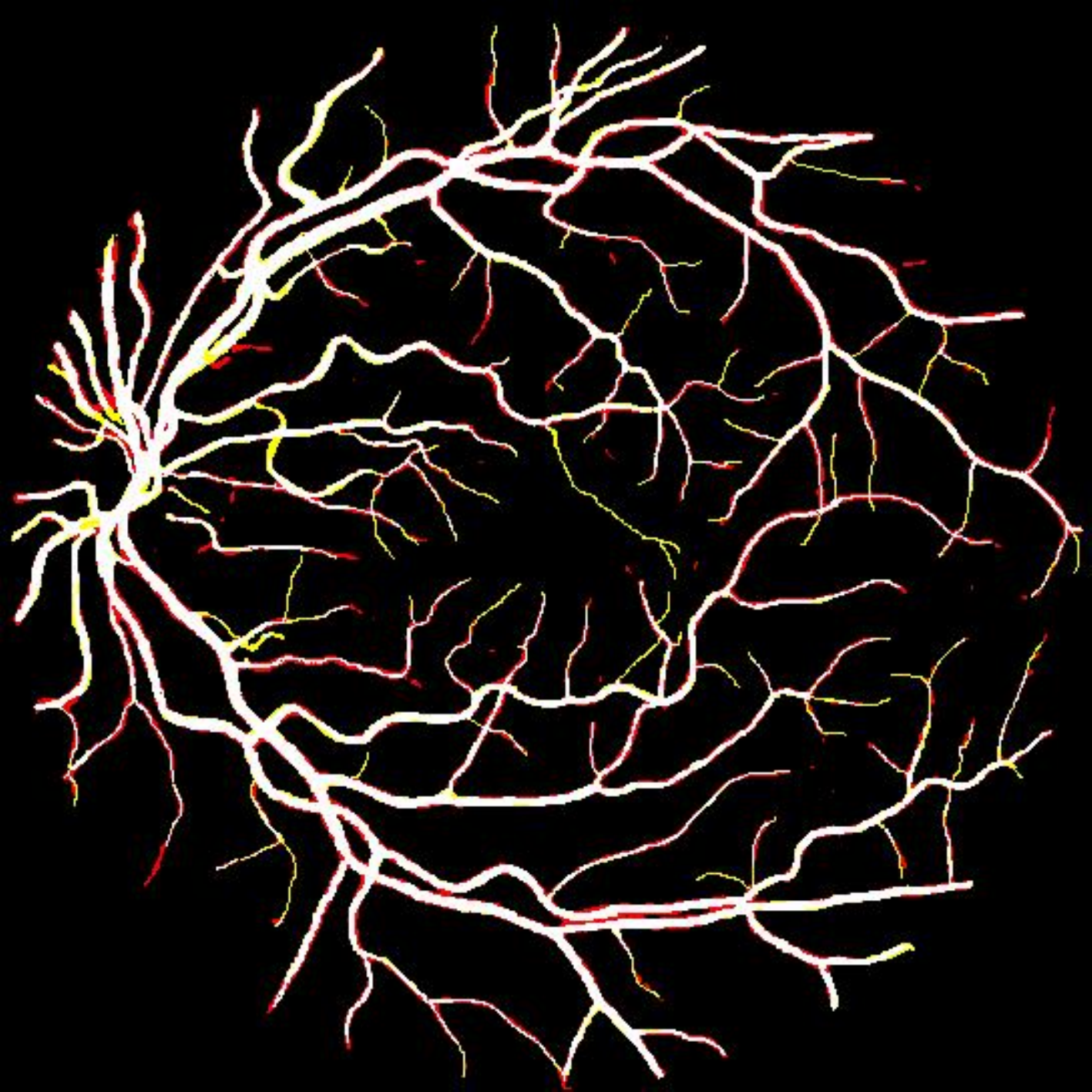} &
        \includegraphics[width=0.2\textwidth]{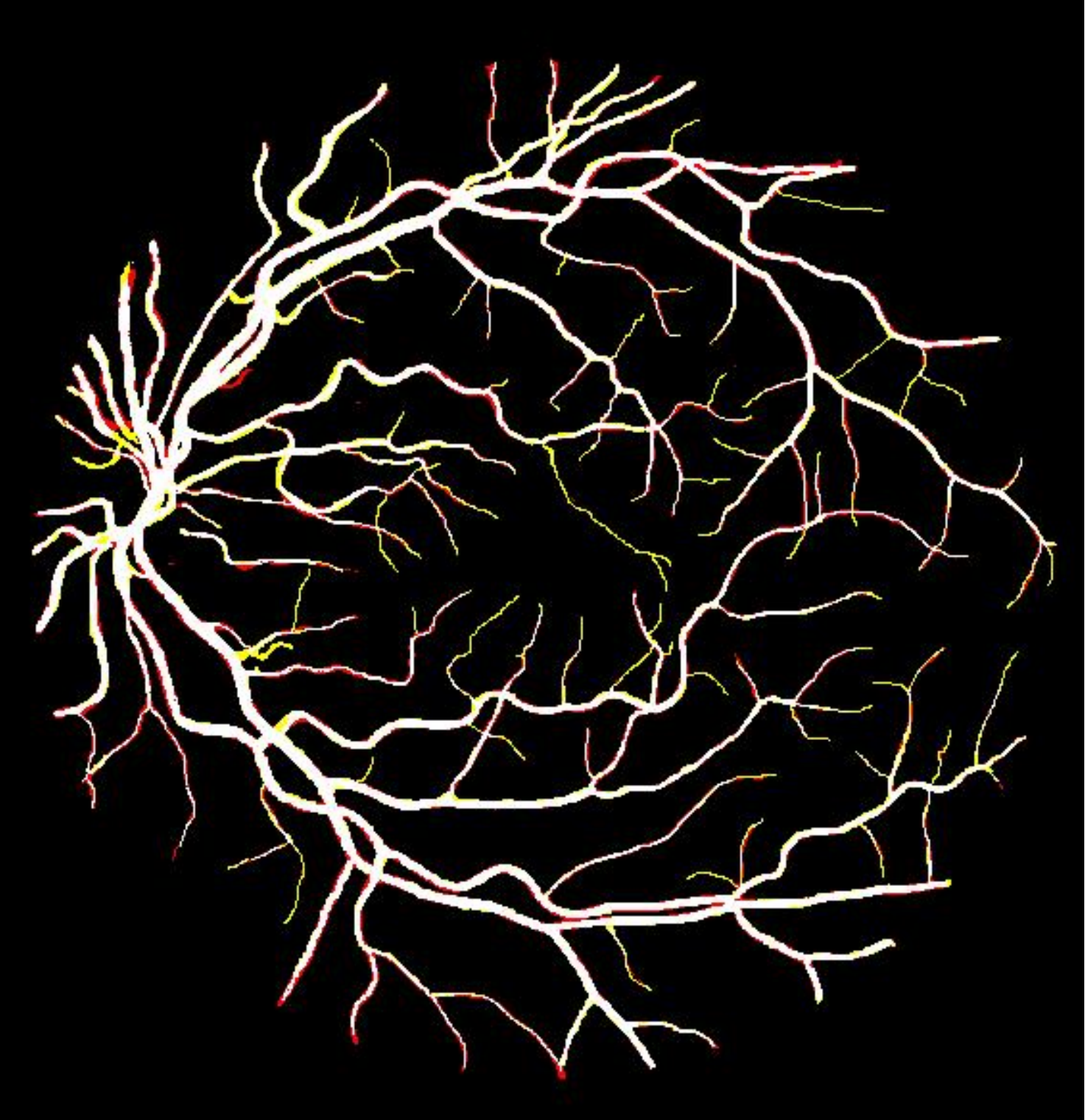} \\
        \includegraphics[width=0.2\textwidth]{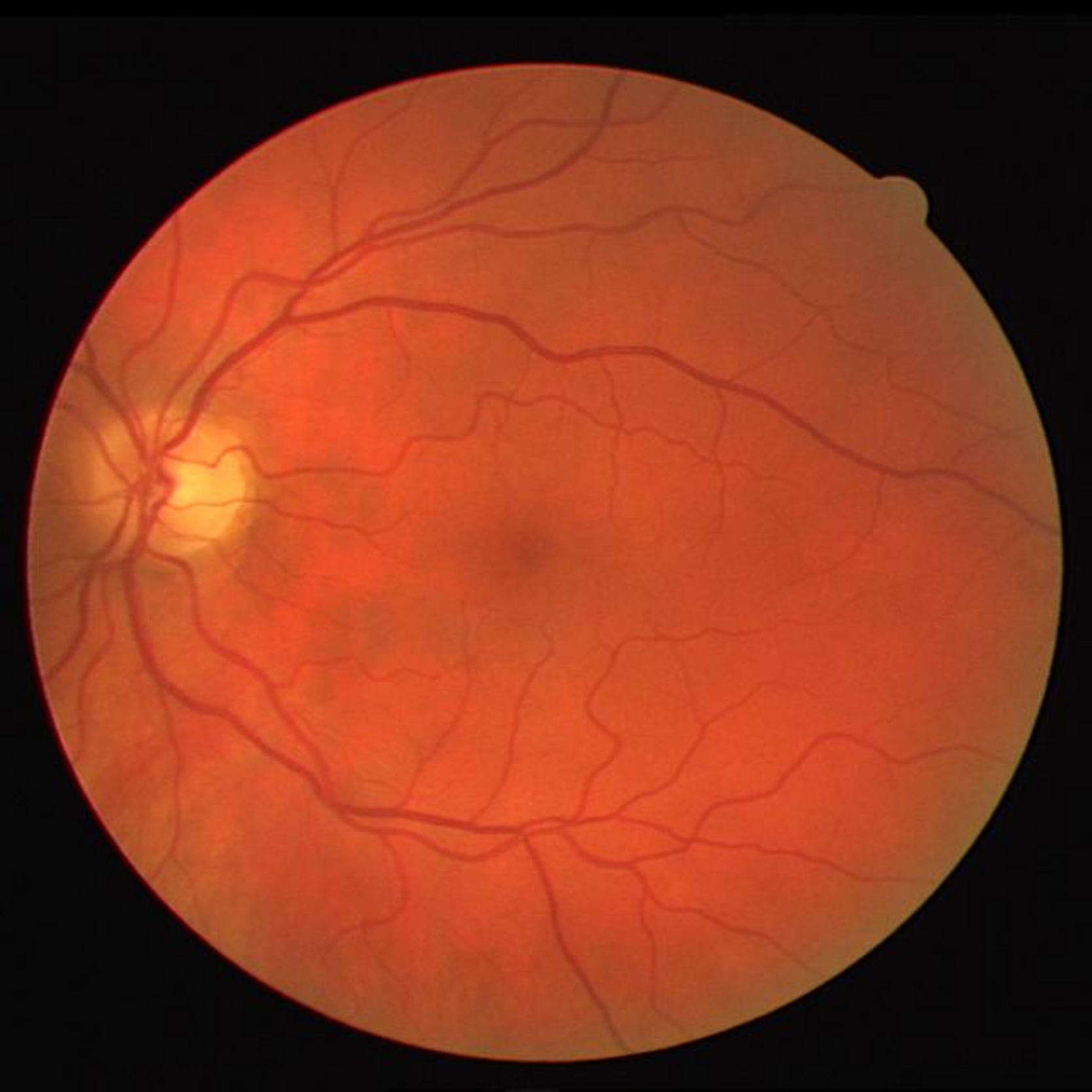} &
        \includegraphics[width=0.2\textwidth]{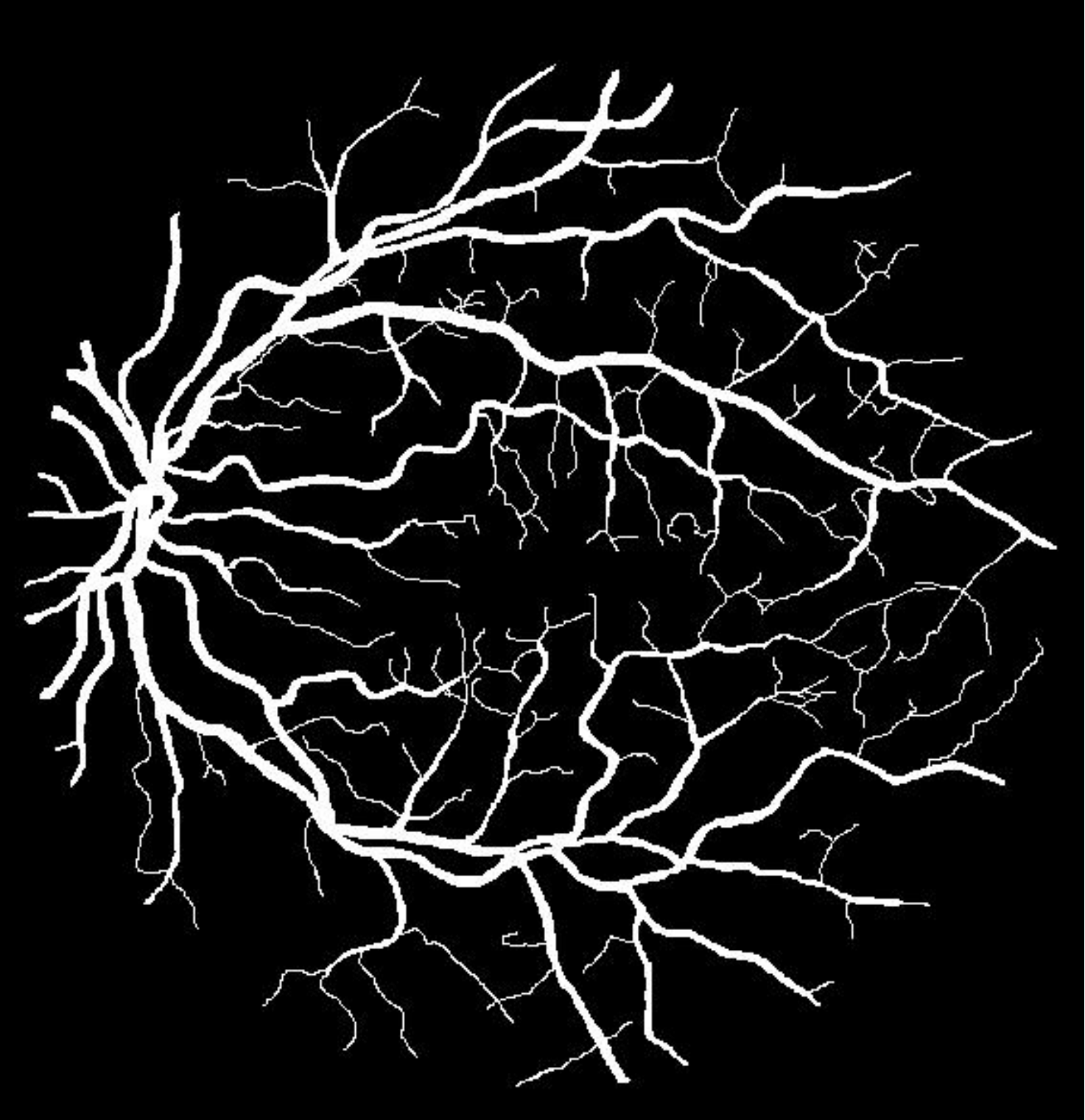} &
        \includegraphics[width=0.2\textwidth]{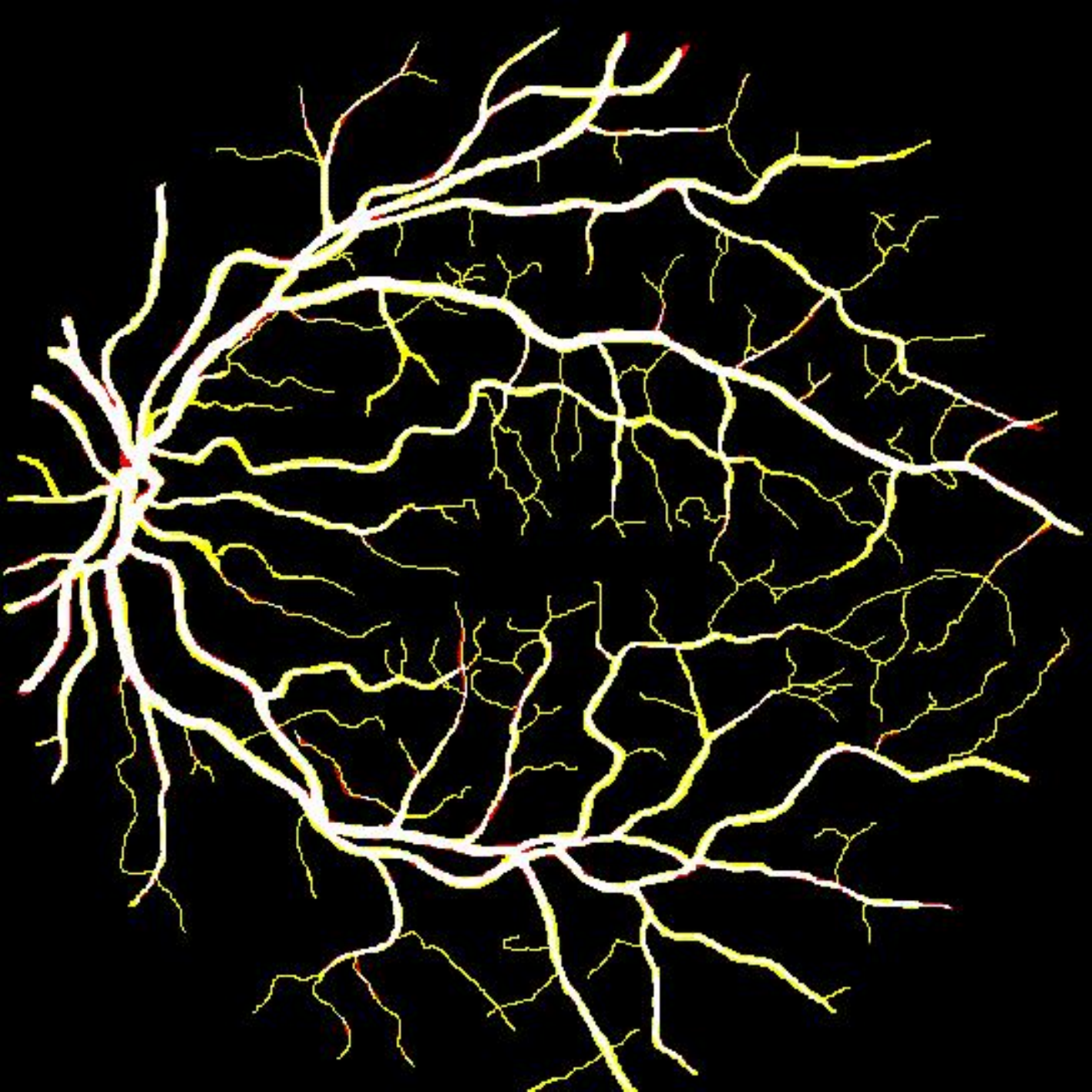} &
        \includegraphics[width=0.2\textwidth]{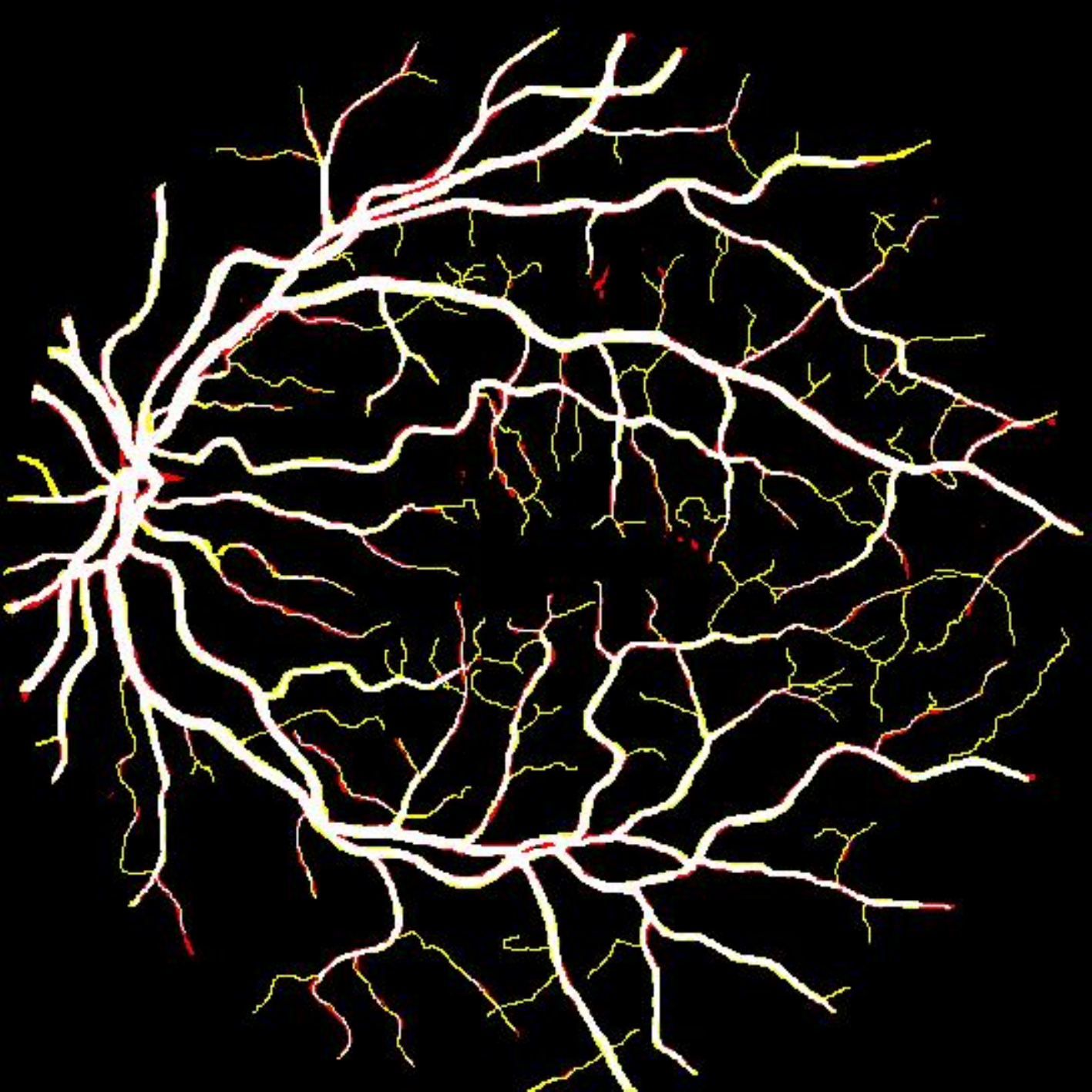} &
        \includegraphics[width=0.2\textwidth]{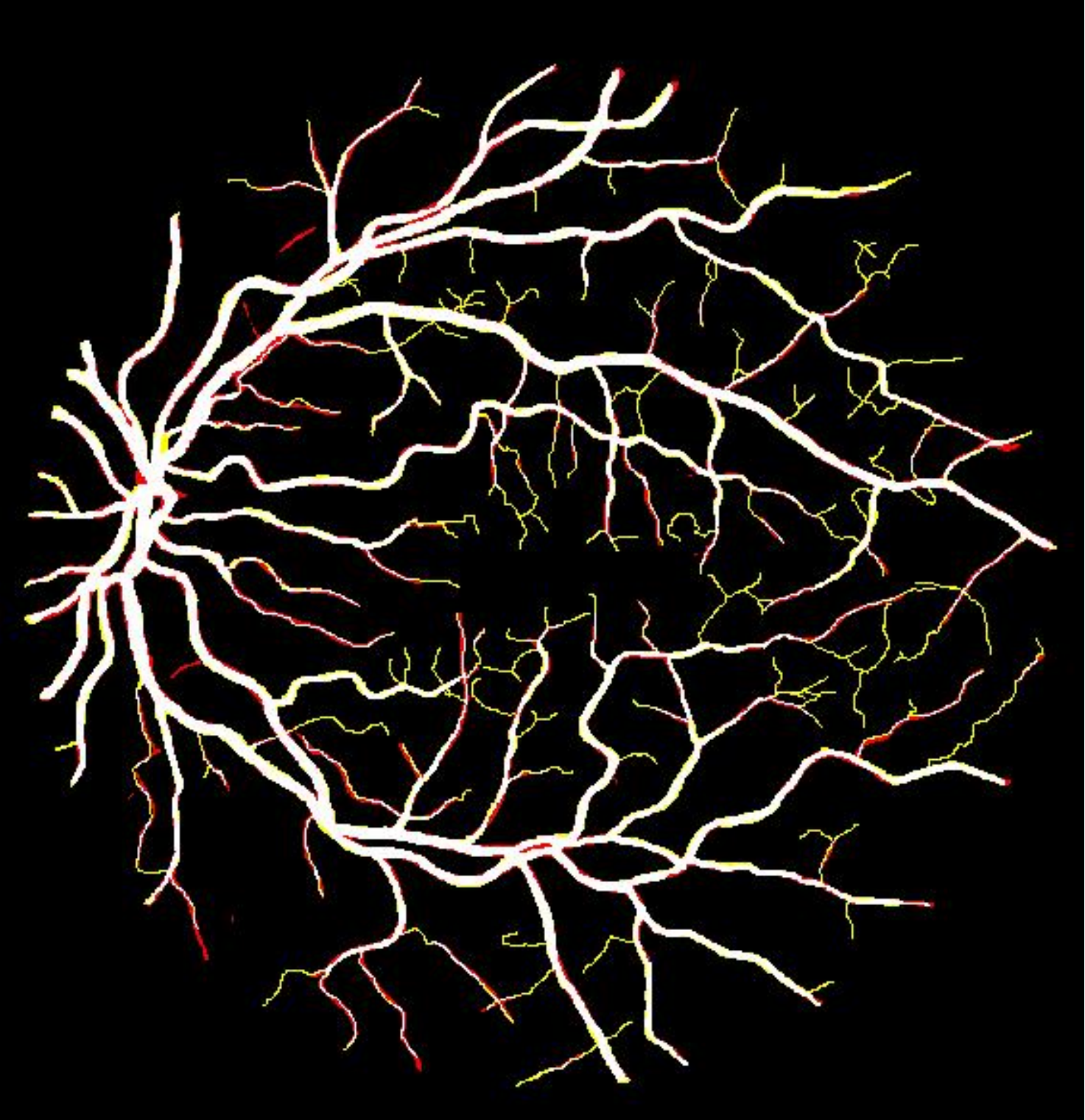} \\
    \end{tabular}}
    \caption{Sample visual results of evaluating our network to two lighter-weight networks on the DRIVE dataset. From left to right: the input images, the gold standard vessel maps manually annotated by an expert, and the results produced by ERFNet and M2U-Net.}
    \label{visualDRIVE_LW}
\end{figure}

\begin{figure}[!t]
    \centering
    \resizebox{1\textwidth}{!}{%
    \begin{tabular}{@{}c@{\ }c@{\ }c@{\ }c@{\ }c@{}}
        \includegraphics[width=0.2\textwidth]{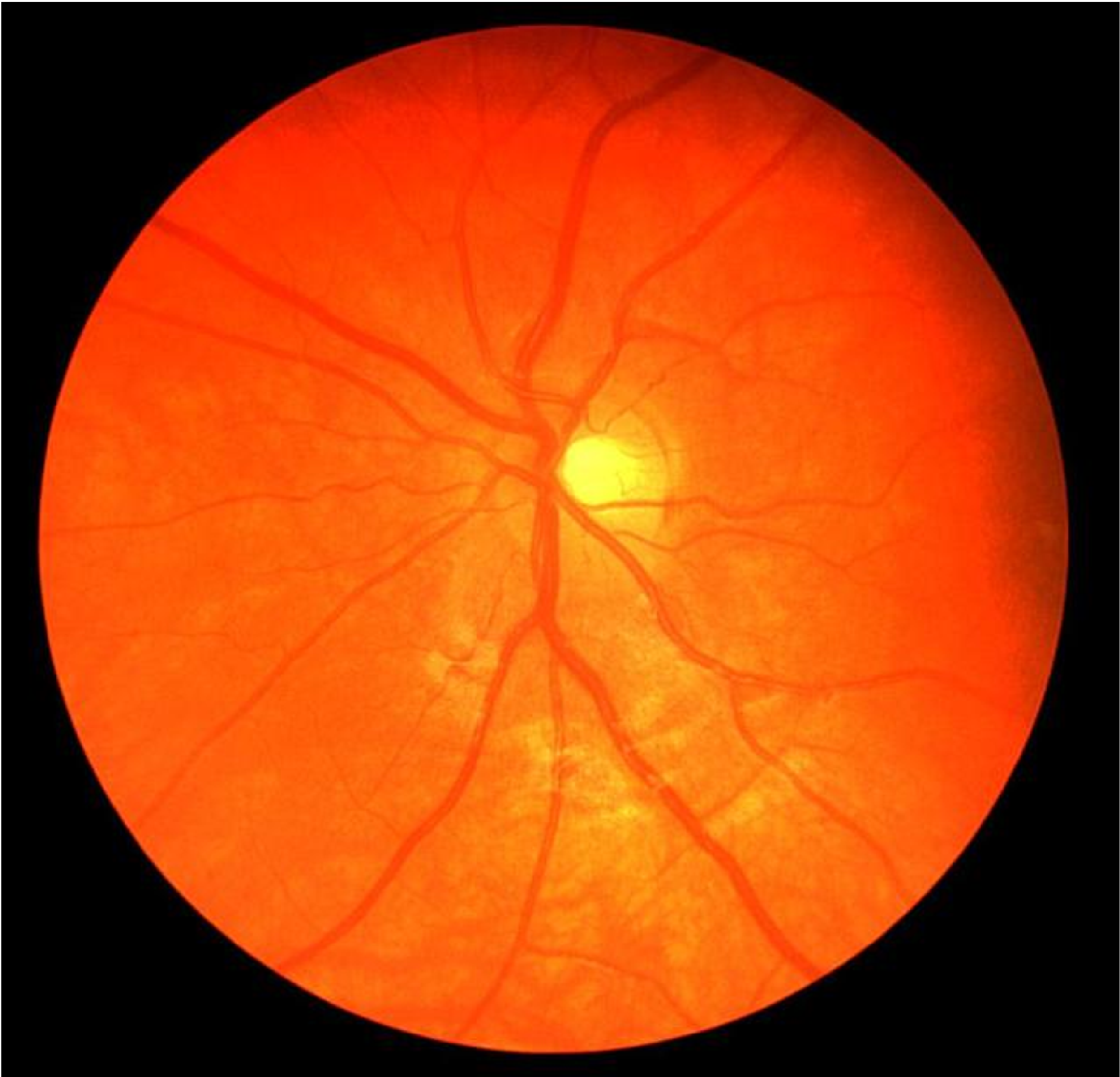} &
        \includegraphics[width=0.2\textwidth]{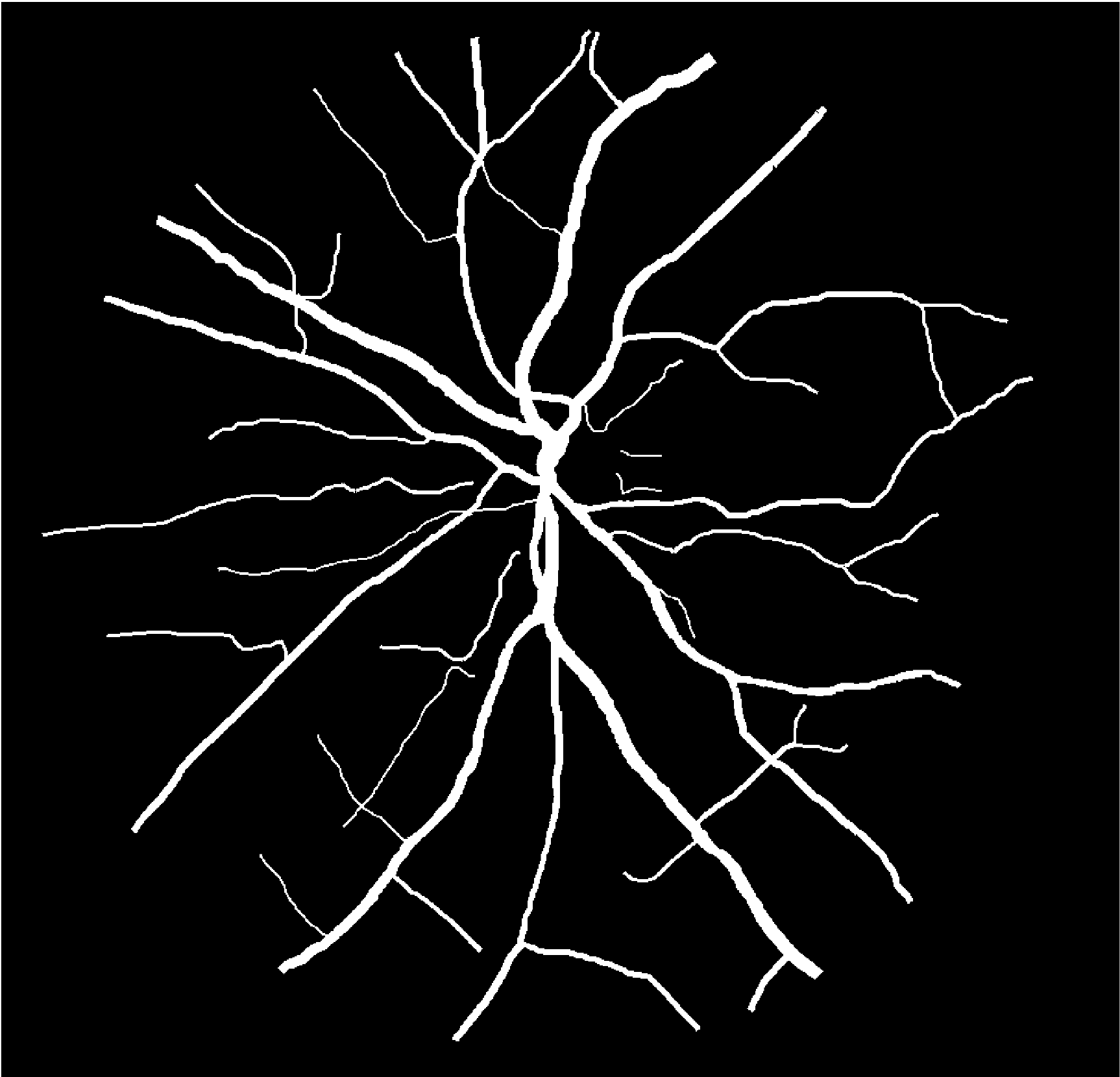} &
        \includegraphics[width=0.2\textwidth]{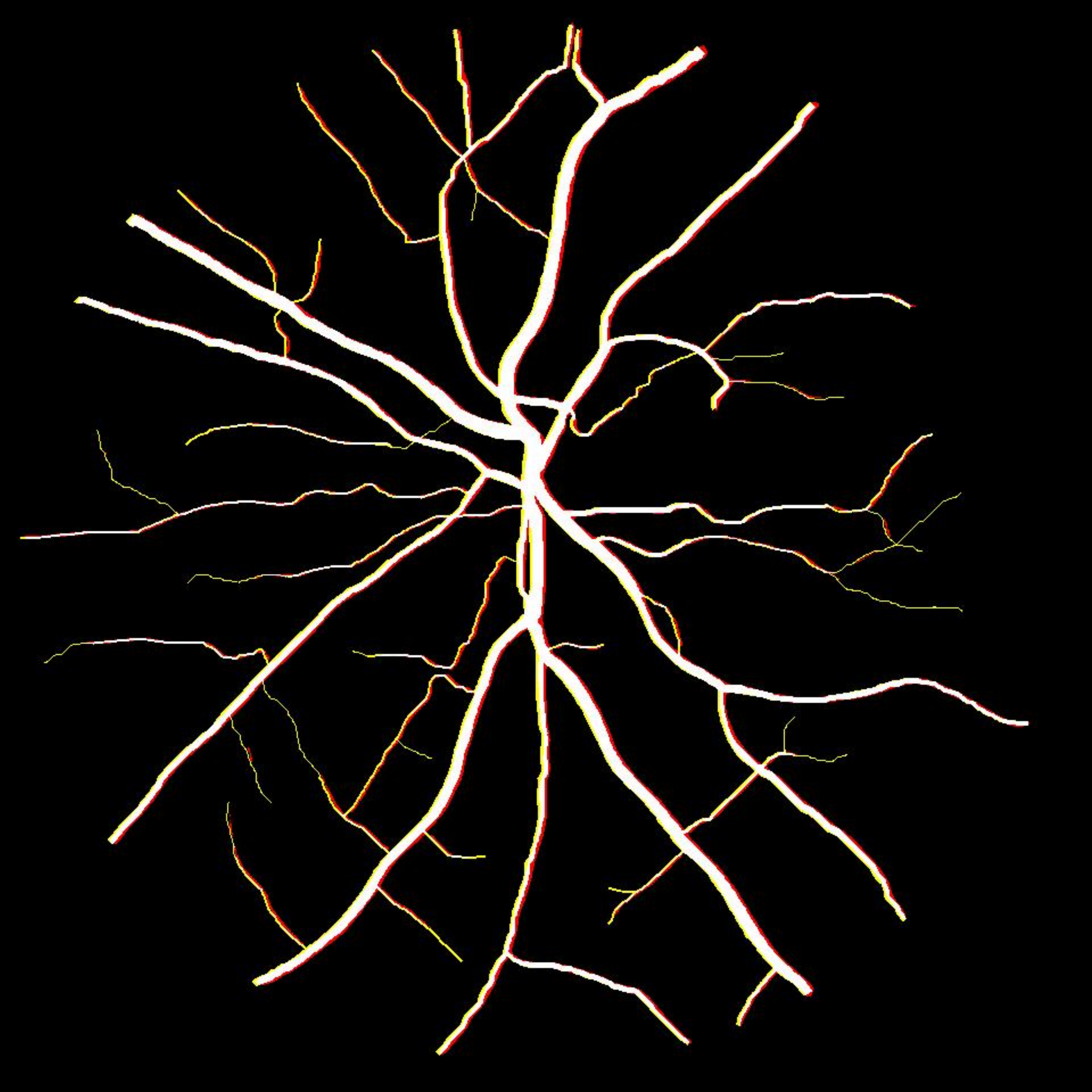} &
        \includegraphics[width=0.2\textwidth]{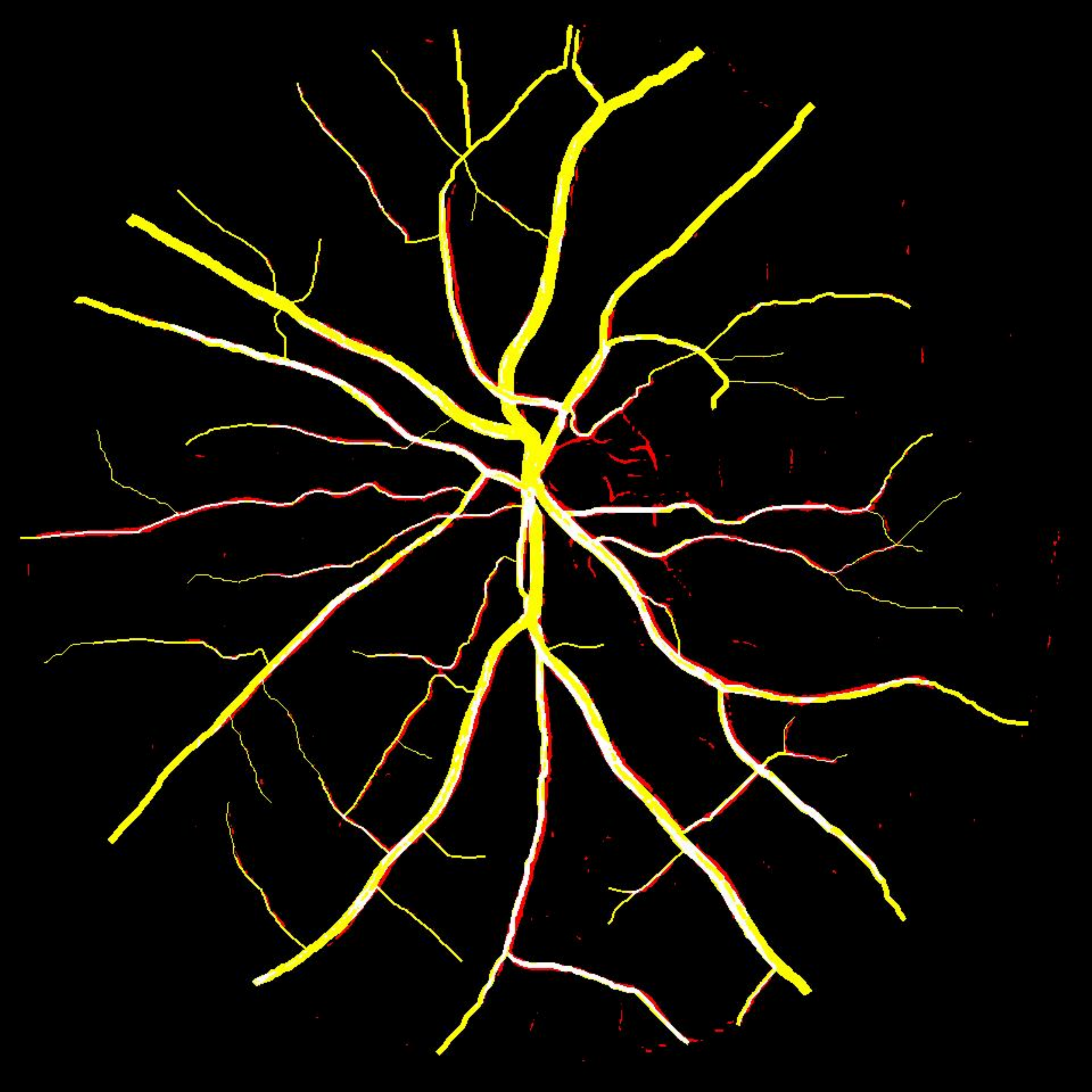} &
        \includegraphics[width=0.2\textwidth]{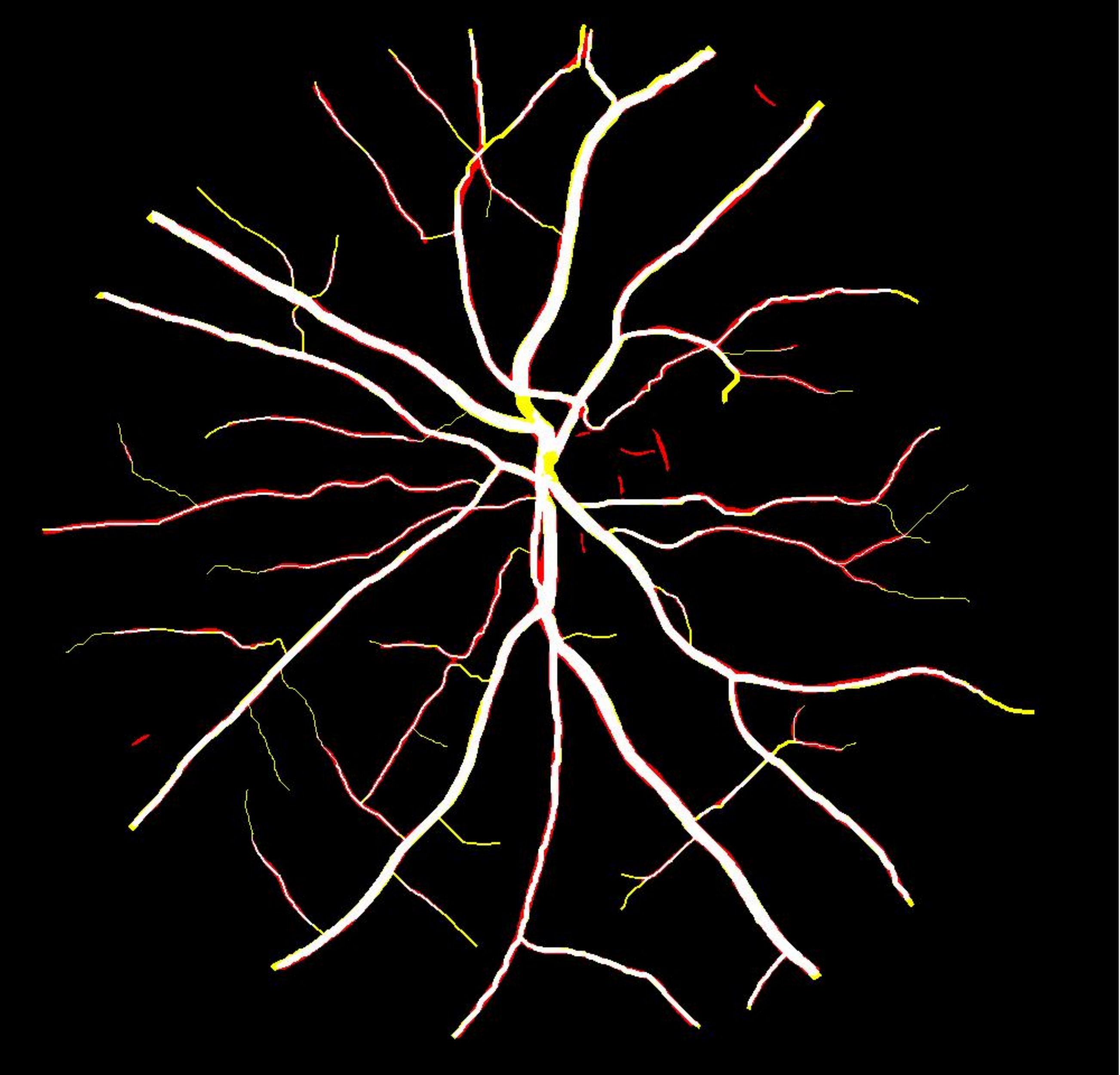} \\
        \includegraphics[width=0.2\textwidth]{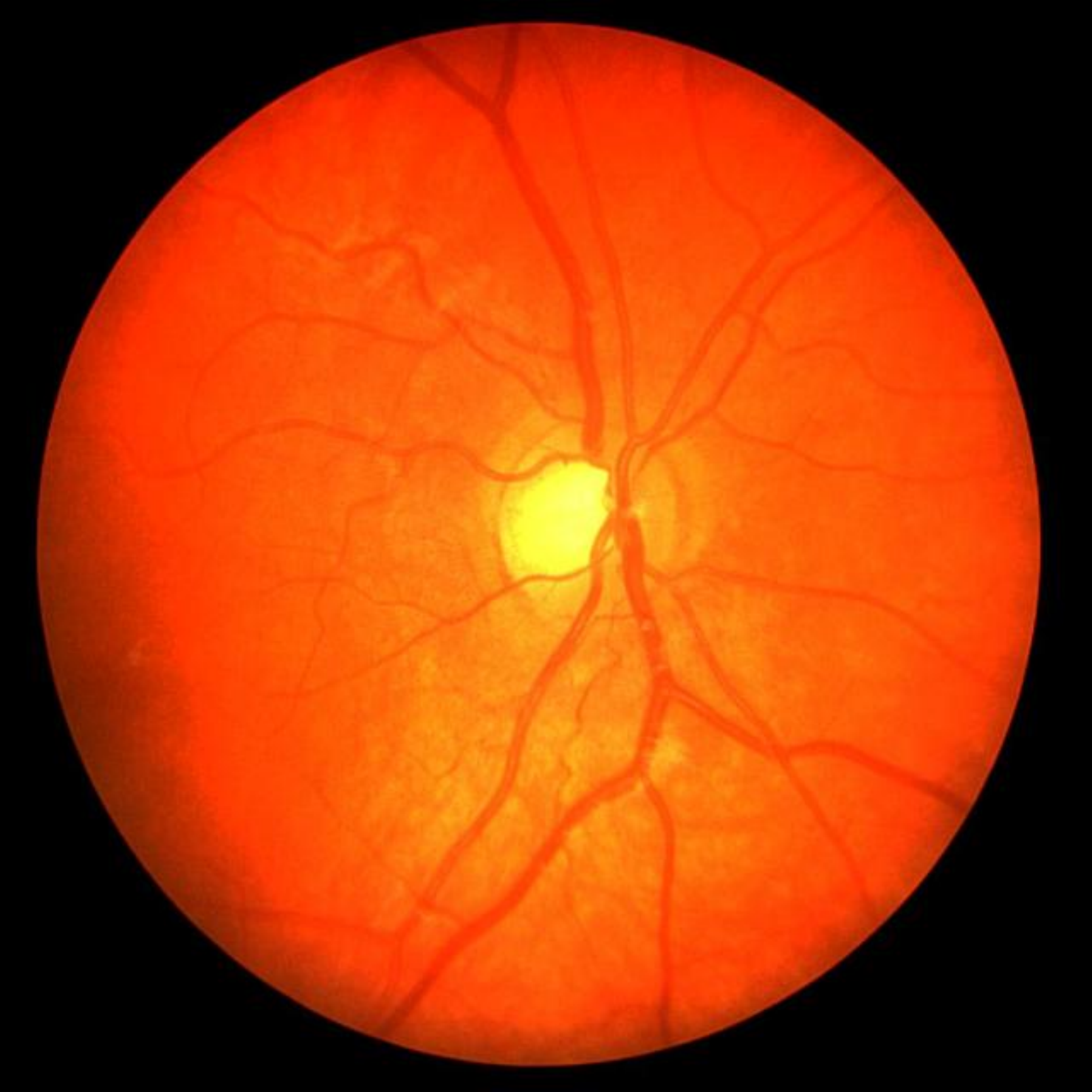} &
        \includegraphics[width=0.2\textwidth]{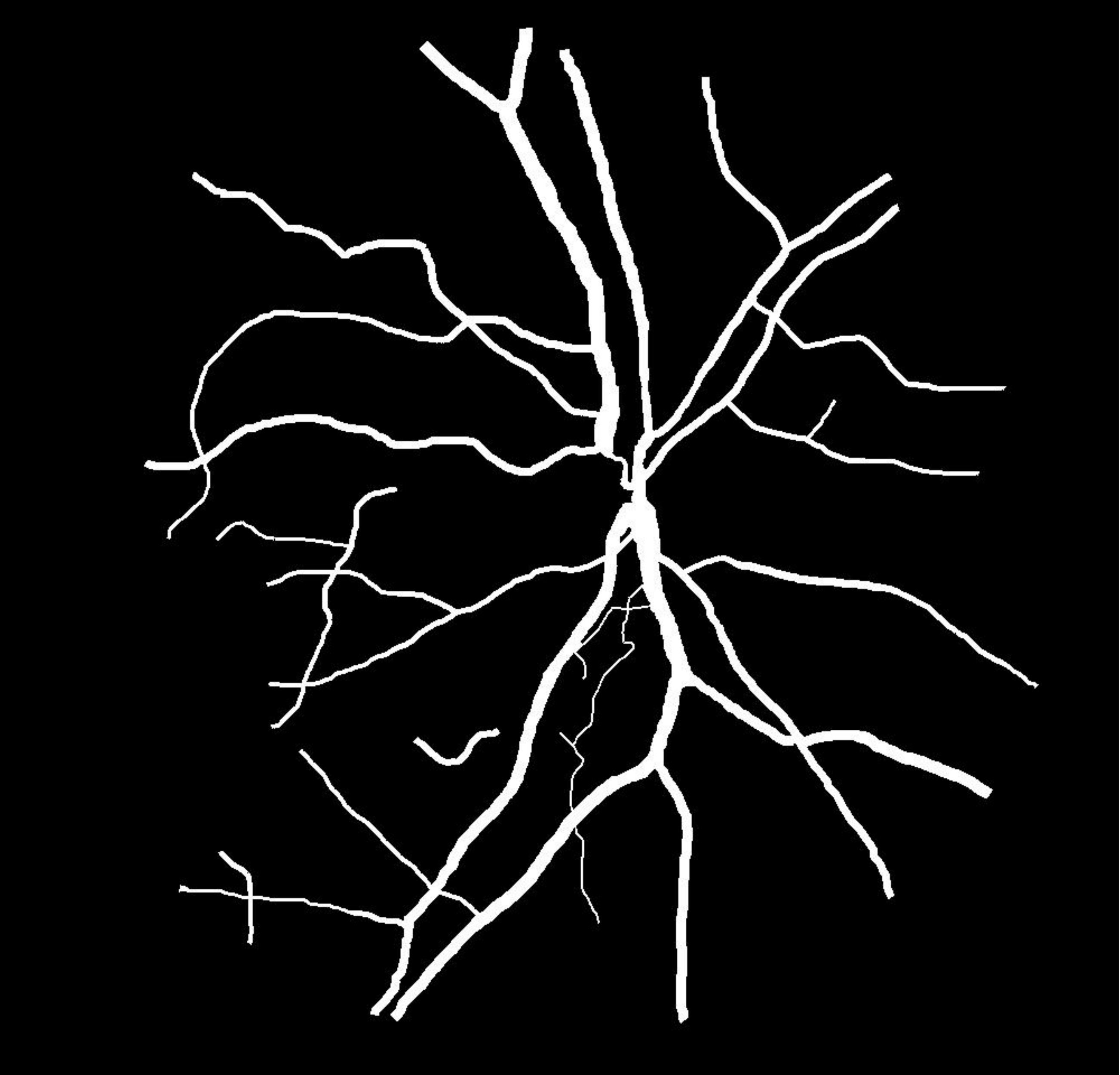} &
        \includegraphics[width=0.2\textwidth]{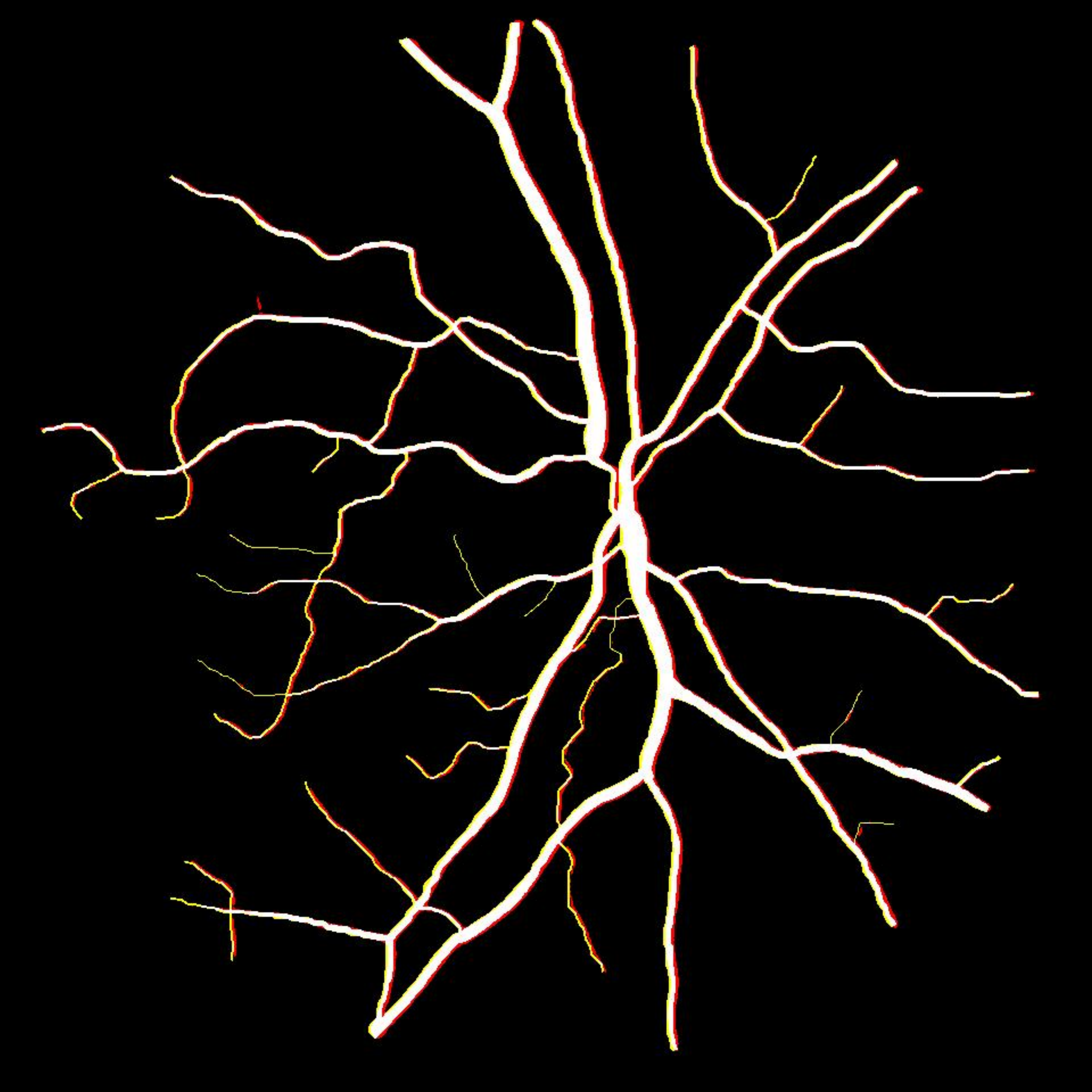} &
        \includegraphics[width=0.2\textwidth]{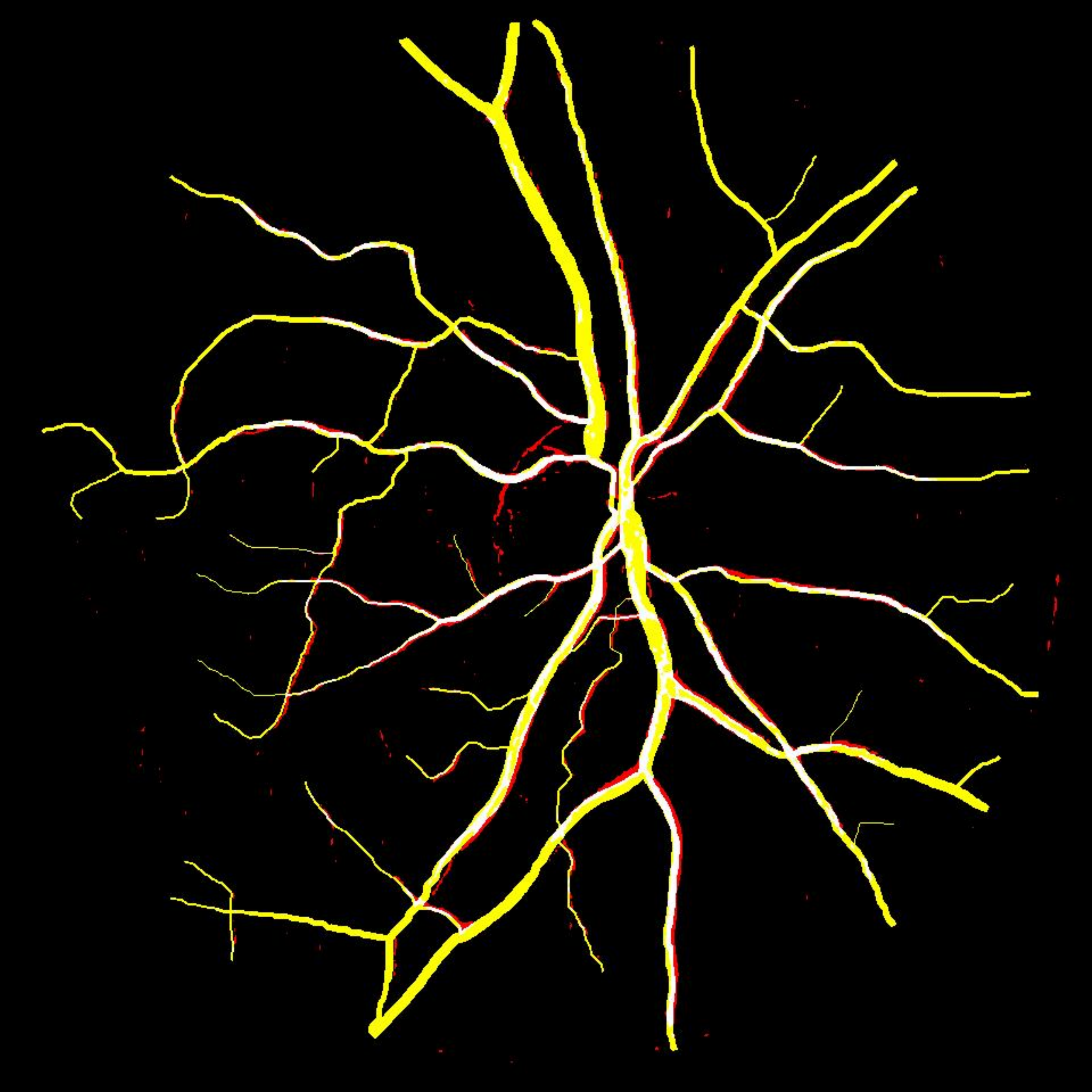} &
        \includegraphics[width=0.2\textwidth]{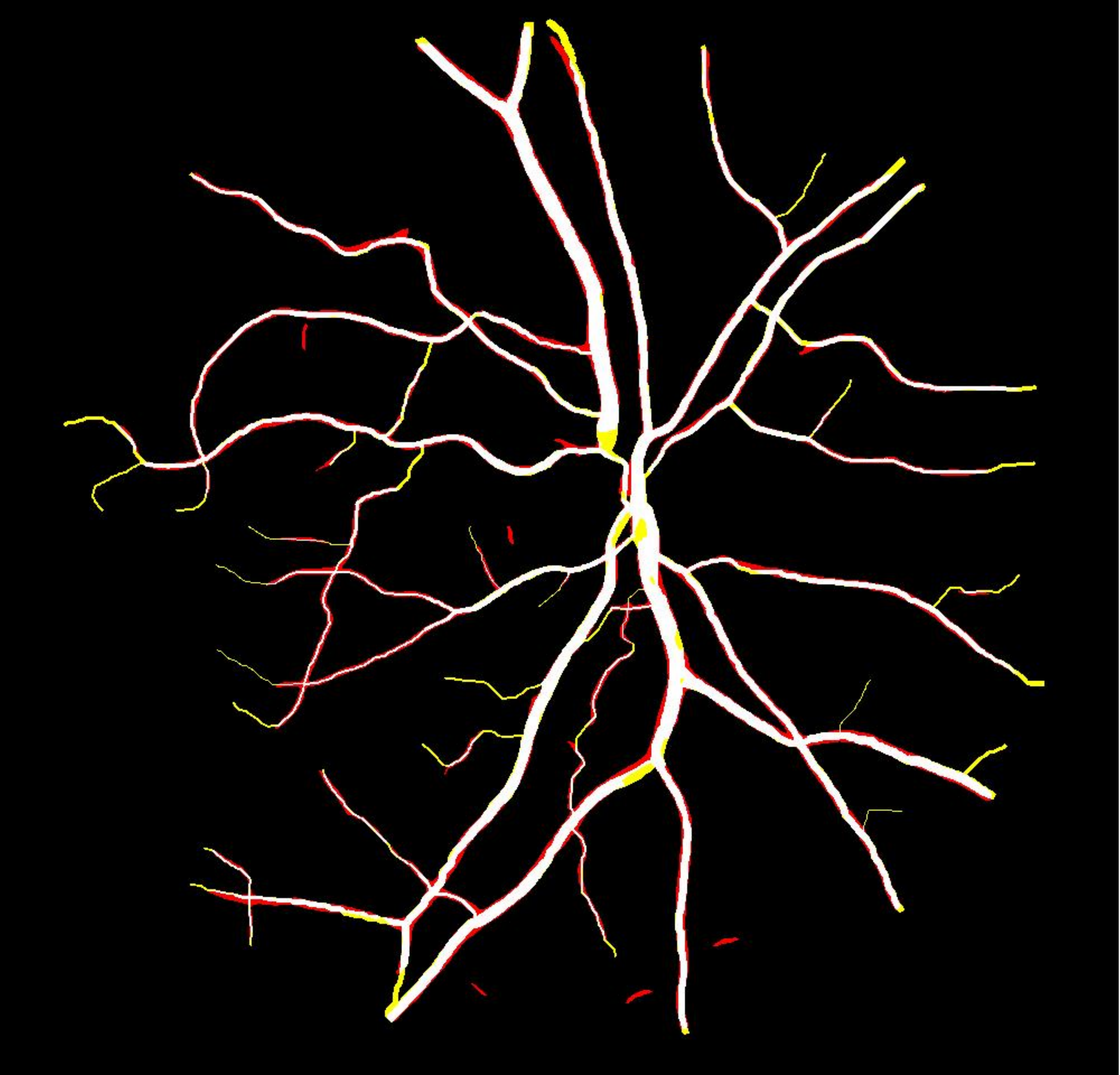} \\
    \end{tabular}}
    \caption{Sample visual results of evaluating our network to two lighter-weight networks on the CHASE dataset. From left to right: the input images, the gold standard vessel maps manually annotated by an expert, and the results produced by ERFNet and M2U-Net.}
    \label{visualCHASE_LW}
\end{figure}

We also compared MKIS-Net both qualitatively and quantitatively to light-weight networks recently presented in literature. To this end, we used the two retinal vessel segmentation datasets DRIVE and CHASE. The qualitative results (Figs.\ \ref{visualDRIVE_LW} and \ref{visualCHASE_LW}) comparing MKIS-Net with ERFNet \cite{Romera_2018} and M2U-Net \cite{Laibacher_2019} are consistent with the findings in the previous subsection, in the sense that our network does deliver good segmentation results, preserving detail and coping with complex, thin vessel structures. We performed a quantitative comparison (Table~\ref{lightWeight_perf}) of these nets in terms of the model size (in MB), number of parameters (Params), multiply and add operations in billions (MAdds), accuracy and F1 score, so as to gauge the computational requirements versus segmentation performance. Note that MKIS-Net yields more than 2\% improvement in segmentation performance while being more than 3.5 times smaller parameter-wise as compared with the second smallest alternative (M2U-Net) in this comparison.

\begin{table*}[!t]
    \centering
    \begin{tabular}{ccccccc}
    \toprule
    \textbf{Method} & \textbf{Se} & \textbf{Sp} & \textbf{Acc} & \textbf{AUC} & \textbf{F1} & \textbf{Params (M)} \\
    \midrule
    VessNet \cite{Arsalan2019}  &   0.8022  & 0.9810  & 0.9655  &0.9820   & N/A & 9\\
    U-Net++ \cite{Zongwei2018} & 0.8116 & 0.9823 & 0.9673 &0.9815& 0.8126 & N/A\\
    MultiResUNet \cite{IBTEHAZ202074} & 0.7900  & 0.9848 & 0.9678 & 0.9784 & 0.8108 & N/A\\
    DRIU \cite{Maninis2016} & 0.7855 & 0.9799 & 0.9552 & 0.9793 & 0.8220 & 7.8 \\
    Patch BTS-DSN \cite{Guo2019} & 0.7891 & 0.9804 & 0.9561 & 0.9806 & 0.8249 & 7.8 \\
    Image BTS-DSN \cite{Guo2019} & 0.7800  & 0.9806 & 0.9551 & 0.9796 & 0.8208 & 7.8 \\
    U-Net \cite{guo2020dpn} & 0.7849 & 0.9802 & 0.9554 & 0.9761 & 0.8175 & 3.4 \\
    Vessel-Net \cite{Wu2019} & 0.8038 & 0.9802 & 0.9578 & 0.9821 & N/A   & 1.7 \\
    MS-NFN \cite{Wu2018} & 0.7844 & 0.9819 & 0.9567 & 0.9807 & N/A   & 0.4 \\
    FCN \cite{oliveira2018retinal} & 0.8039 & 0.9804 & 0.9576 & 0.9821 & N/A   & 0.2 \\
    T-Net\cite{khan2022t} & 0.8262 & \textbf{0.9862} & 0.9697 & \textbf{0.9867} & 0.8269 & \textbf{0.03} \\
    MKIS-Net (Proposed) & \textbf{0.8338} & 0.9828 & \textbf{0.9697} & 0.9827 & \textbf{0.8283} & 0.152 \\
    \bottomrule
    \end{tabular}
    \caption{Results of MKIS-Net and other state-of-the-art nets on the DRIVE dataset. Bold indicates best performance.}
    \label{DRIVE}
\end{table*}

\begin{table}[!t]
  \centering
  \resizebox{1\textwidth}{!}{
    \begin{tabular}{cccccc}
    \toprule
    \textbf{Method} & \textbf{Se} & \textbf{Sp} & \textbf{Acc} & \textbf{AUC} & \textbf{F1} \\
    \midrule
    MS-NFN \cite{Wu2018} & 0.7538 & 0.9847 & 0.9637 & 0.9825 & N/A \\
    Three-stage FCN \cite{8476171} & 0.7641 & 0.9806 & 0.9607 & 0.9776 & N/A \\
    BTS-DSN \cite{Guo2019} & 0.7888 & 0.9801 & 0.9627 & 0.9840 & 0.7983 \\
    Vessel-Net \cite{Wu2019}& 0.8132 & 0.9814 & 0.9661 & 0.9860 & N/A \\
    DEU-Net \cite{Wang2019} & 0.8074 & 0.9821 & 0.9661 & 0.9812 & 0.8037 \\
    SegNet \cite{M.Khan2020} & 0.8190 & 0.9735 & 0.9638 & 0.9780 & 0.7981 \\
    U-Net \cite{guo2020dpn} & \textbf{0.8355} & 0.9698 & 0.9578 & 0.9784 & 0.7792 \\
    T-Net \cite{khan2022t} & 0.8323 & 0.9844& 0.9739 & \textbf{0.9889} & \textbf{0.8143} \\
    MKIS-Net (Proposed) & 0.8266 & \textbf{0.9848} & \textbf{0.9740} & 0.9863 & 0.8137 \\
    \bottomrule
    \end{tabular}
  }
  \caption{Results of MKIS-Net and other state-of-the-art nets on the CHASE dataset. Bold indicates best performance.}
    \label{CHASEDB1}
\end{table}

\begin{figure}[!t]
    \centering
    \resizebox{1\textwidth}{!}{%
    \begin{tabular}{@{}c@{\ }c@{\ }c@{\ }c@{}}
        \includegraphics[width=0.4\textwidth]{DRIVE/Oimg6.eps} &
        \includegraphics[width=0.4\textwidth]{DRIVE/gt6.eps} &
        \includegraphics[width=0.4\textwidth]{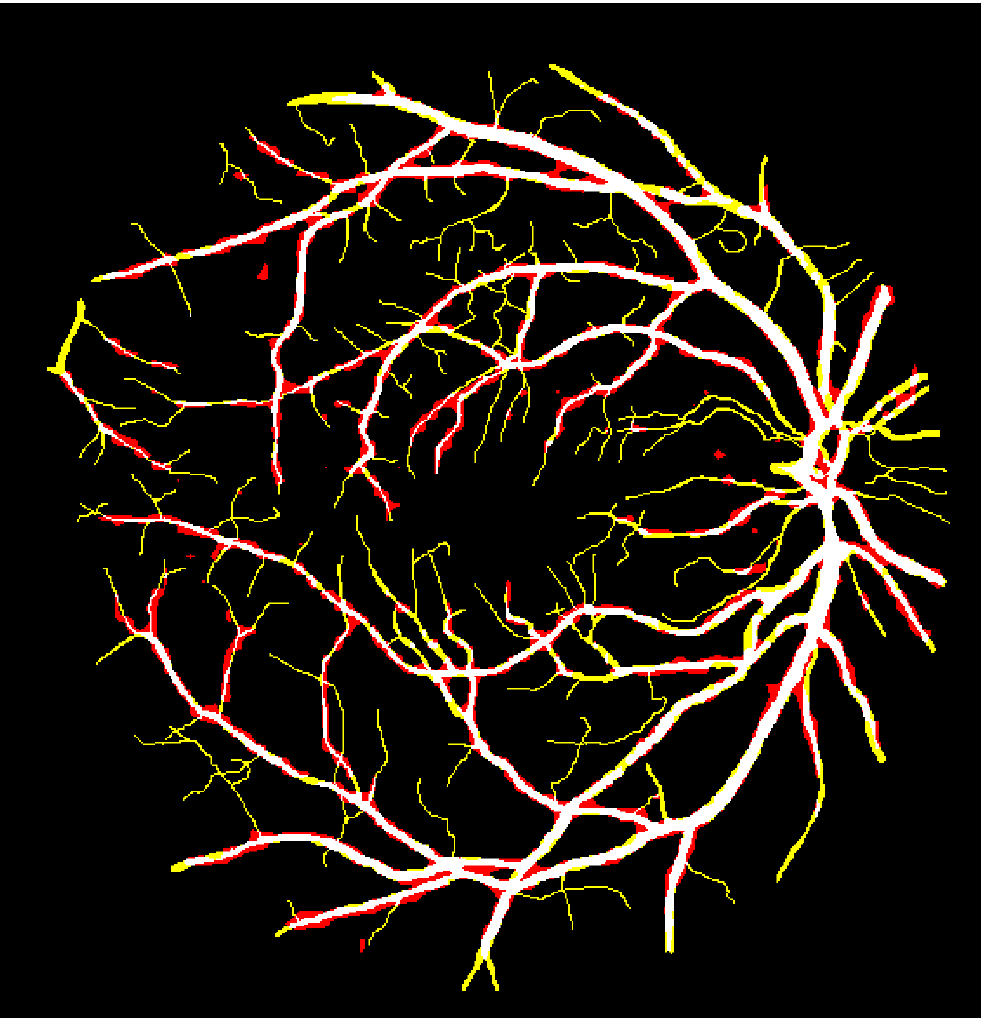} &
        \includegraphics[width=0.4\textwidth]{DRIVE/mmkis6.eps} \\
        \includegraphics[width=0.4\textwidth]{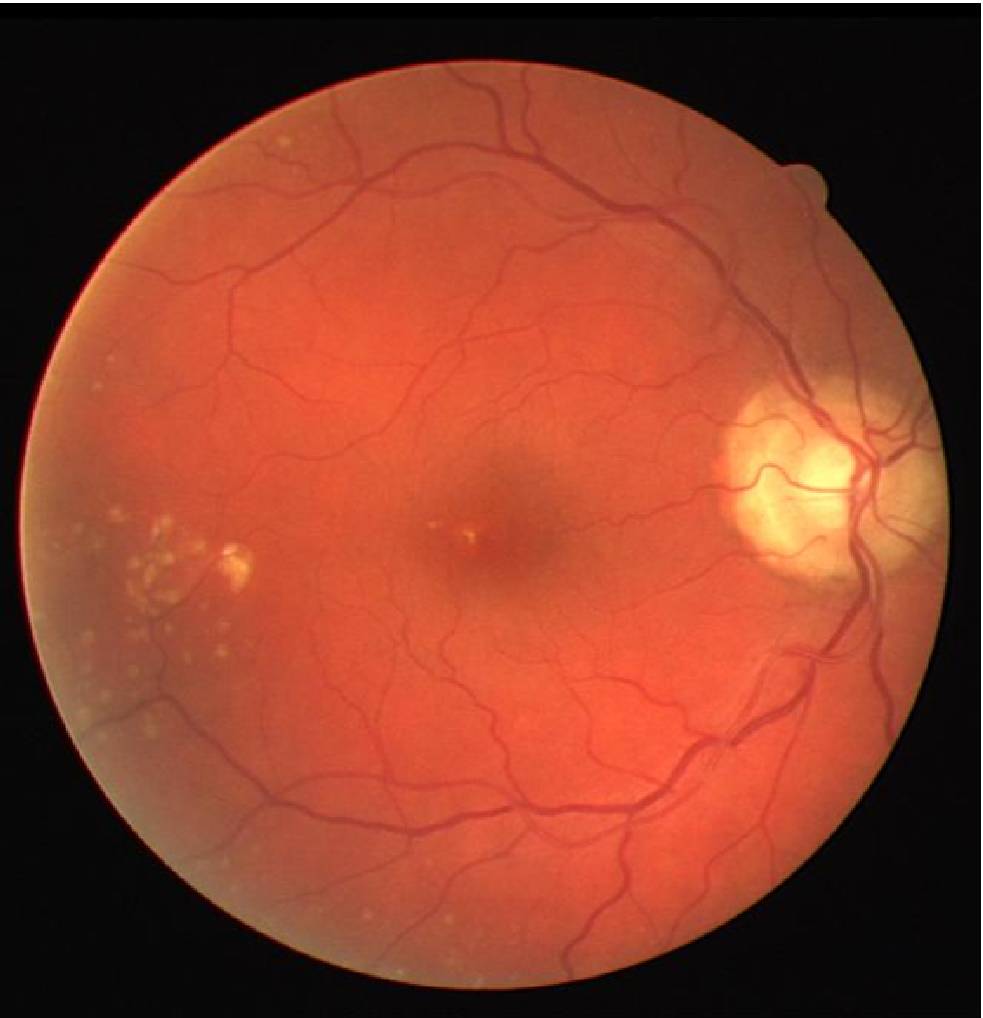} &
        \includegraphics[width=0.4\textwidth]{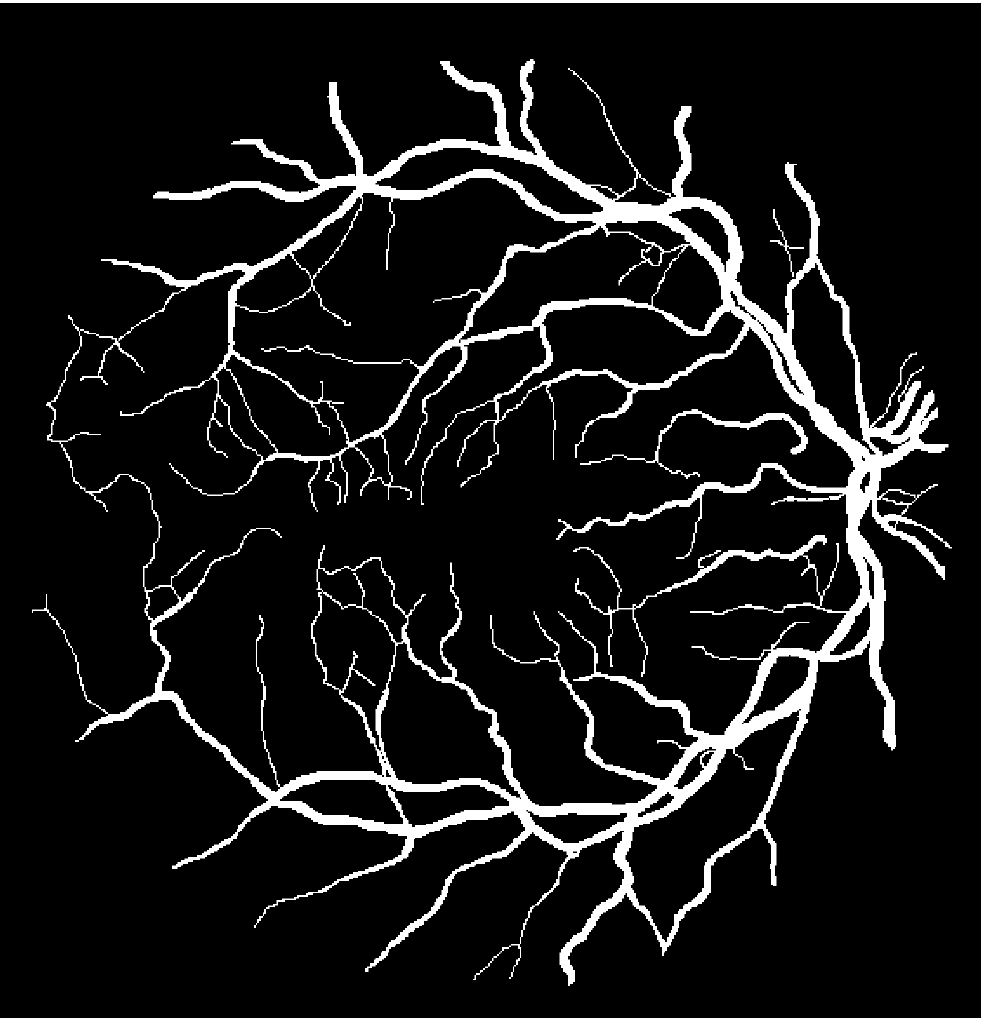} &
        \includegraphics[width=0.4\textwidth]{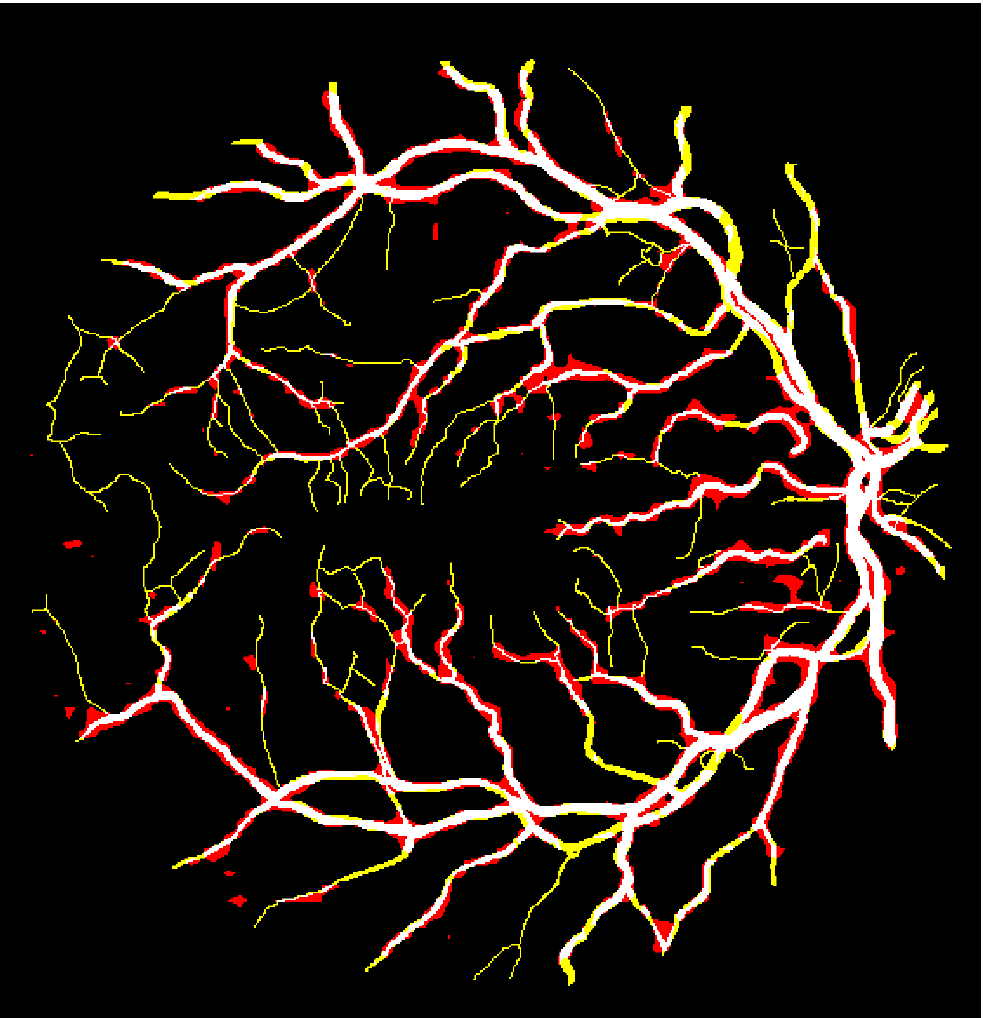} &
        \includegraphics[width=0.4\textwidth]{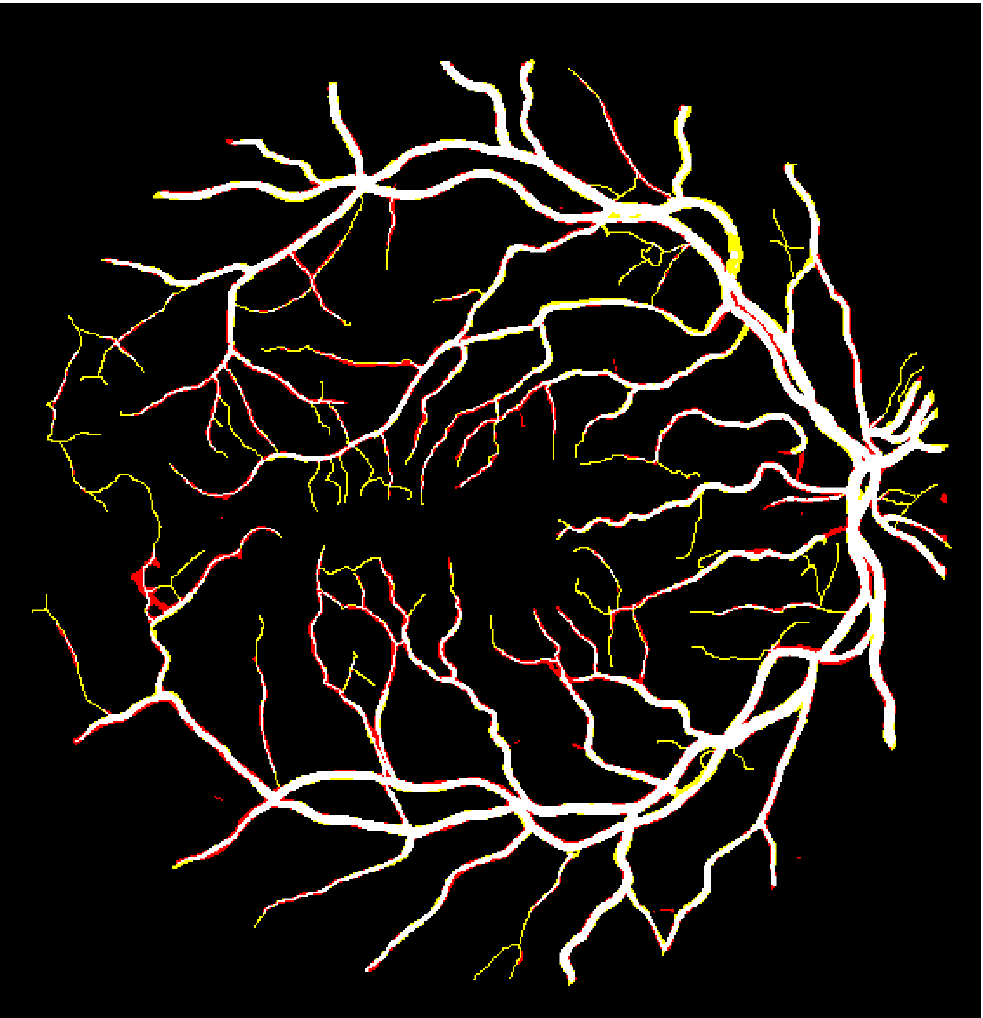} \\
        \includegraphics[width=0.4\textwidth]{DRIVE/Oimg15.eps} &
        \includegraphics[width=0.4\textwidth]{DRIVE/gt15.eps} &
        \includegraphics[width=0.4\textwidth]{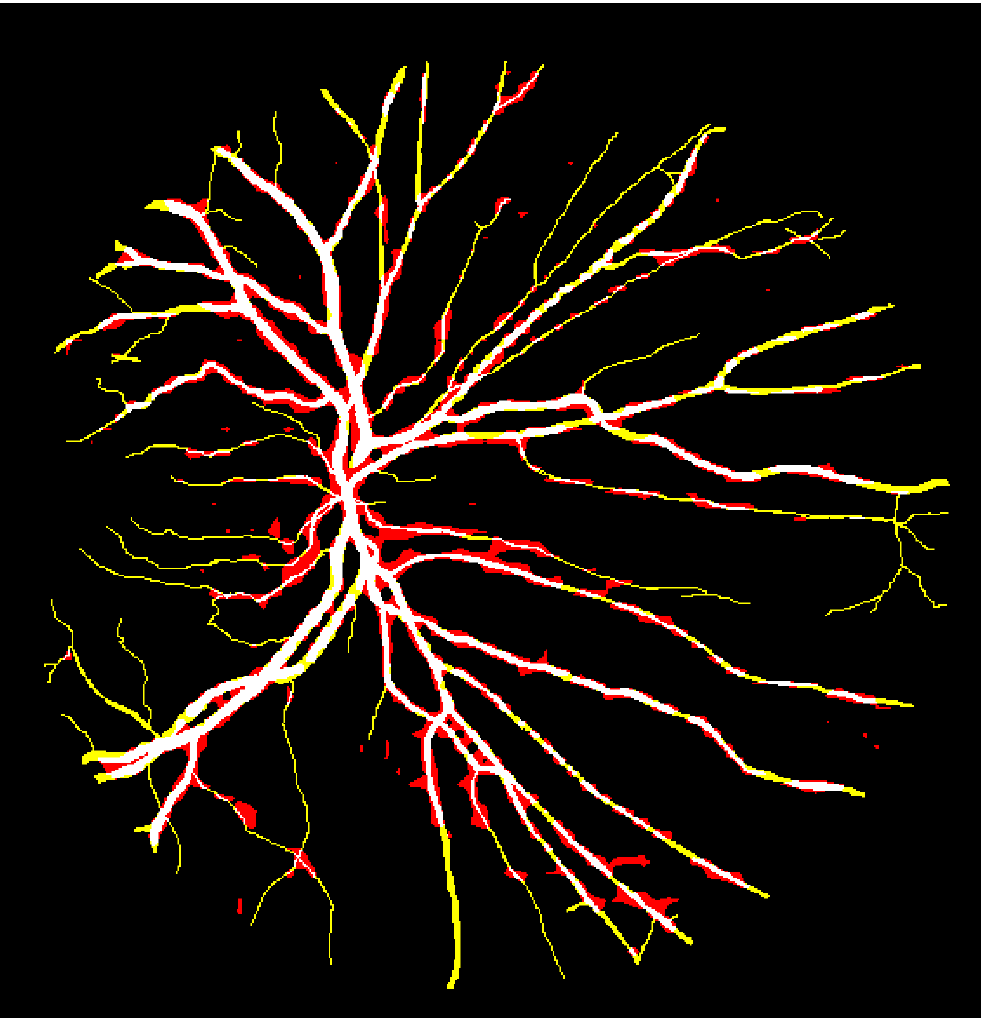} &
        \includegraphics[width=0.4\textwidth]{DRIVE/mmkis15.eps} \\
    \end{tabular}}
    \caption{Sample visual results of our MKIS-Net and MobileNet-V3-Small on the DRIVE dataset. From left to right: the input images, the gold standard vessel maps manually annotated by an expert, and the results generated by MobileNet-V3 and the proposed MKIS-Net.}
    \label{visualDRIVE1}
\end{figure}

\begin{figure}[!t]
    \centering
    \resizebox{1\textwidth}{!}{%
    \begin{tabular}{@{}c@{\ }c@{\ }c@{\ }c@{}}
        \includegraphics[width=0.4\textwidth]{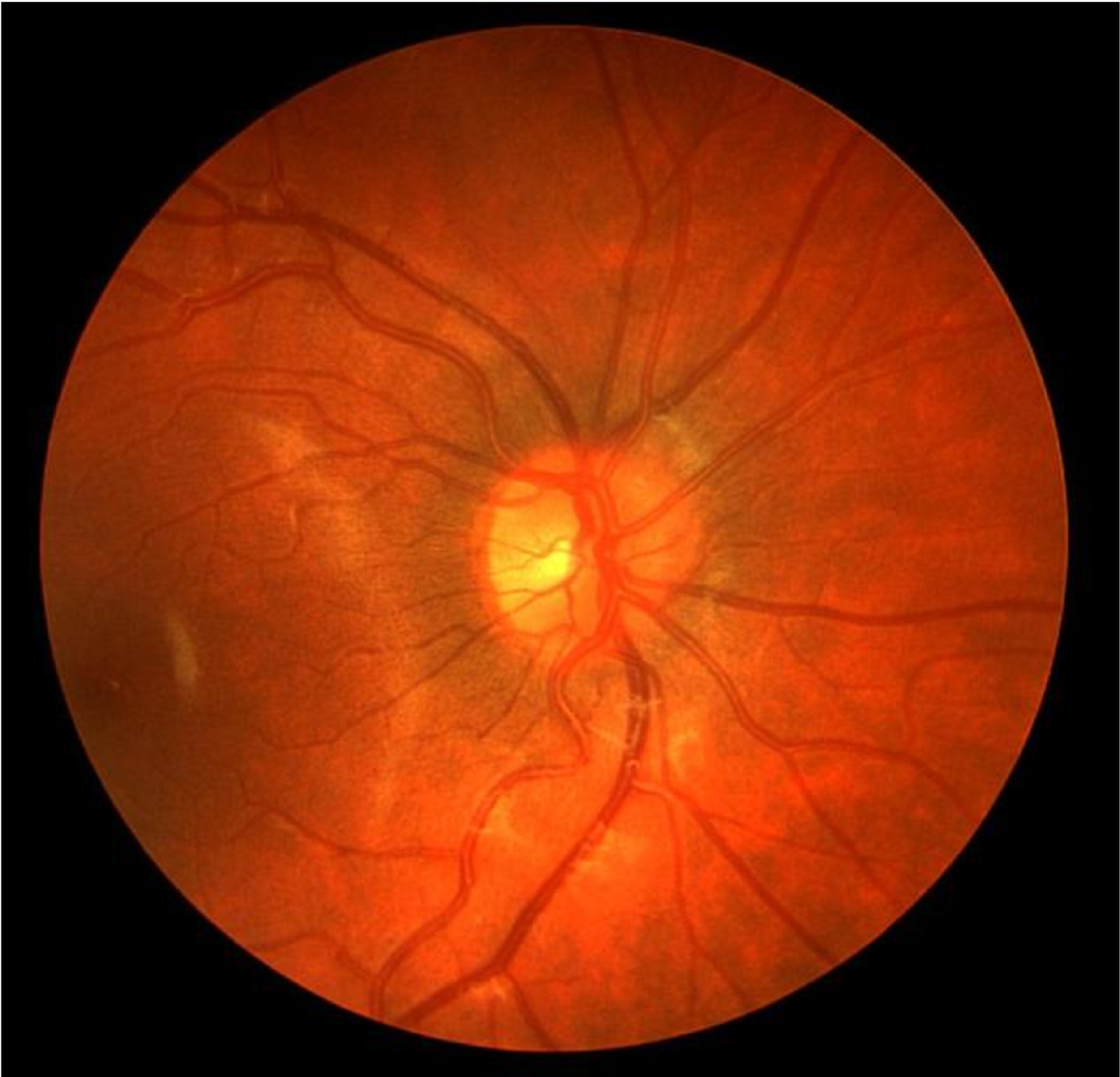} &
        \includegraphics[width=0.4\textwidth]{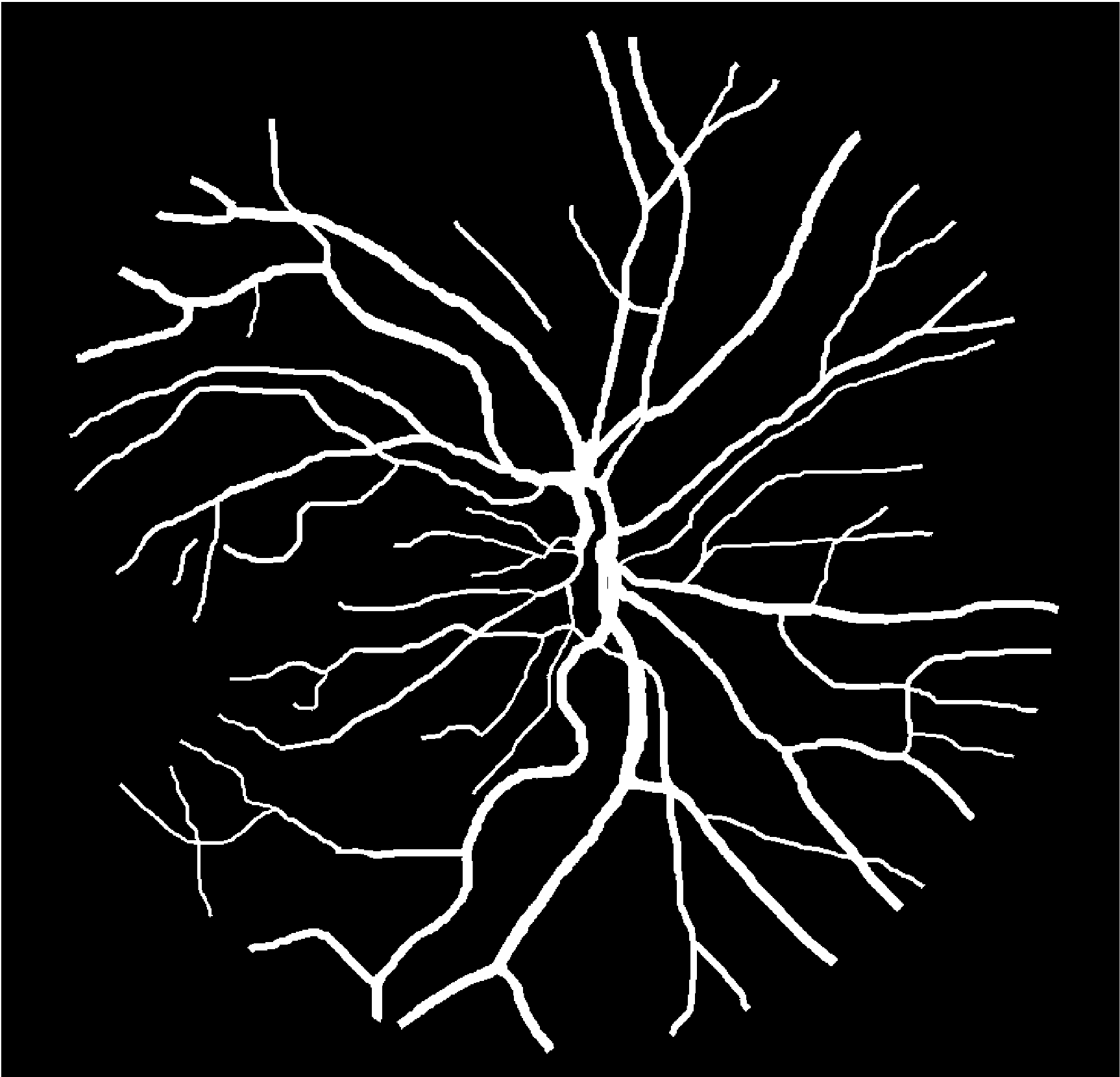} &
        \includegraphics[width=0.4\textwidth]{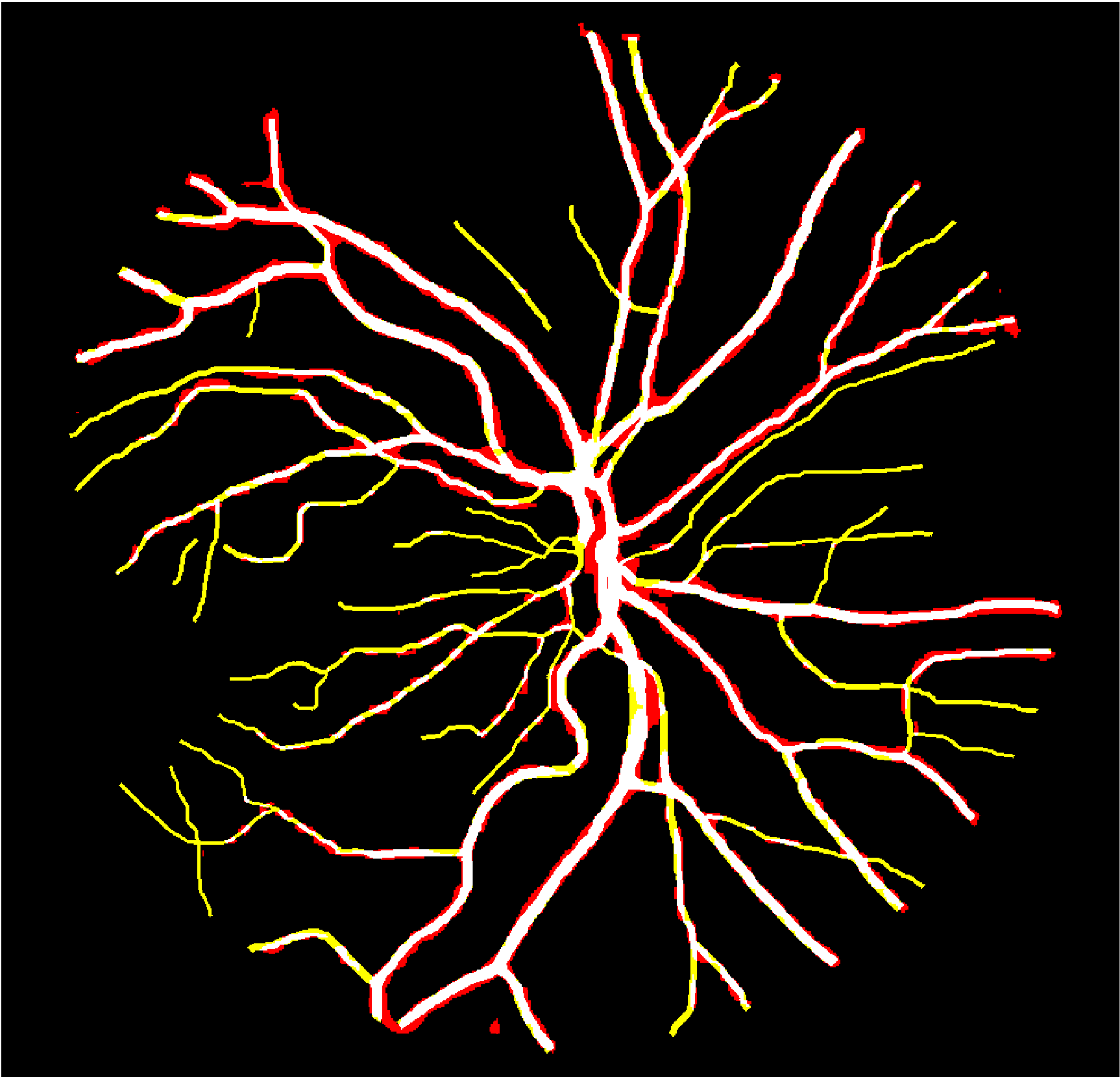} &
        \includegraphics[width=0.4\textwidth]{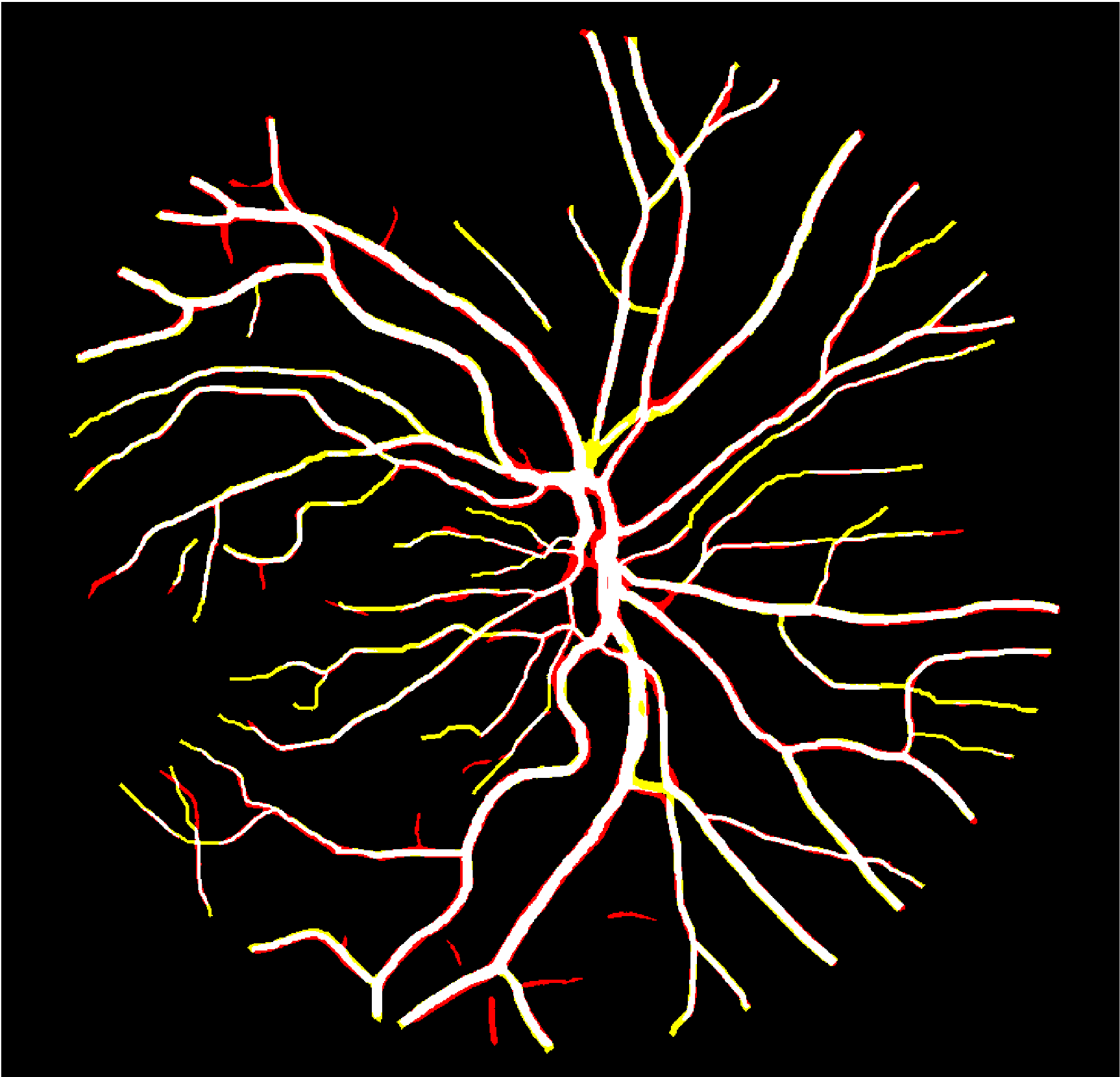} \\
        \includegraphics[width=0.4\textwidth]{CHASE/Oimg6.eps} &
        \includegraphics[width=0.4\textwidth]{CHASE/gt6.eps} &
        \includegraphics[width=0.4\textwidth]{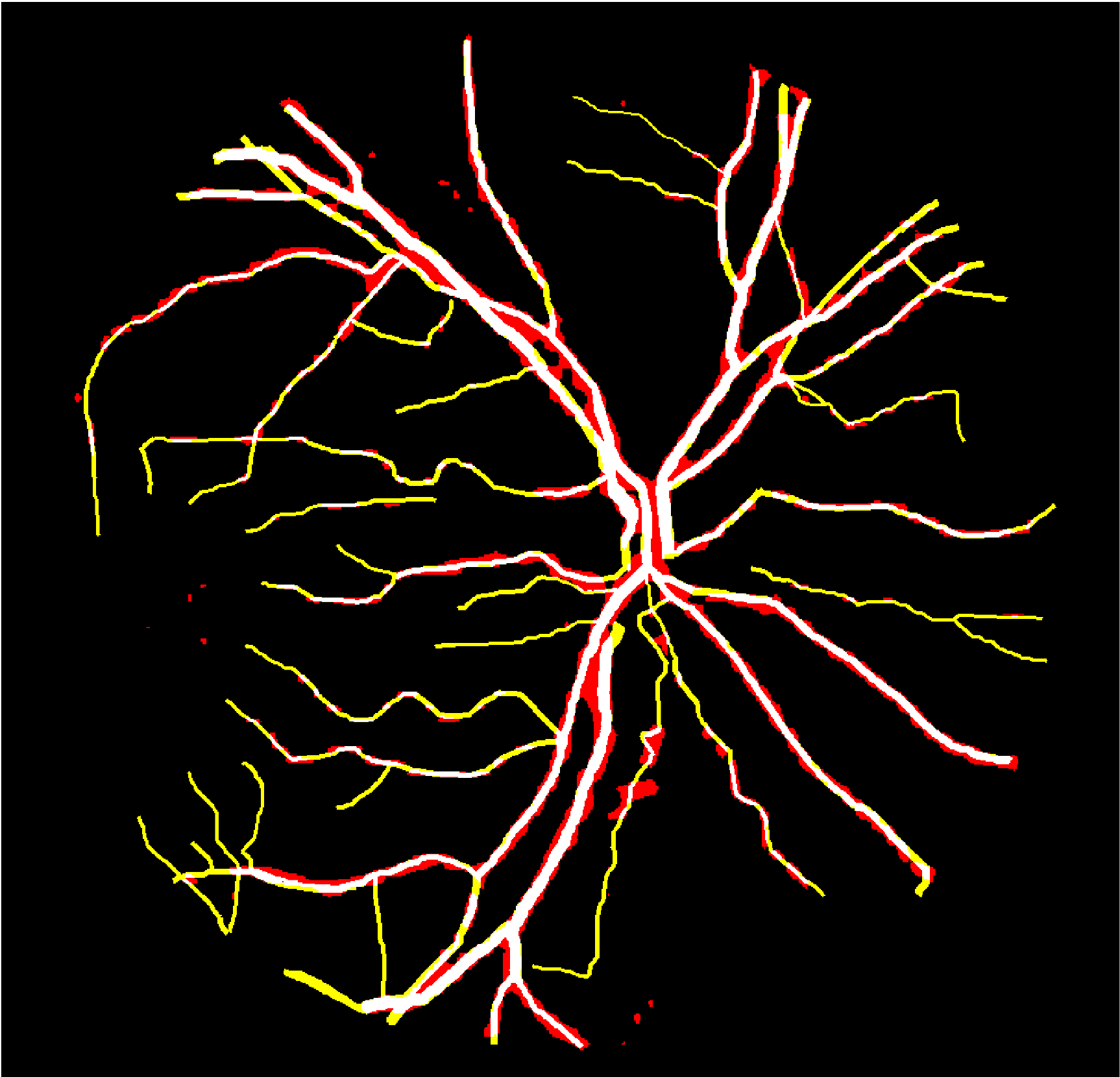} &
        \includegraphics[width=0.4\textwidth]{CHASE/mmkis6.eps} \\
        \includegraphics[width=0.4\textwidth]{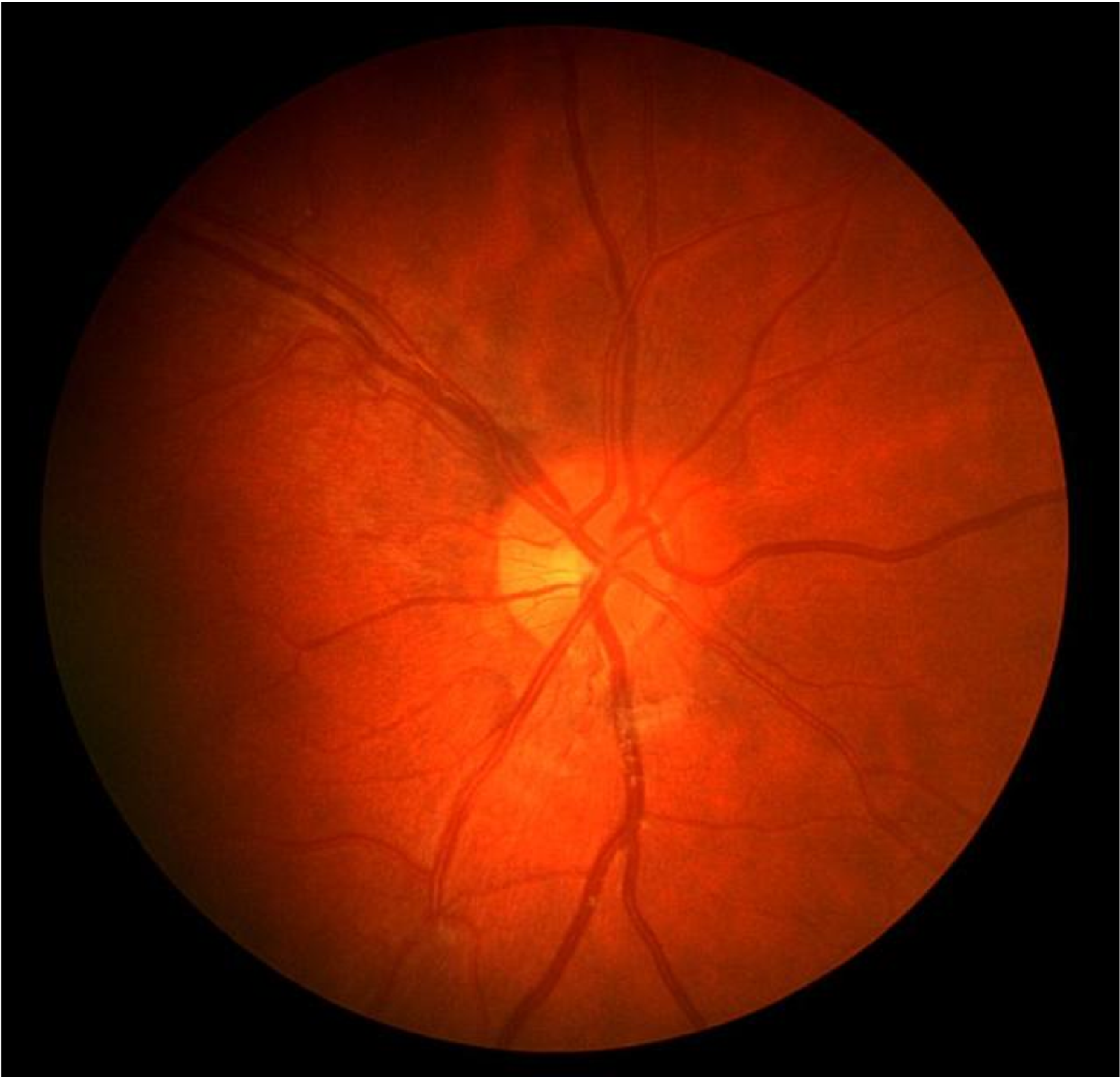} &
        \includegraphics[width=0.4\textwidth]{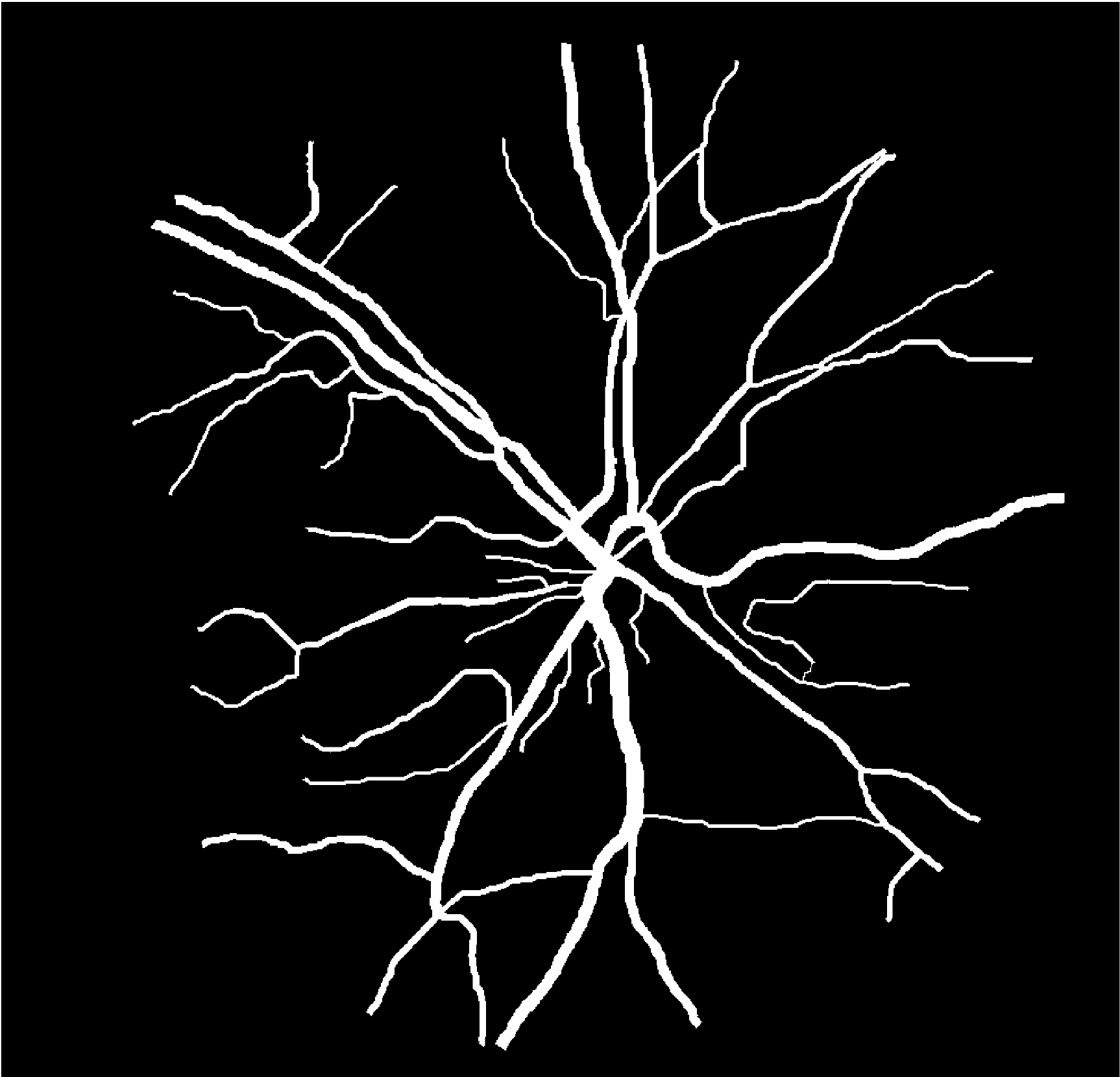} &
        \includegraphics[width=0.4\textwidth]{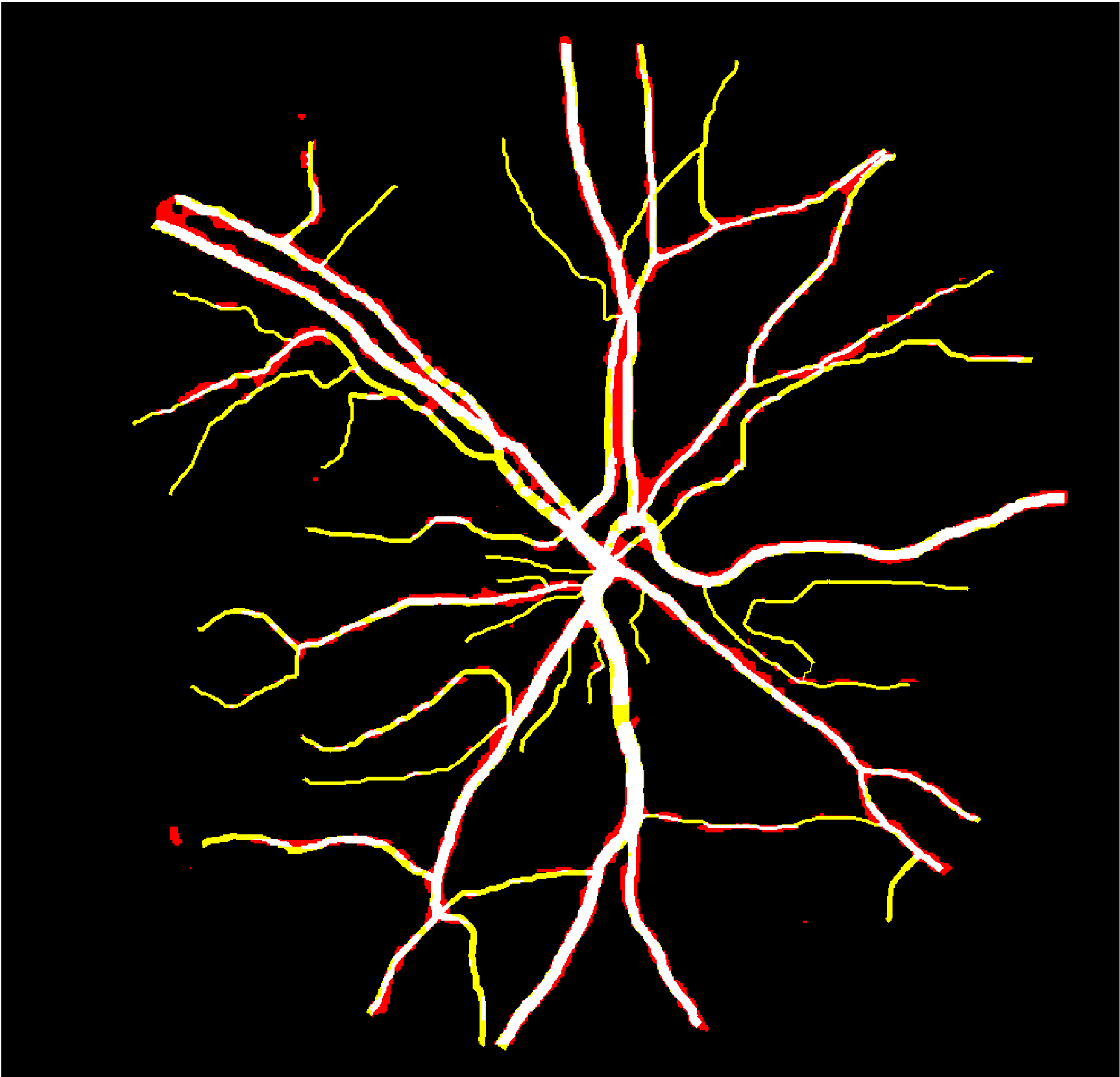} &
        \includegraphics[width=0.4\textwidth]{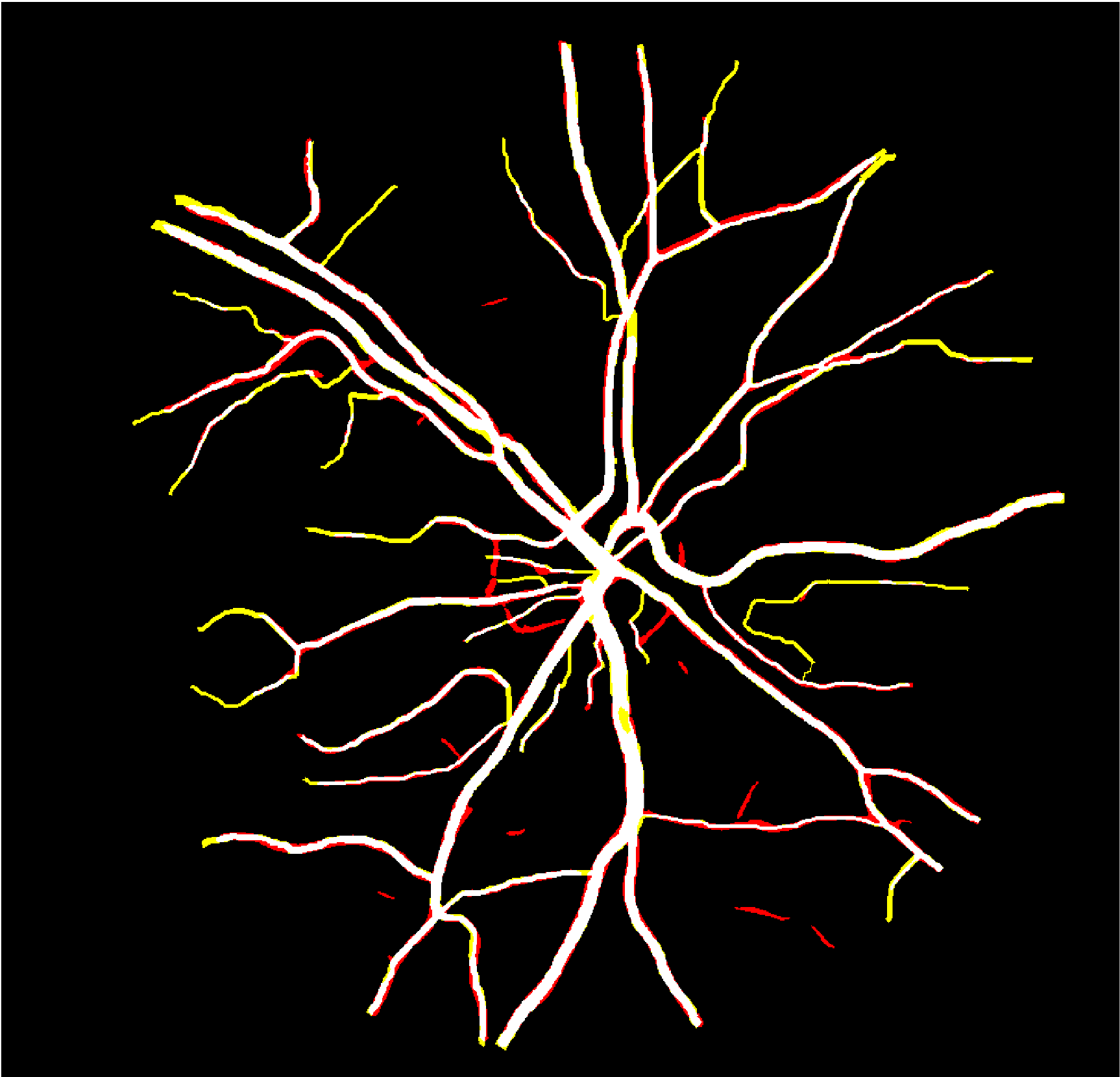} \\
    \end{tabular}}
    \caption{Sample visual results of our MKIS-Net and MobileNet-V3-Small on the CHASE dataset. From left to right: the input images, the gold standard vessel maps manually annotated by an expert, and the results generated by MobileNet-V3 and the proposed MKIS-Net.}
    \label{visualCHASE1}
\end{figure}

\begin{figure}[!t]
    \centering
    \resizebox{1\textwidth}{!}{%
    \begin{tabular}{@{}c@{\ }c@{\ }c@{\ }c@{\ }c@{\ }c@{}}
        \includegraphics[width=0.14\textwidth]{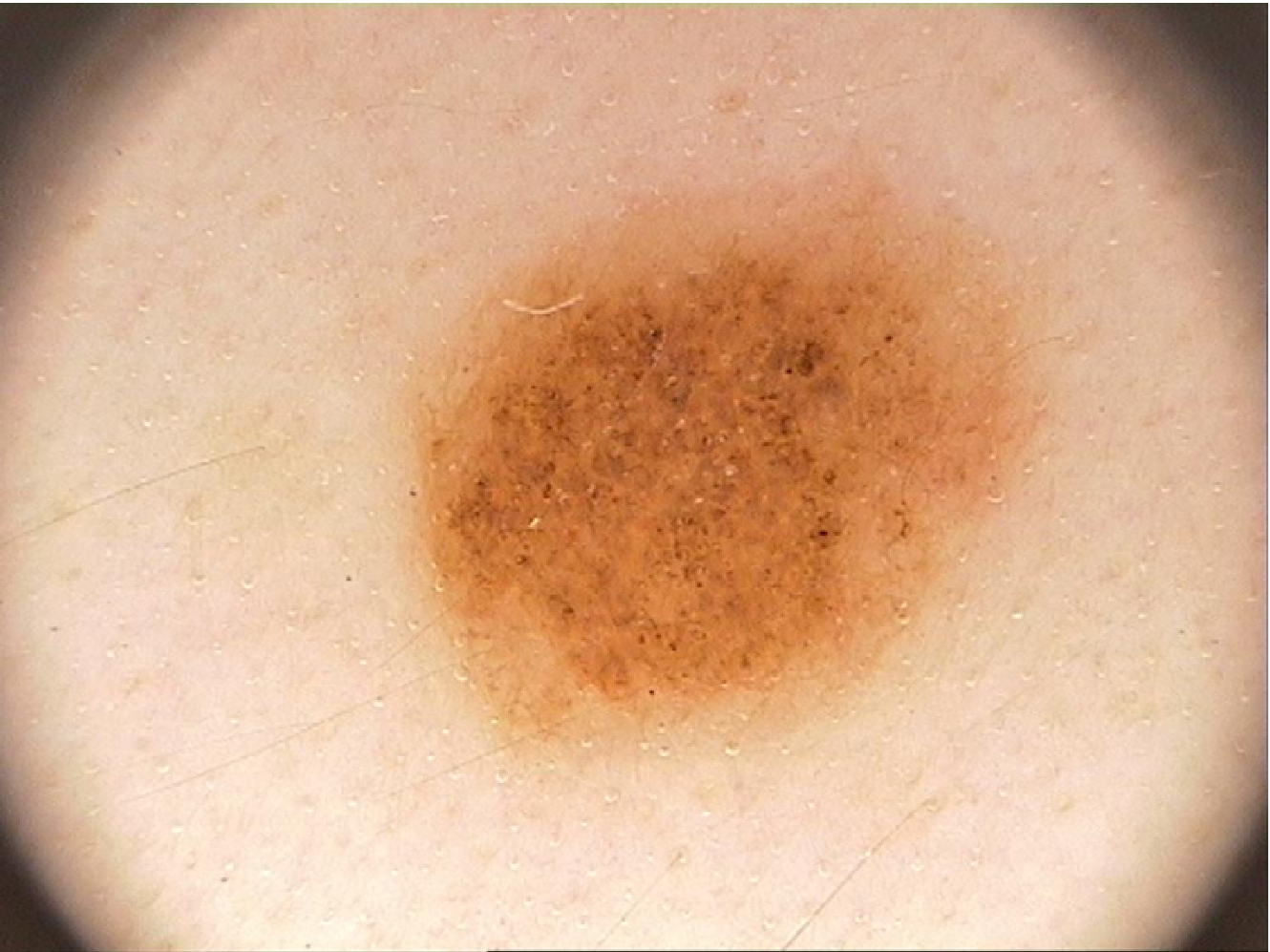} &
        \includegraphics[width=0.14\textwidth]{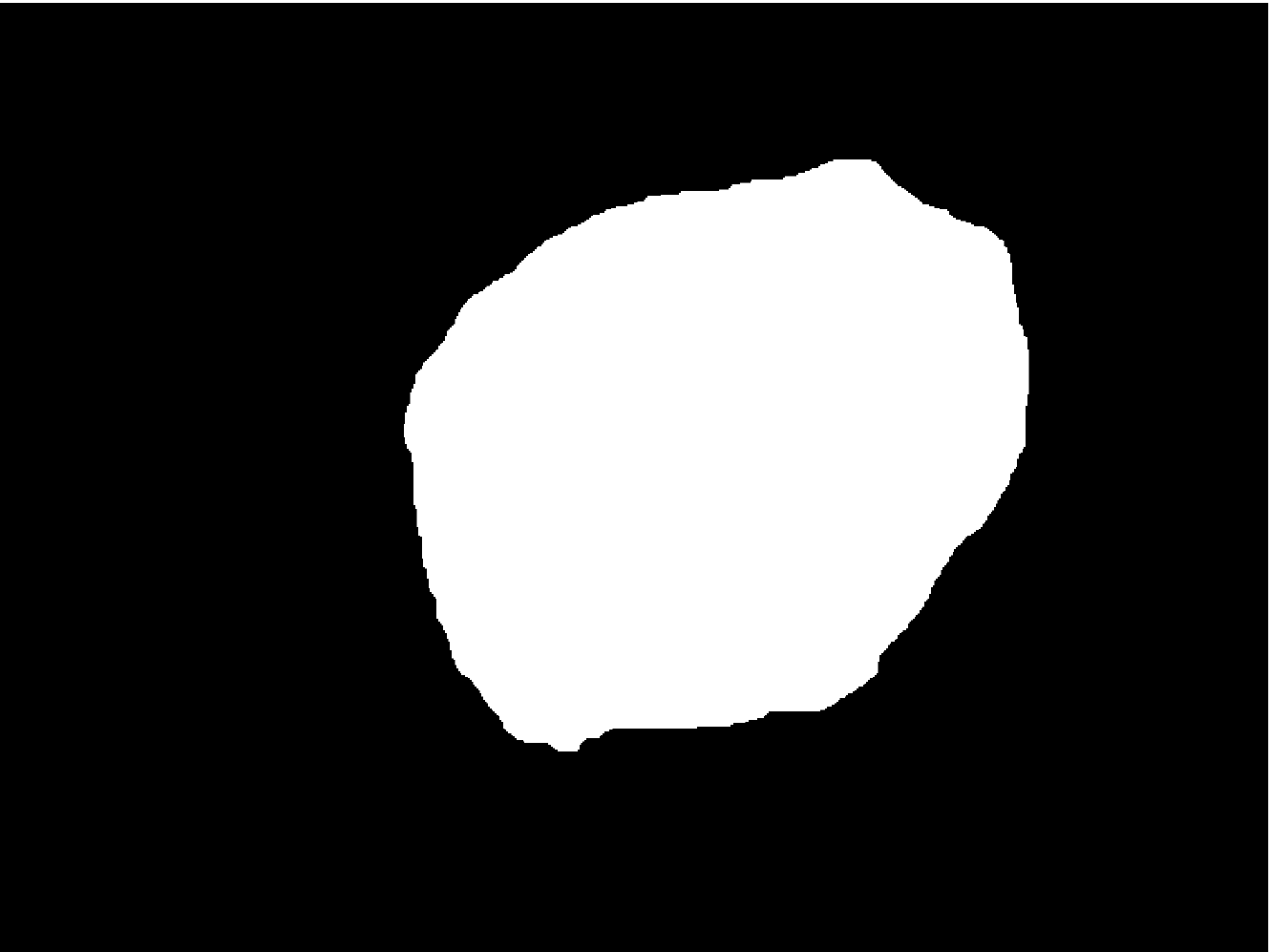} &
        \includegraphics[width=0.14\textwidth]{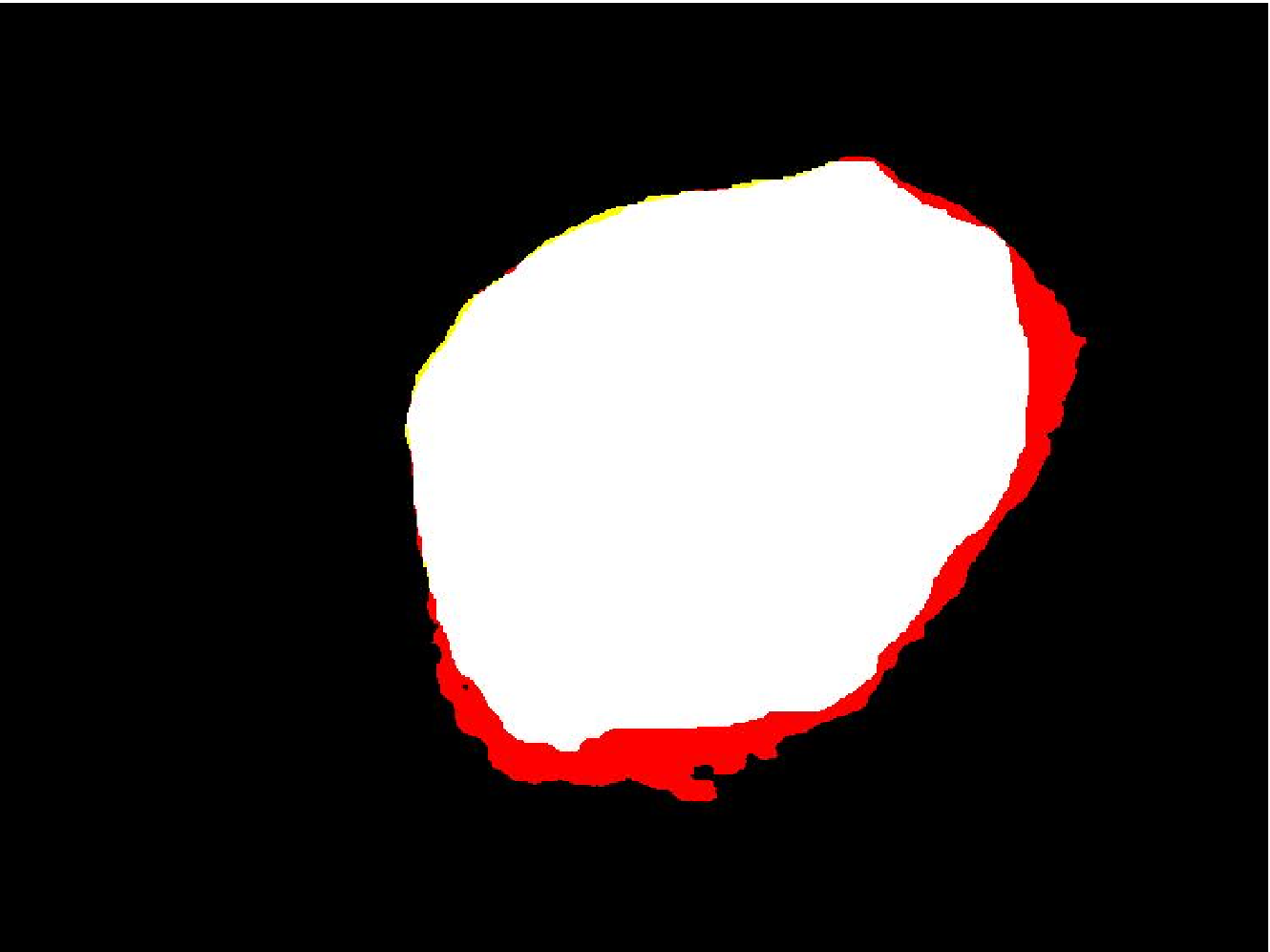} &
        \includegraphics[width=0.14\textwidth]{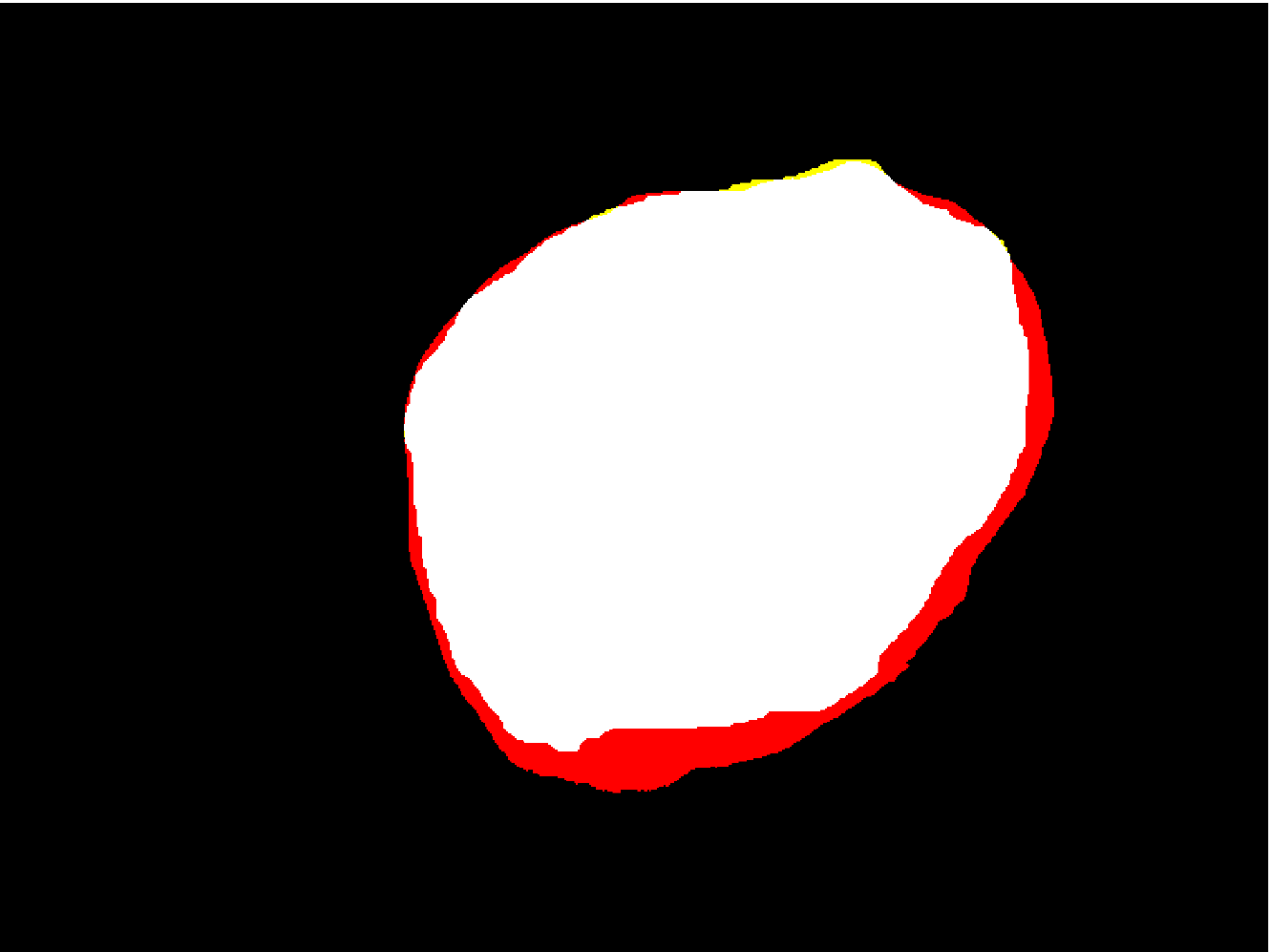} &
        \includegraphics[width=0.14\textwidth]{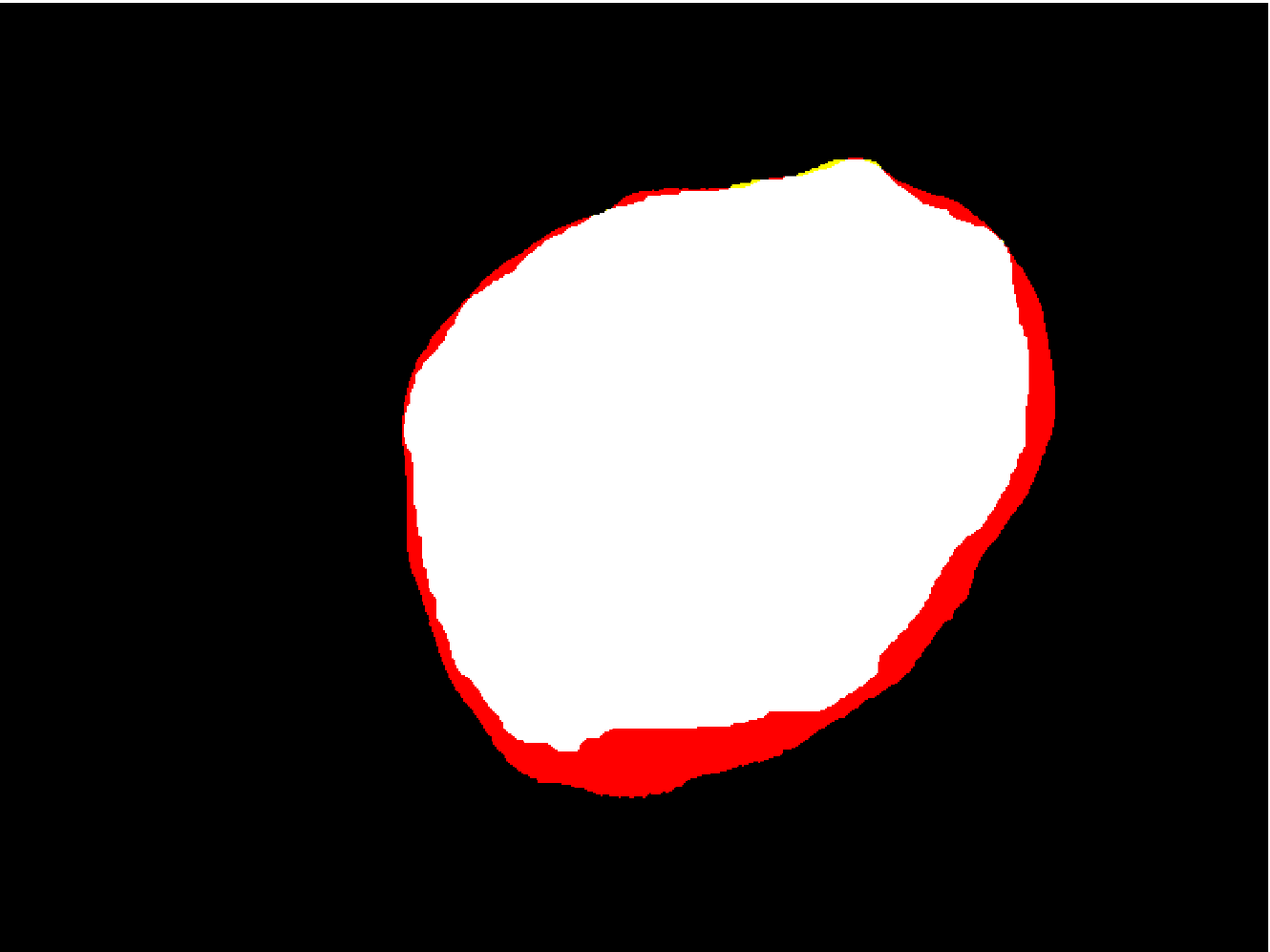} &
        \includegraphics[width=0.14\textwidth]{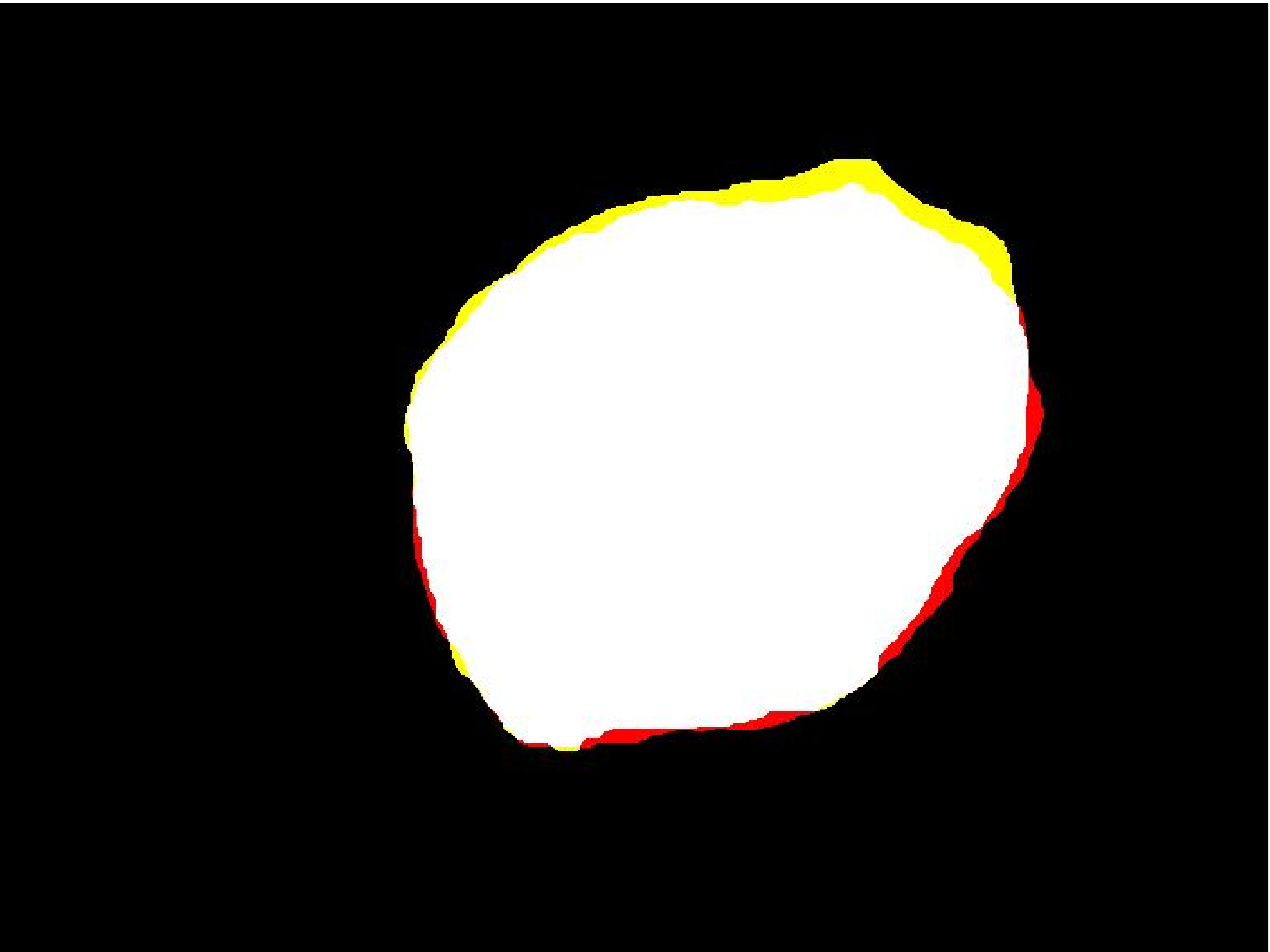} \\
        \includegraphics[width=0.14\textwidth]{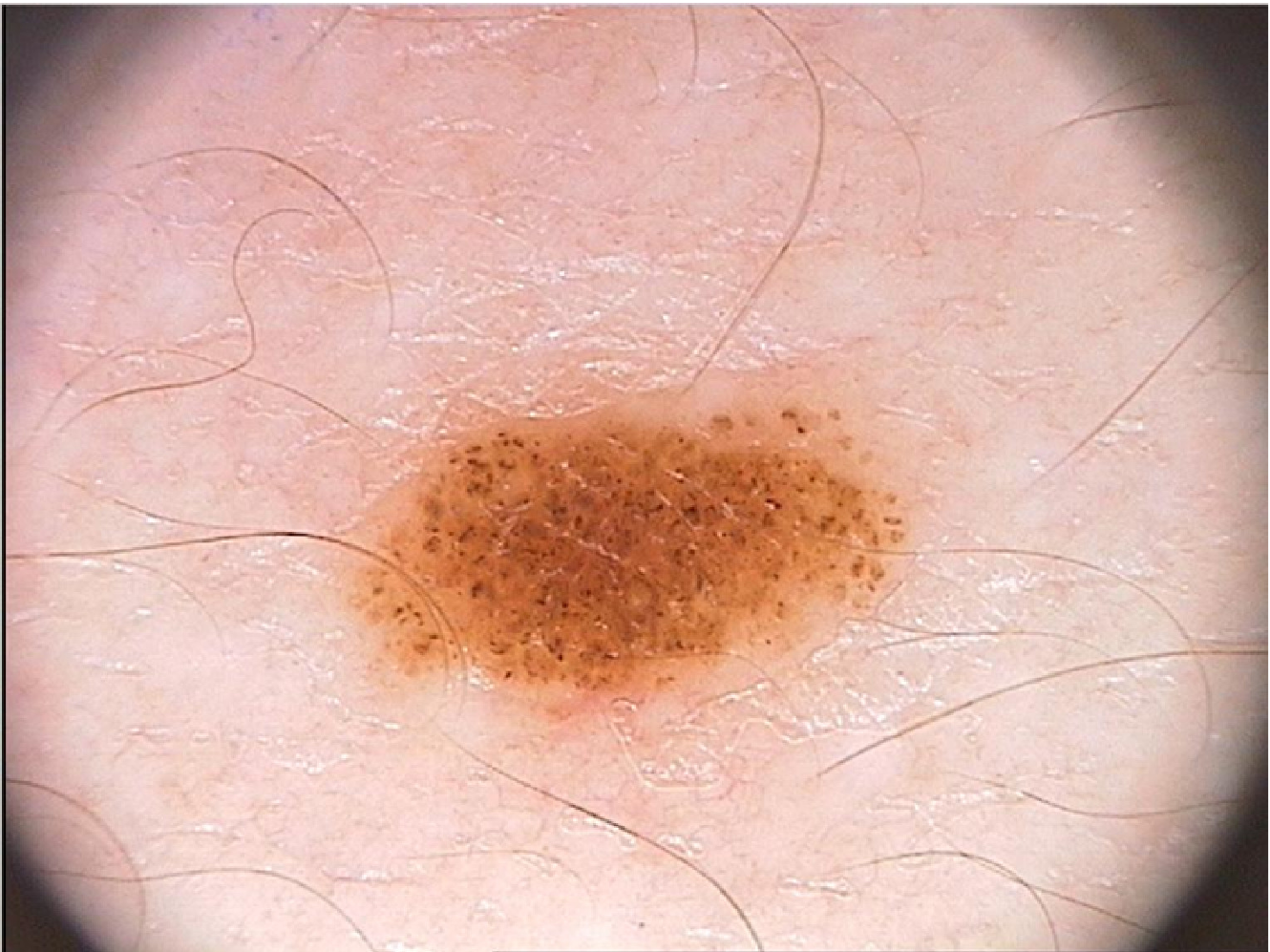} &
        \includegraphics[width=0.14\textwidth]{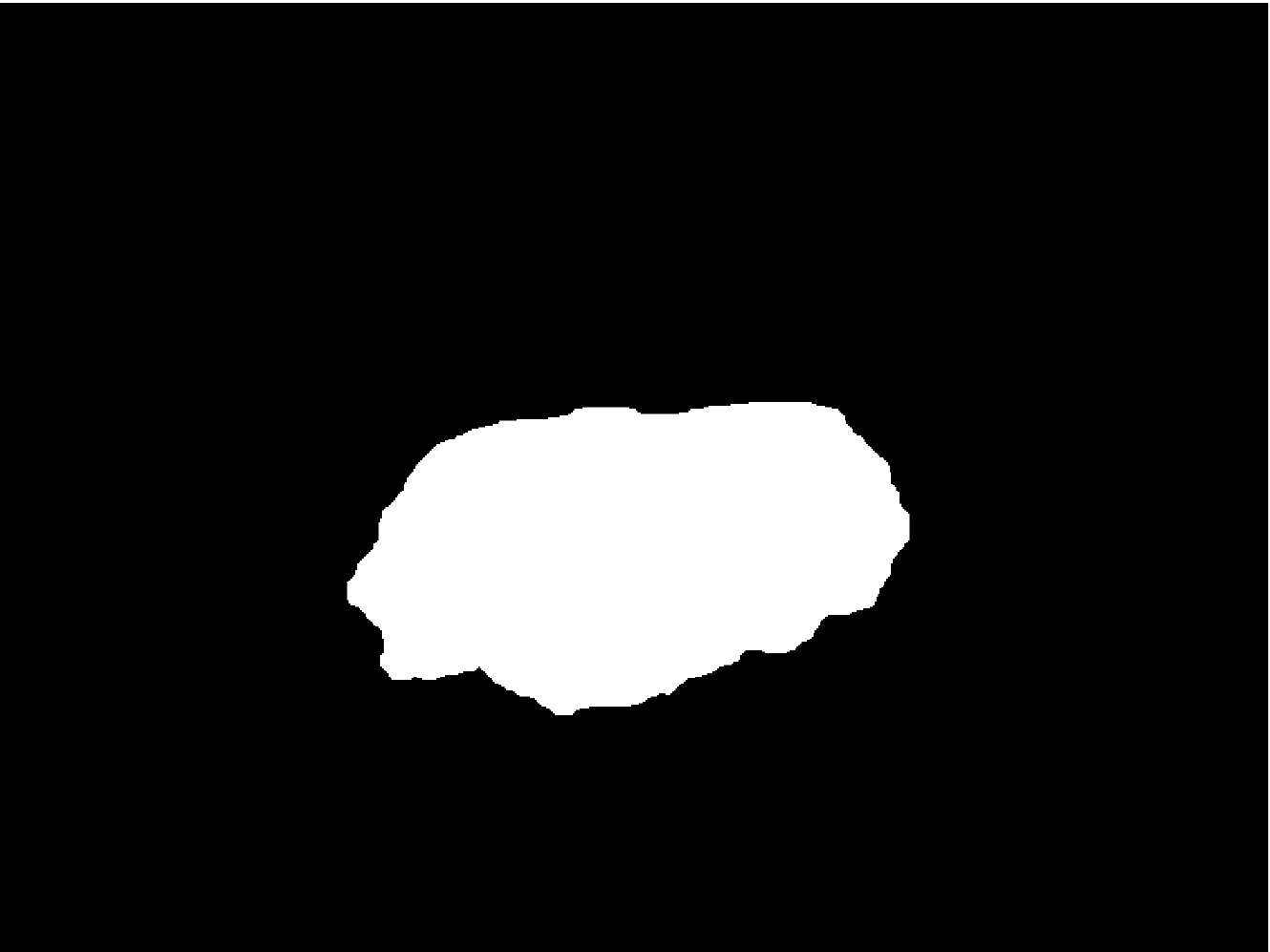} &
        \includegraphics[width=0.14\textwidth]{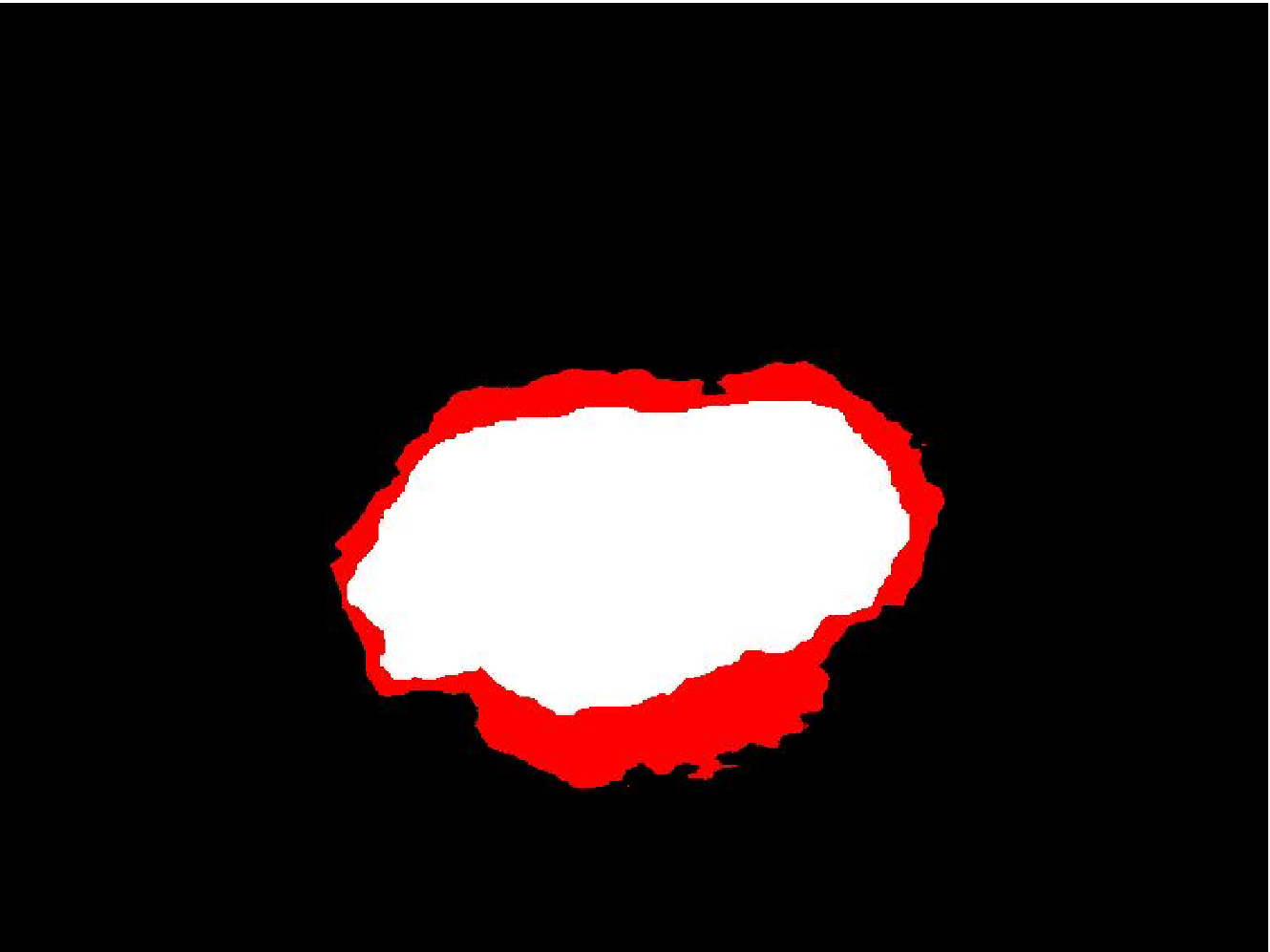} &
        \includegraphics[width=0.14\textwidth]{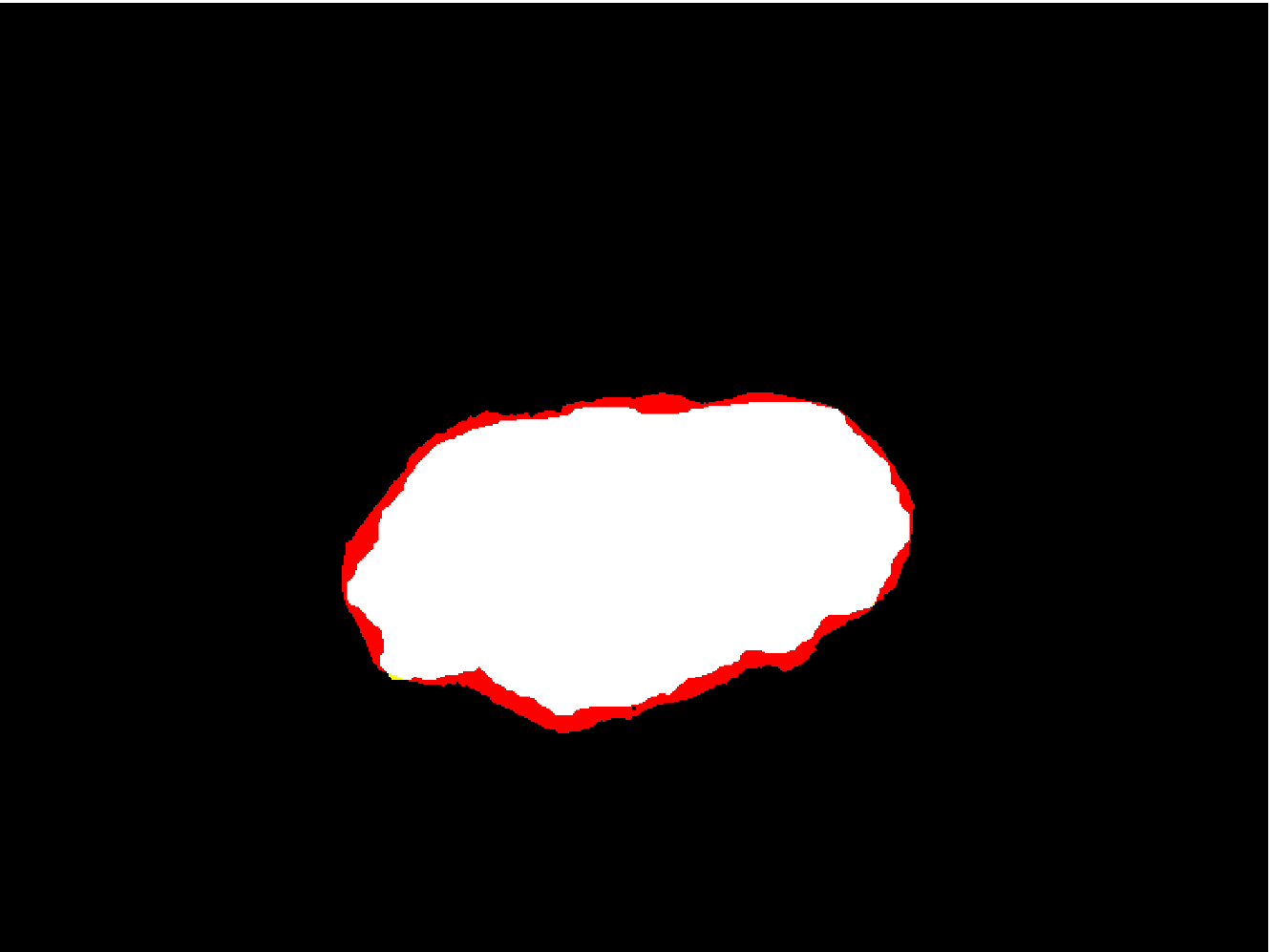} &
        \includegraphics[width=0.14\textwidth]{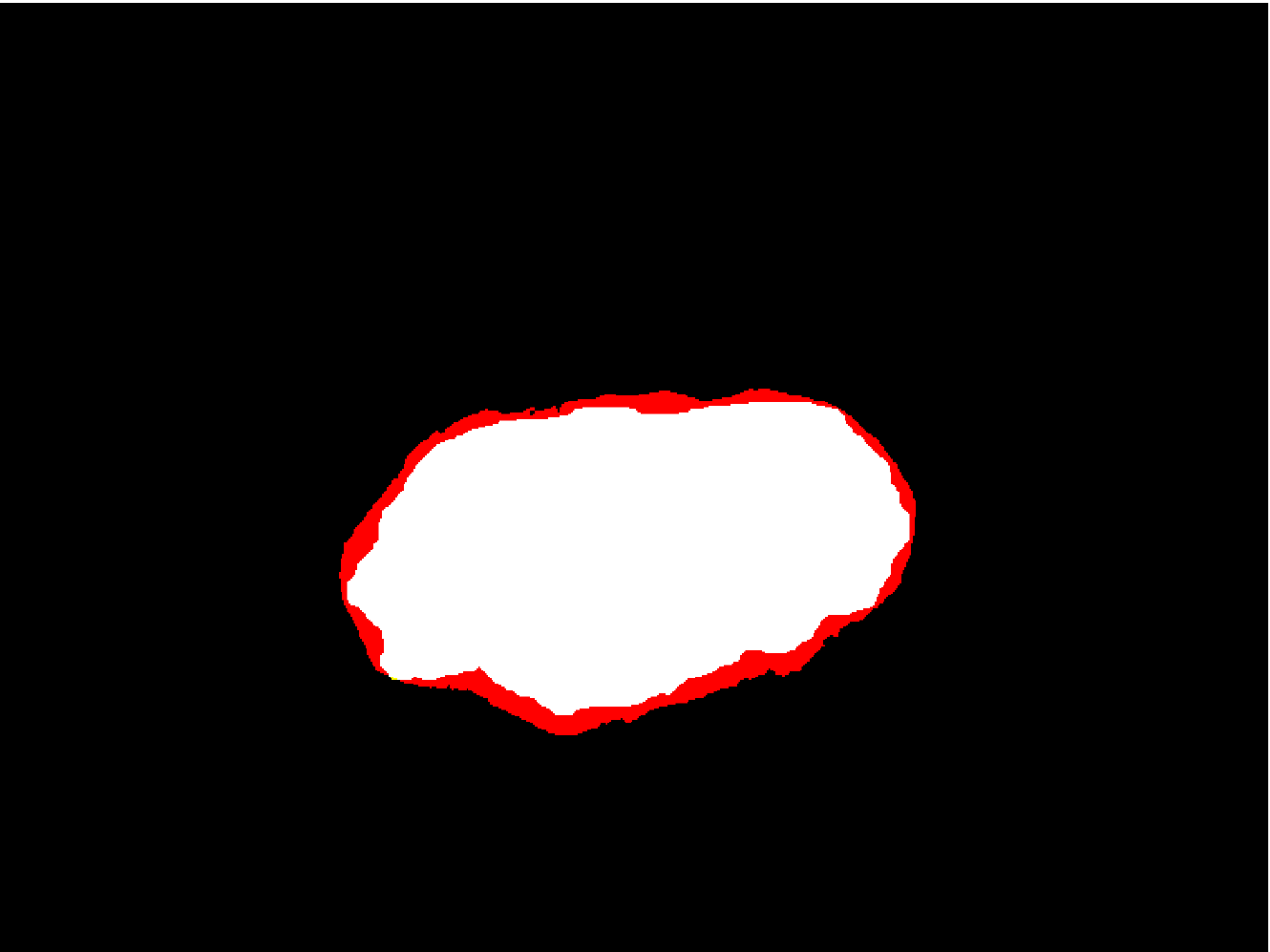} &
        \includegraphics[width=0.14\textwidth]{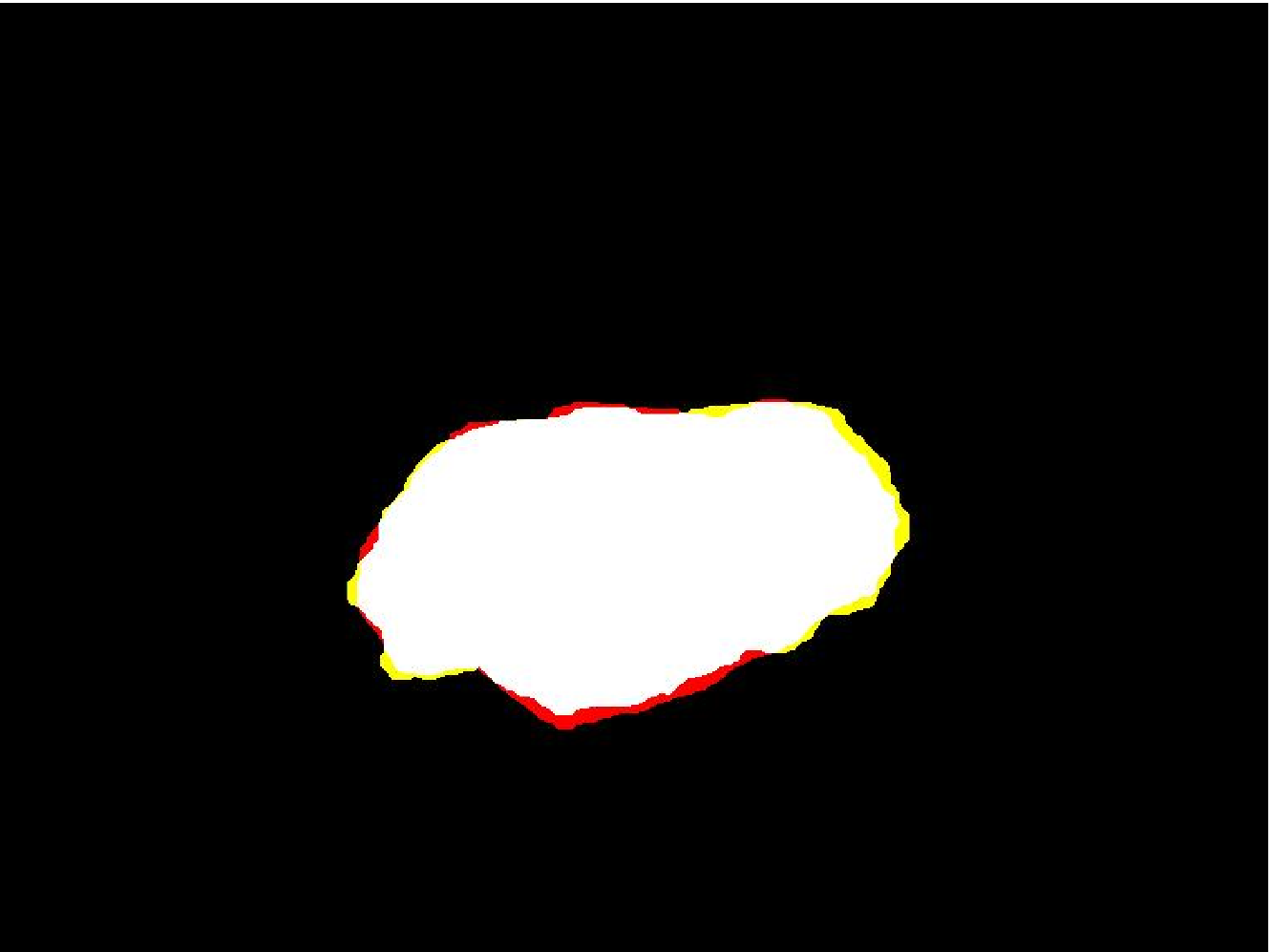} \\
        \includegraphics[width=0.14\textwidth]{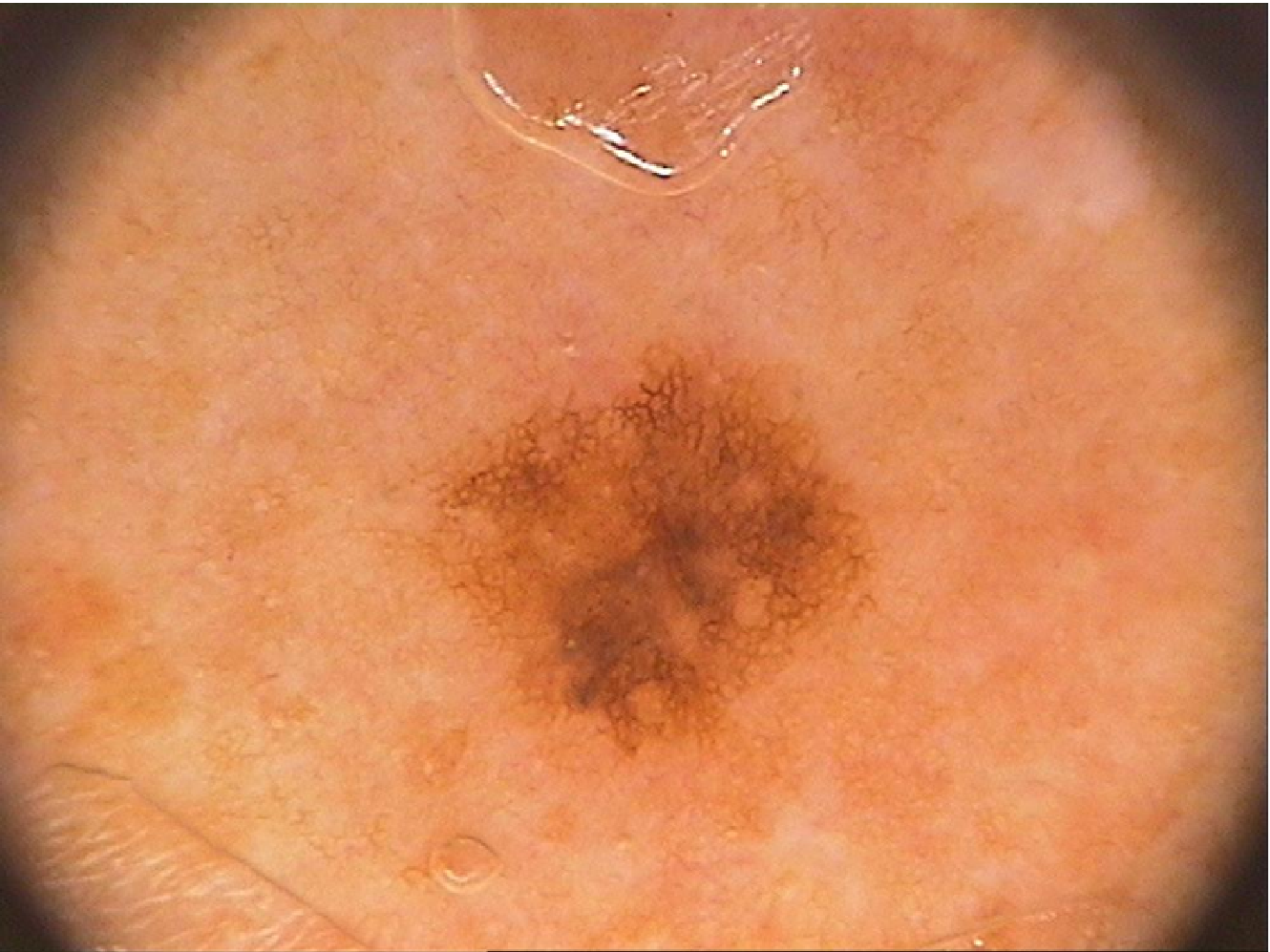} &
        \includegraphics[width=0.14\textwidth]{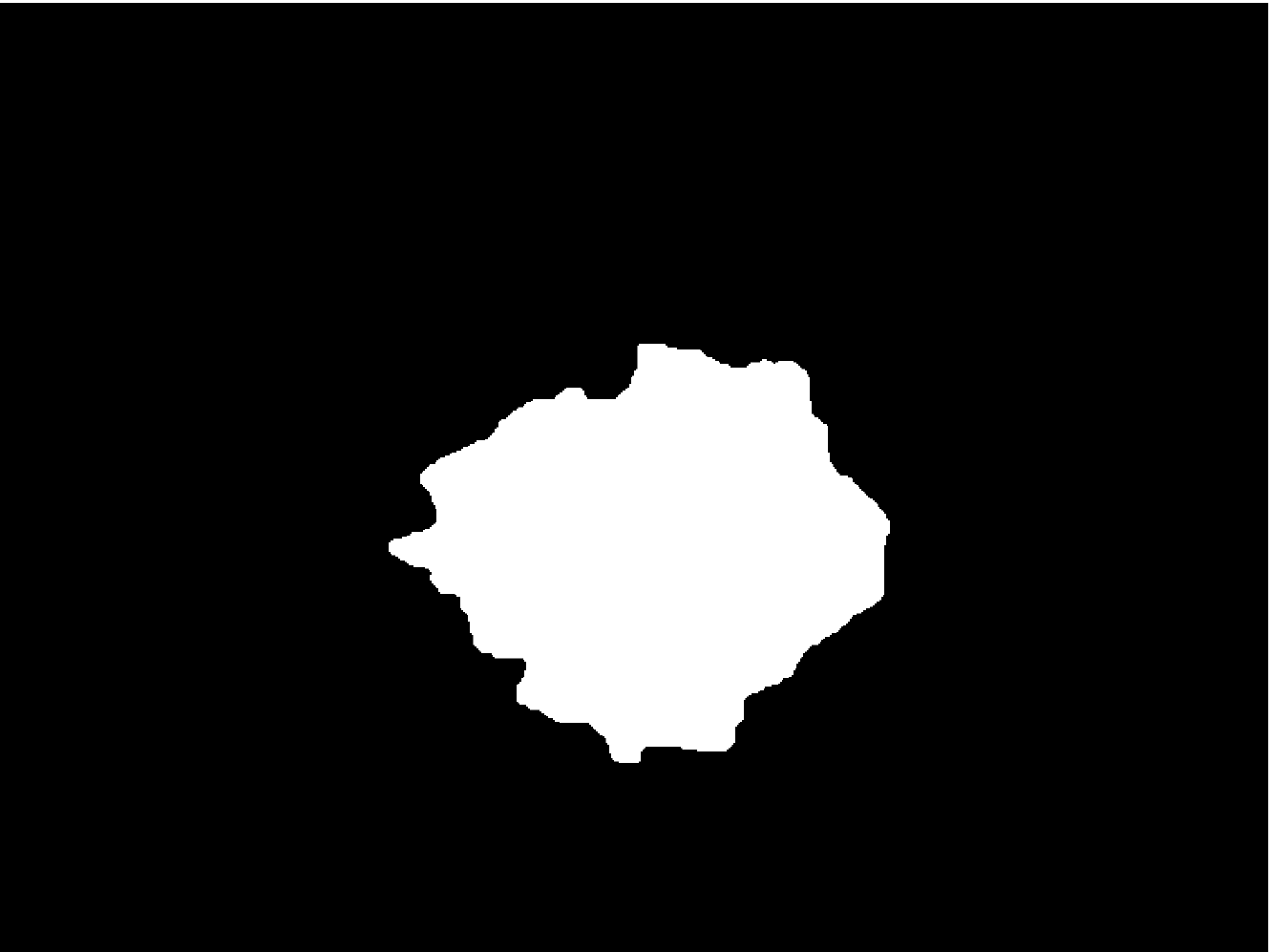} &
        \includegraphics[width=0.14\textwidth]{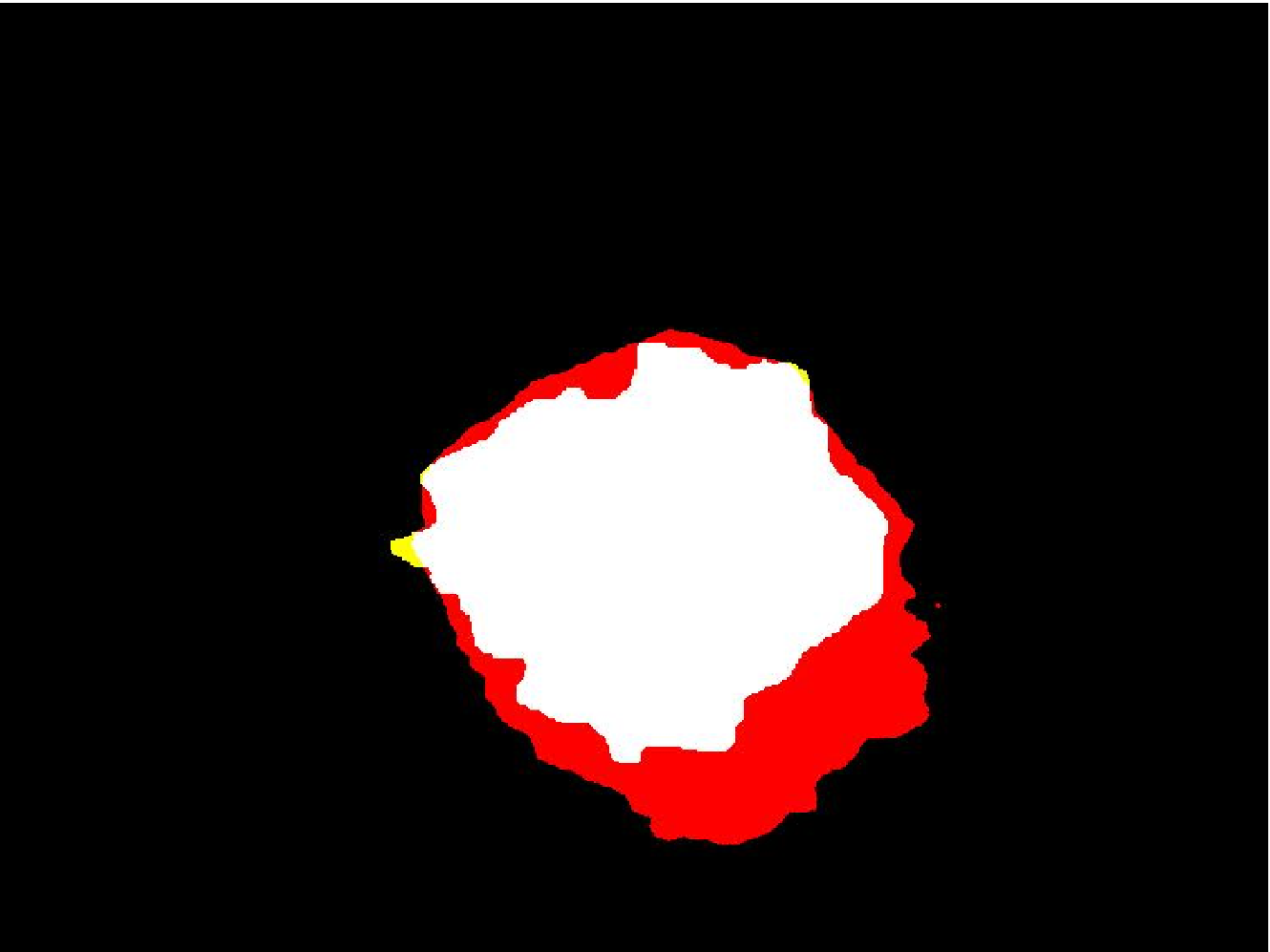} &
        \includegraphics[width=0.14\textwidth]{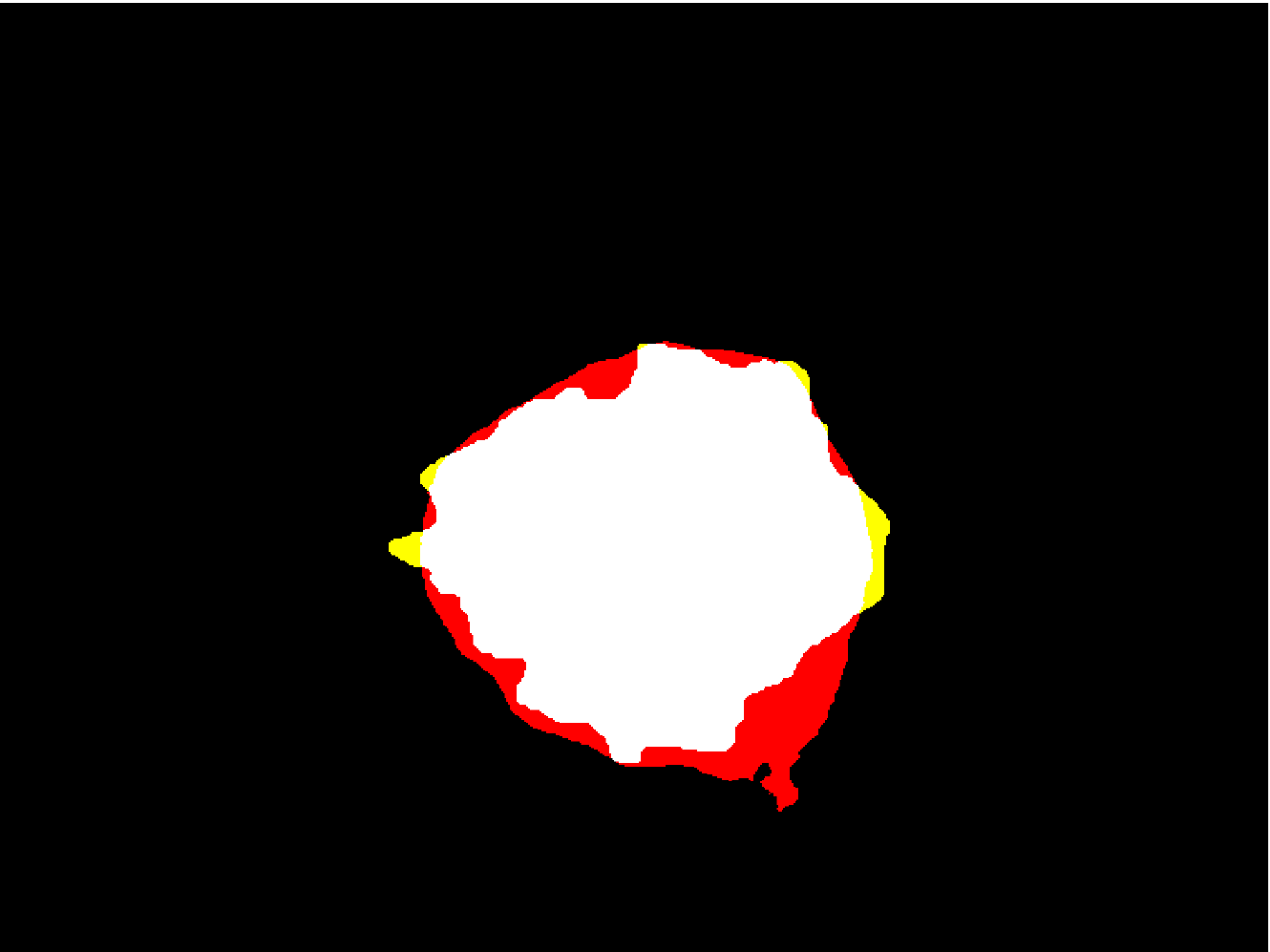} &
        \includegraphics[width=0.14\textwidth]{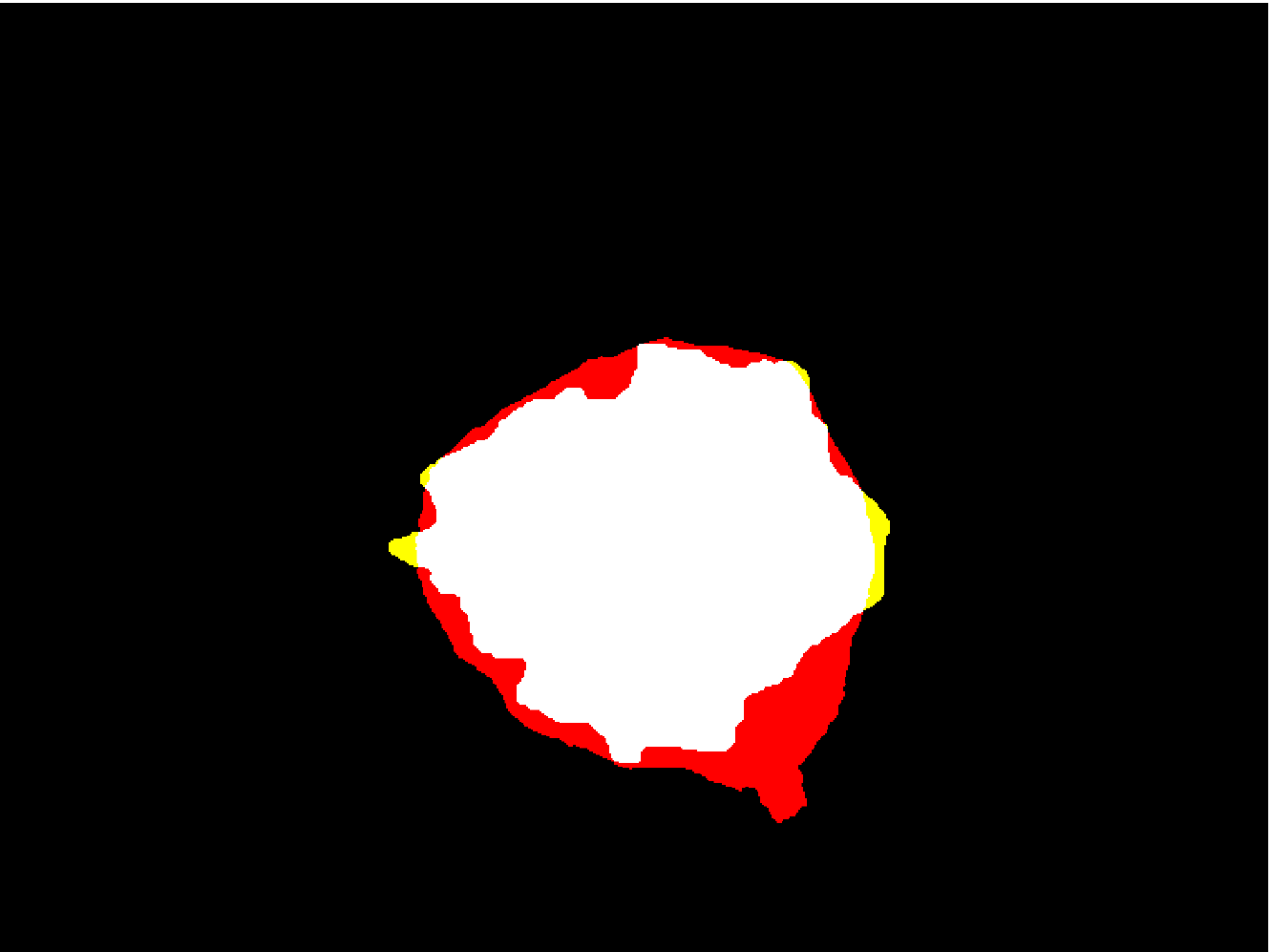} &
        \includegraphics[width=0.14\textwidth]{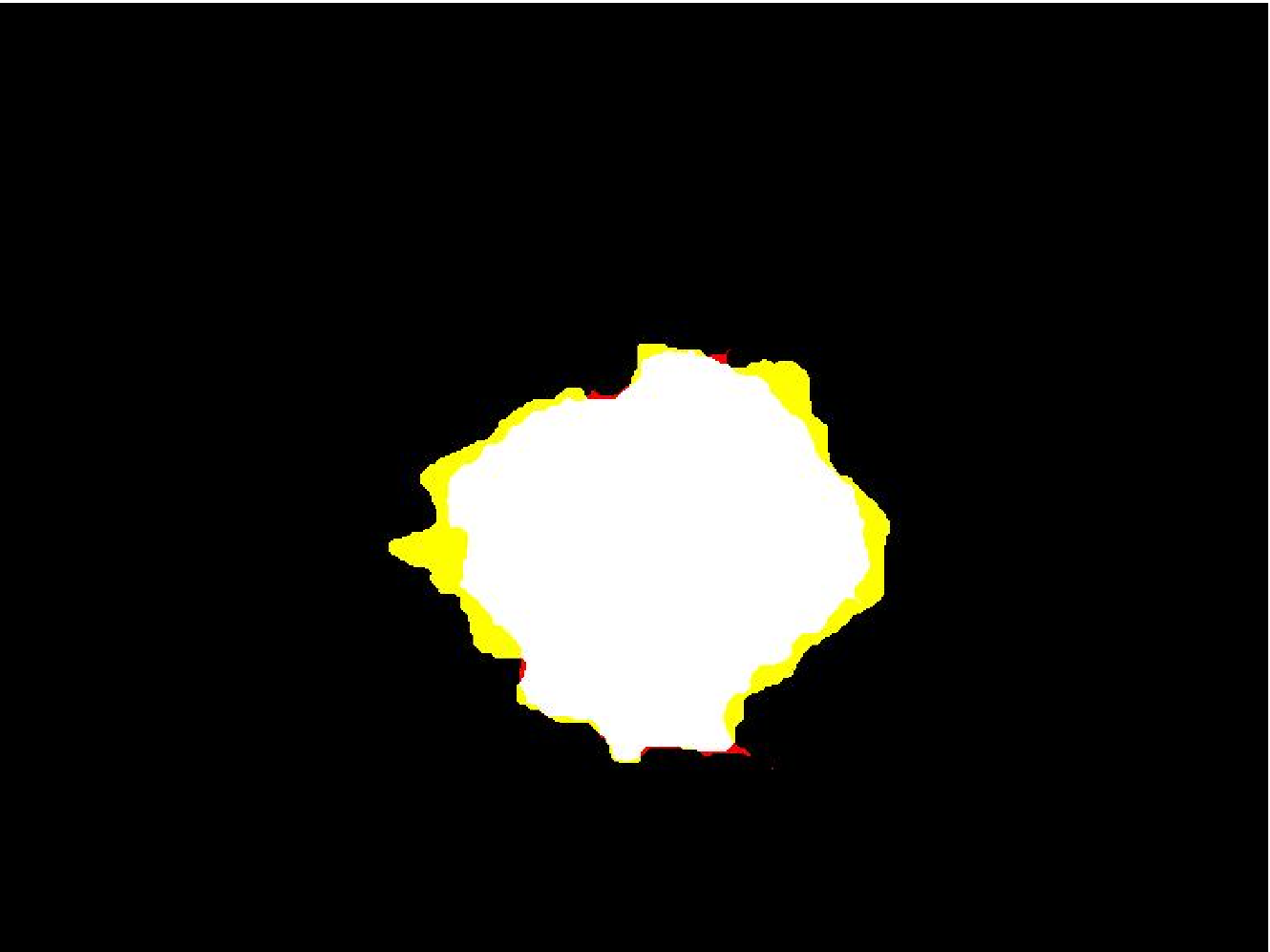} \\
    \end{tabular}}
    \caption{Sample visual results of our MKIS-Net and three alternative networks on the PH2 dataset. From left to right: the input images, the gold standard vessel maps manually annotated by an expert, and the results generated by MobileNet, M2U-Net, ERFNet, and the proposed MKIS-Net.}
    \label{visualSKIN1}
\end{figure}

\begin{figure}[!t]
    \centering
    \resizebox{1\textwidth}{!}{%
    \begin{tabular}{@{}c@{\ }c@{\ }c@{\ }c@{\ }c@{\ }c@{}}
        \includegraphics[width=0.14\textwidth]{Chest/Oimg4.eps} &
        \includegraphics[width=0.14\textwidth]{Chest/gt4.eps} &
        \includegraphics[width=0.14\textwidth]{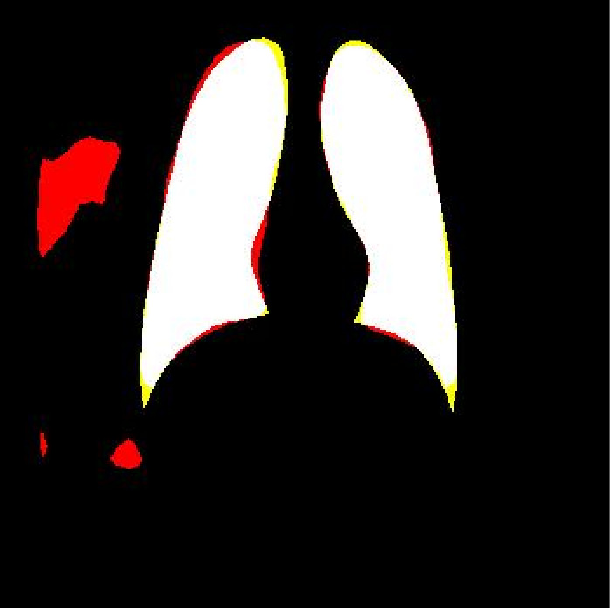} &
        \includegraphics[width=0.14\textwidth]{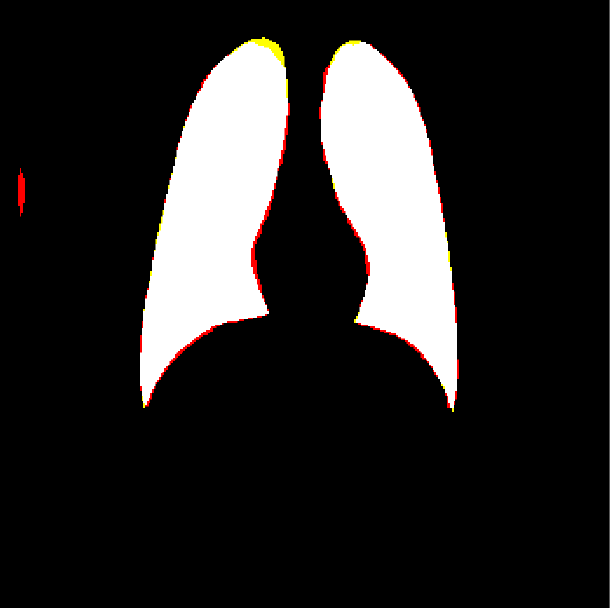} &
        \includegraphics[width=0.14\textwidth]{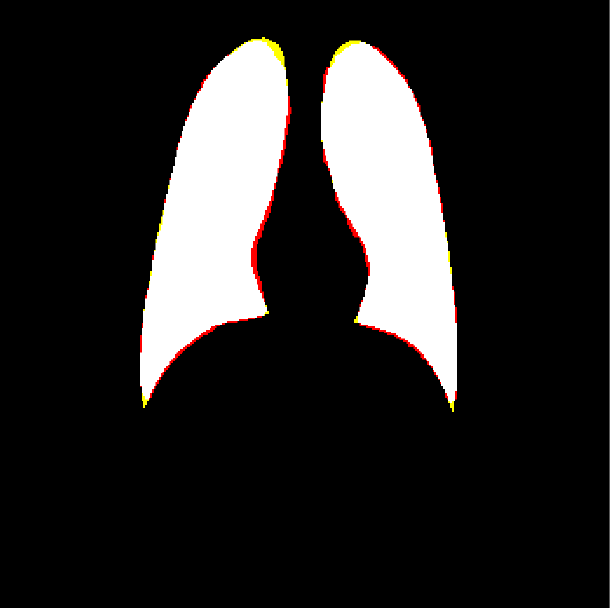} &
        \includegraphics[width=0.14\textwidth]{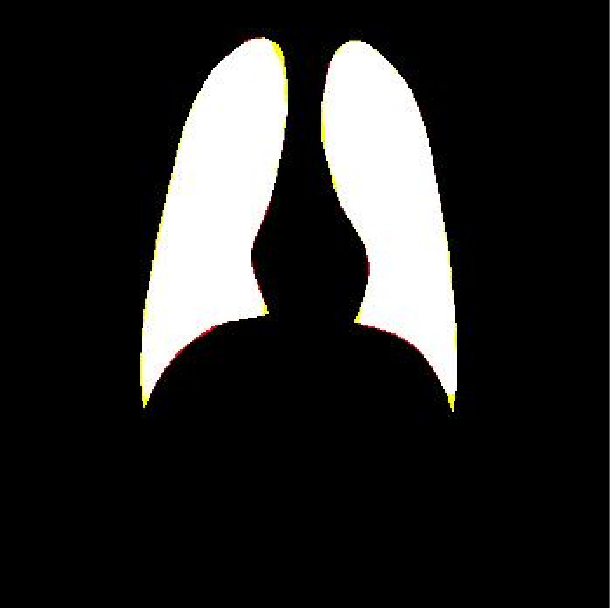} \\
        \includegraphics[width=0.14\textwidth]{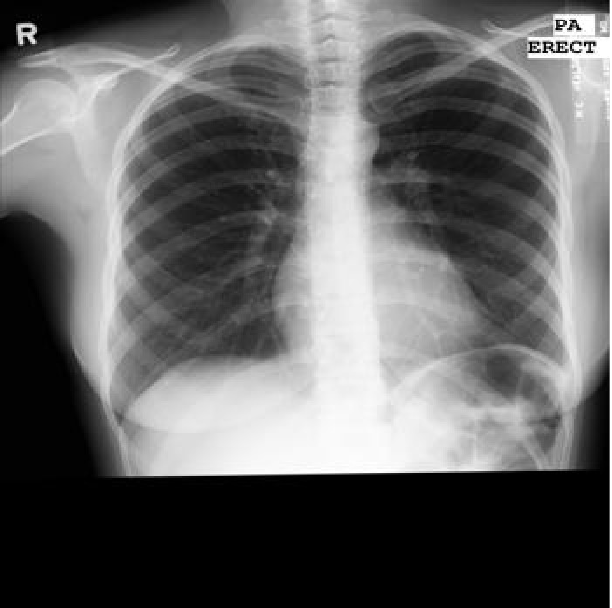} &
        \includegraphics[width=0.14\textwidth]{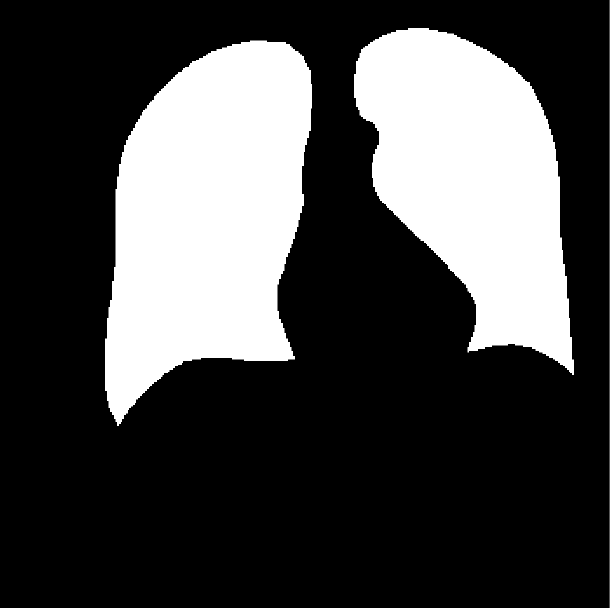} &
        \includegraphics[width=0.14\textwidth]{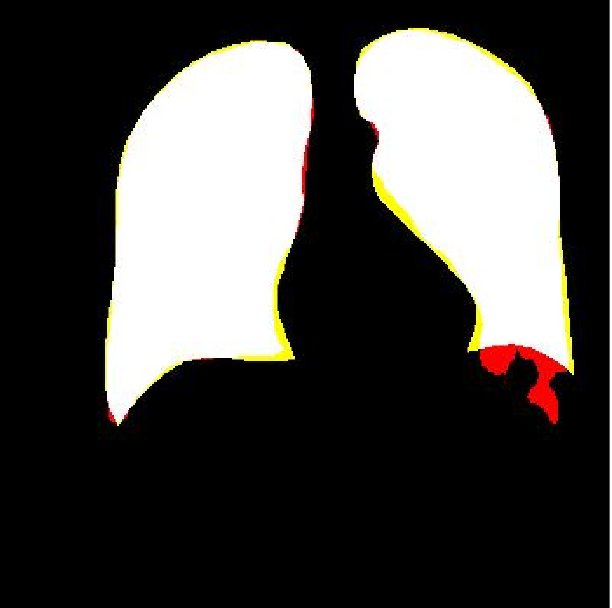} &
        \includegraphics[width=0.14\textwidth]{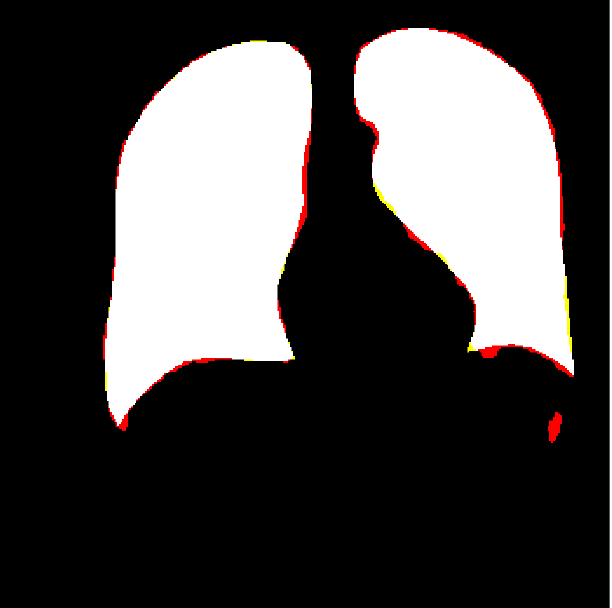} &
        \includegraphics[width=0.14\textwidth]{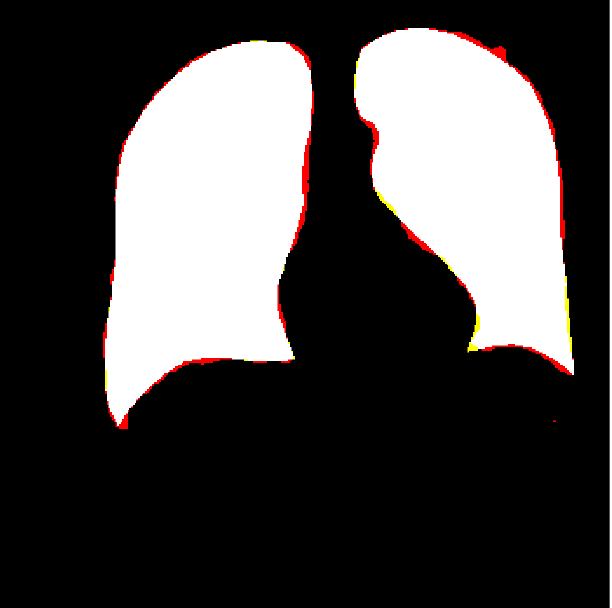} &
        \includegraphics[width=0.14\textwidth]{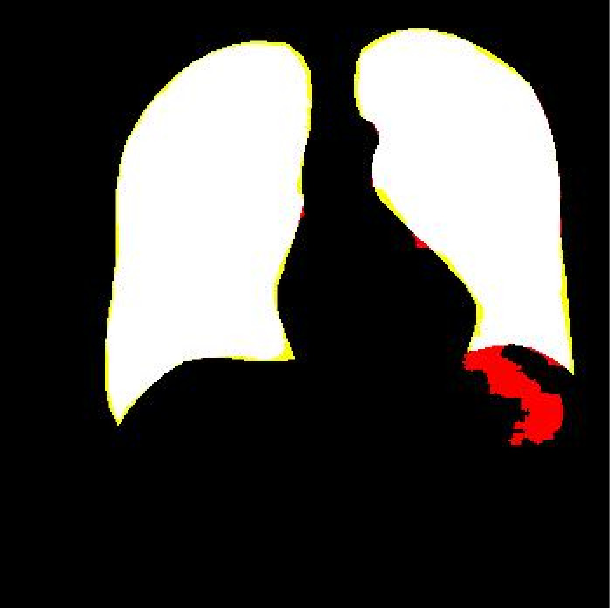} \\
        \includegraphics[width=0.14\textwidth]{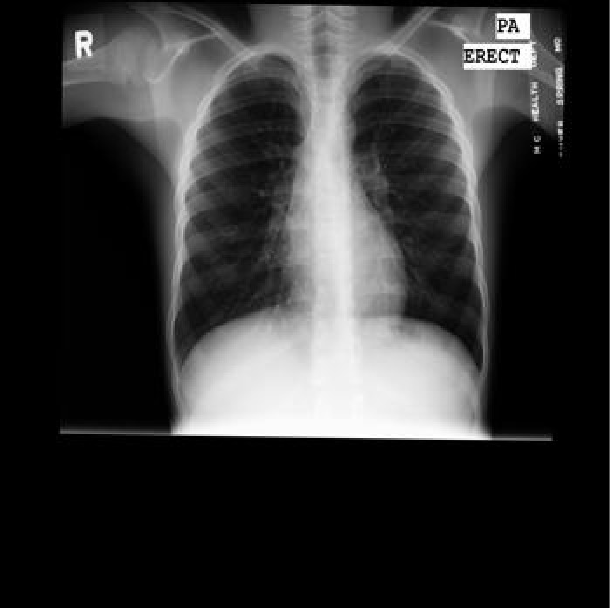} &
        \includegraphics[width=0.14\textwidth]{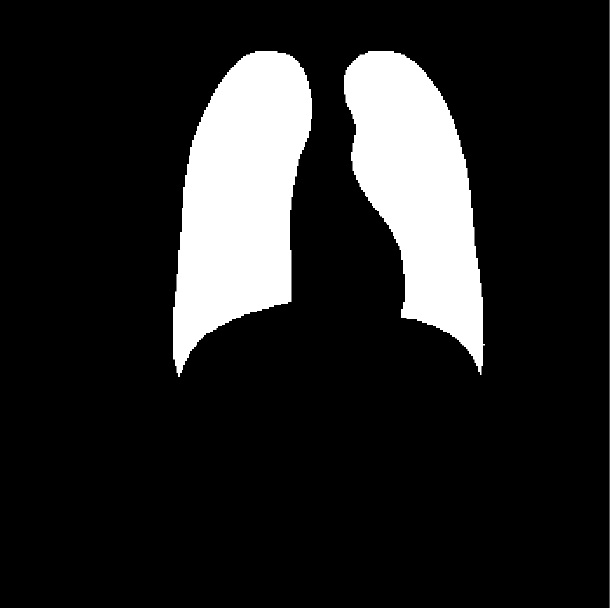} &
        \includegraphics[width=0.14\textwidth]{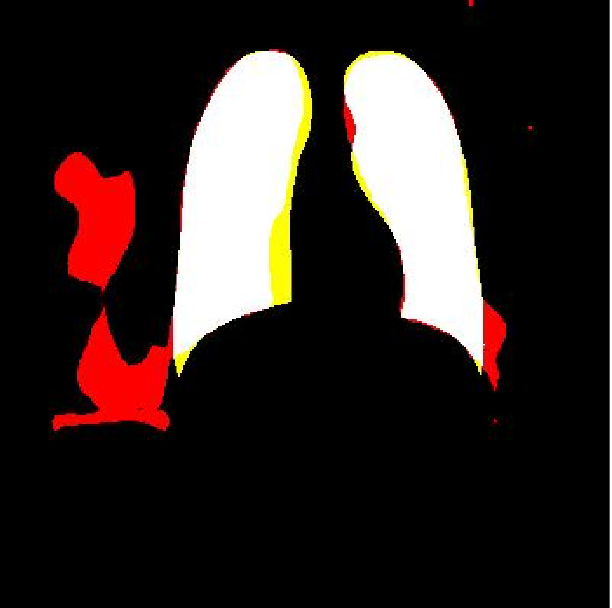} &
        \includegraphics[width=0.14\textwidth]{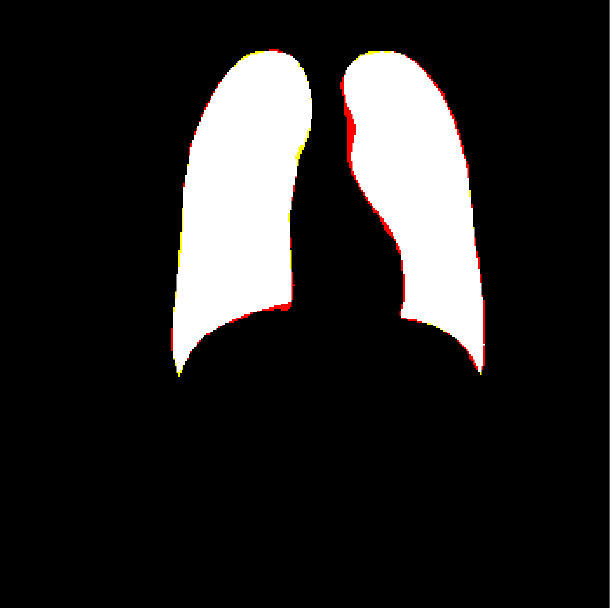} &
        \includegraphics[width=0.14\textwidth]{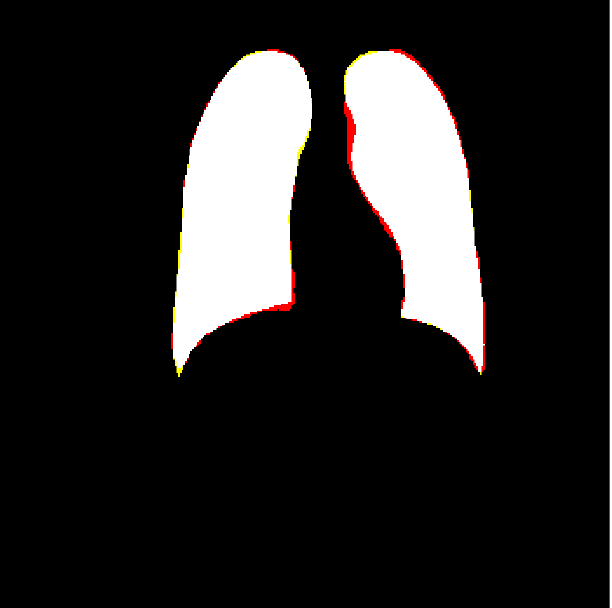} &
        \includegraphics[width=0.14\textwidth]{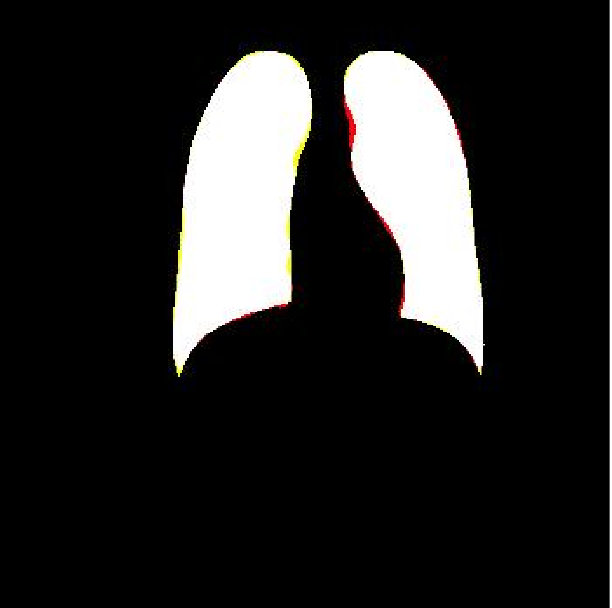} \\
    \end{tabular}}
    \caption{Sample visual results of our MKIS-Net and three alternative networks on the MC dataset. From left to right: the input images, the gold standard vessel maps manually annotated by an expert, and the results generated by MobileNet, M2U-Net, ERFNet, and the proposed MKIS-Net.}
    \label{visualCHESTE}
\end{figure}

Furthermore, we compared MKIS-Net to MobileNet-V3-Small \cite{Howard_2019_ICCV}. In this paper we have so far focused on benchmarks and state-of-the-art segmentation methods primarily used in the medical image segmentation literature. To our knowledge, however, MobileNet-V3-Small has not been applied to medical imaging before. We sequentially present sample visual results (Figs.\ \ref{visualDRIVE1} and \ref{visualCHASE1}) of MKIS-Net and MobileNet-V3-Small on the DRIVE and CHASE datasets, a quantitative comparison of MKIS-Net with MobileNet-V3-Small, T-Net, M2U-Net, and ERFNet on the PH2 dataset (Table~\ref{skin}) with sample visual results yielded by MKIS-Net, MobileNet-V3-Small, M2U-Net, and ERFNet (Fig.\ \ref{visualSKIN1}), and a quantitative comparison of MKIS-Net with MobileNet-V3-Small, M2U-Net, and ERFNet on the MC dataset (Table~\ref{ChestE}) with sample visual results yielded by these networks (Fig.\ \ref{visualCHESTE}). Overall, we conclude that MKIS-Net and T-Net performed comparably well on the DRIVE and CHASE datasets, while MKIS-Net performed better than T-Net on the PH2 and MC datasets.

\begin{table}[!t]
  \centering
    \resizebox{0.9\textwidth}{!}{
    \begin{tabular}{cccc}
    \toprule
    \textbf{Method } & \textbf{F1 } & \textbf{Jacc} & \textbf{Params (M)  } \\
    \midrule
    SCDRR \cite{Bozorgtabar2016} & 0.8600    & 0.7600& N/A\\
     MSCA \cite{7493448} & 0.8157 & 0.7233&N/A  \\
    JCLMM \cite{ROY2017160} & 0.8285 & - & N/A \\
    FCN+BPB+SBE \cite{9157193}& 0.9184 & 0.8430& 8\\
    Multistage FCN\cite{7942129}   & 0.9066 & 0.8399& 10\\
    T-Net\cite{khan2022t} & 0.9282   &   0.8696   & \textbf{0.03}\\
    MKIS-Net (Proposed) & \textbf{0.9301} & \textbf{0.8707}&0.152 \\
    \bottomrule
    \end{tabular}}
  \caption{Results of MKIS-Net and other state-of-the-art nets on the PH2+ISBI 2016 dataset. Bold indicates best performance.}
  \label{PH2+ISBI}
\end{table}

\begin{table}[!t]
  \centering
    \resizebox{1\textwidth}{!}{
    \begin{tabular}{ccccc}
    \toprule
    \textbf{Method} & \textbf{Acc} & \textbf{Jaccard} & \textbf{F1} & \textbf{Params(M)}\\
    \midrule
   BN \cite{8116756} & 0.7700  & N/A    & N/A& N/A \\
     MLP \cite{8116756} & 0.7900  & N/A    & N/A& N/A \\
     RF \cite{8116756} & 0.8100  & N/A    & N/A& N/A \\
     Vote \cite{8116756} & 0.8300  & N/A    & N/A& N/A \\
    Souza et al. \cite{Souza2019} & 0.9697 & 0.8870 & 0.9697& N/A \\
    X-RayNet-2 \cite{jcm9030871} & 0.9872 & 0.9496 & 0.9740& 2.39 \\
    X-RayNet-1 \cite{jcm9030871} & \textbf{0.9911} & \textbf{0.9636} & \textbf{0.9814} & 9.2 \\
    MKIS-Net (Proposed) & 0.9908 & 0.9627 & 0.9810 & \textbf{0.152} \\
    \bottomrule
    \end{tabular}}
  \caption{Results of MKIS-Net and other state-of-the-art nets on the MC dataset. Bold indicates best performance.}
  \label{CHEST}
\end{table}

\section{Conclusions}
In this paper, we have presented MKIS-Net, a small convolutional neural network for medical image segmentation compared to alternatives in the scientific literature. It employs a multiscale kernel structure to produce an effective receptive field and improved segmentation efficiency. Our network is ideal for devices with limited resources while providing support for high-resolution medical image datasets. We have illustrated its utility for retinal vessel, skin lesion, and chest X-ray segmentation tasks. As the experimental results show, MKIS-Net is quite competitive, often outperforming much larger, state-of-the-art networks. Our network also outperforms other light-weight network alternatives.

\begin{table*}[!t]
\centering
\begin{tabular}{cccccccccc}
\hline
               &               &                 & \multicolumn{3}{c}{\textbf{DRIVE}}            &  & \multicolumn{3}{c}{\textbf{CHASE}}             \\
\cline{4-6}\cline{8-10}
\textbf{Model} & \textbf{Size (MB)} & \textbf{Params (M)} & \textbf{MAdds (B)} & \textbf{F1} & \textbf{Acc} &  & \textbf{MAdds (B)} & \textbf{F1} & \textbf{Acc}  \\
\hline
ERFNet   & 8.0    & 2.06  & 2.8   & 0.8022   & 0.9598    &  & 51.3    & 0.7994   & 0.9716      \\
M2U-Net  & 2.2    & 0.55  & 1.4   & 0.8091   & 0.9627    &  & 4.4     & 0.8006   & 0.9446      \\
MKIS           & \textbf{0.6}         & \textbf{0.152}    & \textbf{0.05}   & \textbf{0.8283}        & \textbf{0.9697}       &  & \textbf{0.05}          & \textbf{0.8137}        & \textbf{0.9740}        \\
\hline
\end{tabular}
\caption{Comparison of the results yielded by MKIS-Net with those delivered by alternative light-weight networks on the DRIVE and CHASE datasets. Best results are highlighted in bold.}
\label{lightWeight_perf}
\end{table*}

\begin{table}[!t]
\centering
\resizebox{1\textwidth}{!}{%
\begin{tabular}{cccc}
\multicolumn{1}{l}{} & \multicolumn{1}{l}{} & \multicolumn{1}{l}{} & \multicolumn{1}{l}{}  \\
\hline
\textbf{Method}   & \textbf{F1}  & \textbf{Jacc}  & \textbf{Params (M)} \\
\hline
MobileNet-V3-Small \cite{Howard_2019_ICCV} & 0.8675 & 0.7762  & 0.47  \\
M2U-Net \cite{Laibacher_2019}   &   0.9179 &0.8527 & 0.55  \\
ERFNet \cite{Romera_2018}   &  0.9083   &   0.8370   &   2.06   \\
T-Net\cite{khan2022t}  &  0.9282   &   \textbf{0.8696}   &   \textbf{0.03}   \\
MKIS-Net   & \textbf{0.9301} & 0.8607 & 0.152    \\
\hline
\end{tabular}}
\caption{Comparison of MKIS-Net with alternative light-weight networks on the PH2 dataset. Best results are highlighted in bold.}
\label{skin}
\end{table}

\begin{table}[!t]
\centering
\resizebox{1\textwidth}{!}{%
\begin{tabular}{cccccc}
\multicolumn{1}{l}{} & \multicolumn{1}{l}{} & \multicolumn{1}{l}{} & \multicolumn{1}{l}{} & \multicolumn{1}{l}{} & \multicolumn{1}{l}{}  \\
\cline{1-5}
\textbf{Method}      & \textbf{Acc}         & \textbf{Jacc}        & \textbf{F1}        & \textbf{Params (M)}  &                       \\
\cline{1-5}
MobileNet-V3-Small \cite{Howard_2019_ICCV}   & 0.9826                & 0.9236                & 0.9576               & 0.47                 &                       \\
M2U-Net \cite{Laibacher_2019}  & 0.9906     &     0.9615                 &            0.9803          & 0.55                  &                       \\
ERFNet  \cite{Romera_2018}             &     0.9900                 &   0.9590                   &      0.9791                &    2.06                  &                       \\
MKIS-Net             & \textbf{0.9908}               & \textbf{0.9627}               & \textbf{0.981}                & \textbf{0.152}                &                       \\
\cline{1-5}
\end{tabular}
}
\caption{Comparison of MKIS-Net with alternative light-weight networks on the MC dataset. Best results are highlighted in bold.}
\label{ChestE}
\end{table}

\section*{Acknowledgment}
The work presented here was done while Antonio Robles-Kelly was with Deakin University, Waurn Ponds, Australia.

\bibliographystyle{elsarticle-num}
\bibliography{egbib}

\end{document}